\documentclass{aa}  
\usepackage{xcolor}
\usepackage{graphicx}
\usepackage{multirow}
\usepackage{subcaption}
\usepackage{caption}
\usepackage{comment}
\usepackage{threeparttable}
\usepackage{array}
\usepackage{placeins}

\usepackage{ulem}
\usepackage{txfonts}
\usepackage{hyperref}
\newcommand{\Lya}{\mbox{Ly-$\alpha$}}
\begin{document} 
 
   \title{Eruptive events with exceptionally bright emission in \ion{H}{i}~\Lya\;observed by the Metis coronagraph} 

    \author{
        G. Russano\inst{\ref{INAF-OAC}}
        \and
        V. Andretta\inst{\ref{INAF-OAC}}
        \and
        Y. De Leo\inst{\ref{MPS},\ref{UniCt}}
        \and
        L. Teriaca\inst{\ref{MPS}}
        \and 
        M. Uslenghi\inst{\ref{INAF-IASF}}
        \and
        S. Giordano\inst{\ref{INAF-OATo}}
        \and
        D. Telloni\inst{\ref{INAF-OATo}}
         \and
        P. Heinzel\inst{\ref{CAS},\ref{UniWro}}
        \and
        S. Jej\v ci\v c\inst{\ref{CAS},\ref{UniLj1},\ref{UniLj2}}
        \and
        L. Abbo\inst{\ref{INAF-OATo}}
        \and
        A. Bemporad\inst{\ref{INAF-OATo}}
        \and
        A. Burtovoi\inst{\ref{INAF-OAA}}
        \and
        G. E. Capuano\inst{\ref{UniCt},\ref{INAF-OACt},\ref{CNR-IMM}}
        \and 
        F. Frassati\inst{\ref{INAF-OATo}}
        \and
        S. Guglielmino\inst{\ref{INAF-OACt}}
        \and
        G. Jerse\inst{\ref{INAF-OATs}}
        \and
        F. Landini\inst{\ref{INAF-OATo}}
        \and
        A. Liberatore\inst{\ref{JPL}}
        \and
        G. Nicolini\inst{\ref{INAF-OATo}}
        \and
        M. Pancrazzi\inst{\ref{INAF-OATo}}
        \and
        P. Romano\inst{\ref{INAF-OACt}}
        \and
        C. Sasso\inst{\ref{INAF-OAC}}
        \and
        R. Susino\inst{\ref{INAF-OATo}}
        \and
        L. Zangrilli\inst{\ref{INAF-OATo}}
        \and
        V. Da Deppo\inst{\ref{CNR-IFN},\ref{INAF-OAPd}}
        \and
        S. Fineschi\inst{\ref{INAF-OATo}} 
        \and
        C. Grimani\inst{\ref{UniUrb},\ref{INFN}} 
        \and
        J.D. Moses\inst{\ref{NASAH}}
        \and
        G. Naletto\inst{\ref{UniPd}}
        \and
        M. Romoli\inst{\ref{UniFi}}
        \and
        D. Spadaro\inst{\ref{INAF-OACt}}
        \and
        M. Stangalini\inst{\ref{ASI}}
.}
   \institute{
    National Institute for Astrophysics (INAF), Astronomical Observatory of Capodimonte, Napoli, Italy\\
              \email{giuliana.russano@inaf.it}
              \label{INAF-OAC}
        \and
        Max-Planck-Institut f\"ur Sonnensystemforschung, Göttingen, Germany
        \label{MPS}
        \and
        Universit\`a di Catania – Dip. Fisica e Astronomia “Ettore Majorana”, Catania, Italy
        \label{UniCt}
        \and
        INAF – Istituto di Astrofisica Spaziale e Fisica Cosmica, Milan, Italy
        \label{INAF-IASF}
        \and
        National Institute for Astrophysics, Astrophysical Observatory of Torino, Pino Torinese (TO), Italy
        \label{INAF-OATo}
        \and
        Astronomical Institute of the Czech Academy of Sciences, Ond\v{r}ejov, Czech Republic
        \label{CAS}
        \and
        University of Wroc{\l}aw, Center of Scientific Excellence - Solar and Stellar Activity, Wroc{\l}aw, Poland
        \label{UniWro}
        \and
        Faculty of Education, University of Ljubljana, Slovenia
        \label{UniLj1}
        \and
        Faculty of Mathematics and Physics, University of Ljubljana, Slovenia
        \label{UniLj2}
        \and
        National Institute for Astrophysics, Astrophysical Observatory of Arcetri, Firenze, Italy
        \label{INAF-OAA}
        \and
        National Institute for Astrophysics, Astrophysical Observatory of Catania, Catania, Italy
        \label{INAF-OACt}
        \and
        Institute for Microelectronics and Microsystems (CNR-IMM), Catania, Italy
        \label{CNR-IMM}
        \and
        National Institute for Astrophysics, Astronomical Observatory of Trieste, Trieste, Italy
        \label{INAF-OATs}
        \and
        Jet Propulsion Laboratory, California Institute of Technology, Pasadena, CA 91109, USA.
        \label{JPL}
        \and
        Institute for Photonics and Nanotechnologies (CNR-IFN), Padova, Italy
        \label{CNR-IFN}  
        \and
        National Institute for Astrophysics, Astrophysical Observatory of Padova, Padova, Italy 
        \label{INAF-OAPd}
        \and
        University of Urbino, Dipartimento di Scienze Pure e Applicate, Urbino, Italy
        \label{UniUrb}
        \and
        National Institute for Nuclear Physics (INFN), Section in Florence, Sesto Fiorentino (FI), Italy
        \label{INFN}
        \and
        NASA Headquarters, Washington DC 20546-0001, USA
        \label{NASAH}
        \and
       University of Padova, Department of Physics and Astronomy ``Galileo Galilei'', Padova, Italy
        \label{UniPd}
        \and
        University of Florence, Sesto Fiorentino (FI), Italy
        \label{UniFi}
        \and
        Italian Space Agency, Roma, Italy
        \label{ASI}
             }
   \date{\today}

 
  \abstract
   {Ultraviolet emission from coronal mass ejections can provide information on the evolution of plasma dynamics, temperature and elemental composition, as demonstrated by the UV Coronagraph Spectrometer (UVCS) on board the SOlar and Heliospheric Observatory (SOHO). Metis, the coronagraph on board Solar Orbiter (SolO), provides for the first time coronagraphic imaging in the ultraviolet \ion{H}{i}~\Lya\ line and, simultaneously, in polarized visible light, thus providing a host of information on the properties of coronal mass ejections and solar eruptions like their overall dynamics, time evolution, mass content, and outflow propagation velocity in the expanding corona.}
   {We analyzed in this work six coronal mass ejections observed by Metis between April and October 2021, which are characterized by a very strong \ion{H}{i}~\Lya\ emission. We studied in particular the morphology, kinematics, and the temporal and radial evolution of the emission of such events, focusing on the brightest UV features.}
   {The kinematics of the eruptive events under consideration were studied by determining the height-time profiles of the brightest parts on the Metis plane of the sky.  Furthermore, the 3D position in the heliosphere  
   of the coronal mass ejections were determined by employing co-temporal images, when available, from two other coronagraphs: LASCO/C2 onboard SOHO, and COR2 onboard STEREO-A. In three cases, the most likely source region on the solar surface could be identified. Finally, the radiometrically calibrated Metis images of the bright UV features were analyzed to provide estimates of their volume and density. From the kinematics and radiometric analyses, we obtained indications of the temperatures of the bright UV cores of these events. These results were then compared with previous studies with the UVCS spectrocoronagraph.}
   {The analysis of these strong UV-emitting features associated with coronal mass ejections demonstrates the capabilities of the current constellation of space coronagraphs, Metis, LASCO/C2, and COR2, in providing a complete characterization of the structure and dynamics of eruptive events in their propagation phase from their inception up to several solar radii. Furthermore, we show how the unique capabilities of the Metis instrument to observe these events in both \ion{H}{i} Ly-$\alpha$ line and polarized VL radiation allow plasma diagnostics on the thermal state of these events. 
   }
   {}
   
   \maketitle
   
\section{Introduction}
\label{intro}

Coronal Mass Ejections (CMEs) are large-scale, often spectacular, eruptions of mass and magnetic flux propagating from the solar surface through the corona and into interplanetary space \citep[see review by][]{Chen_CME_review_2011LRSP....8....1C,Webb_CME_review_2012LRSP....9....3W}. The bulk of information currently available about CMEs comes from coronagraphs working in visible light (VL) and measuring radiation from the so-called K-corona, i.e. emission due to Thomson scattering of photospheric radiation by coronal electrons.  Therefore, VL coronagraphs mainly provide the evolution of the plasma electron density and kinematic properties. On the other hand, ultraviolet (UV) spectral line emission from CMEs has been studied in great detail for many events thanks to the UV Coronagraph Spectrometer \citep[UVCS,][]{Kohl_UVCS_1995SoPh..162..313K} on board the SOlar and Heliospheric Observatory \citep[SOHO,][]{SOHO_1995SoPh..162....1D}. 
UVCS observations provided information on CME plasma properties like temperature, flow velocities and thermal energy content. The UVCS instrument field-of-view (FoV) could cover the same
height range as the coronagraph LASCO-C2 \citep{LASCO_1995SoPh..162..357B} on board SOHO (out to $\sim$\,10\,R$_{\sun}$), but it was limited to the spectrometer entrance slit: see Fig.\,3 from \cite{Giordano_UVCS_2013}. Nevertheless, a wealth of new findings were possible through those observations in the UV. 
A comprehensive review of CME observations with UVCS can be found in \cite{kohl_2006}, and a catalog is provided by \cite{Giordano_UVCS_2013}.

Metis instrument, on board the Solar Orbiter (SolO) spacecraft \citep{SolO_2020A&A...642A...1M}, on the other hand, is the first coronagraph capable of imaging the solar corona simultaneously in both VL broadband (580-640\,nm) and UV narrow band \ion{H}{i}~\Lya\;at 121.6\,$\pm$\,10\,nm, from $\rm 1.7\; R_{\sun}$ to $\rm 9\; R_{\sun}$ depending on the spacecraft
heliocentric distance, with high-spatial resolution and time cadence \citep[see][]{metis_instr}. The combination of images in the two channels allows the investigation of the thermodynamic evolution of CMEs in their propagation into the solar corona, providing information about its physical conditions like temperatures, densities and flow velocity \citep{first_light_2021A&A...656A..32R,Bemporad_second_cme_2022A&A...665A...7B,Telloni_solar_wind_2021ApJ...920L..14T,Capuano_2021A&A...652A..85C}. 

During the Solar Orbiter cruise phase, Metis carried out long stretches of synoptic observations, at the beginning of Solar Cycle 25. During those synoptic programs, several CMEs were detected, some of which appeared to be associated with relatively compact  features characterized by strong \ion{H}{i}~\Lya\ emission. 
Six of such eruptive events were observed on April 25$^\mathrm{th}$, September 11$^\mathrm{th}$, October 2$^\mathrm{nd}$, October 25$^\mathrm{th}$, and October 28$^\mathrm{th}$ 2021.  No other events with similarly bright UV emission were observed by Metis during that orbital phase nor, in fact, in the initial phase of the nominal mission, until the end of 2022.
Those bright features were associated with eruptive prominences. 

Prominences, also named filaments when observed on the disk, start as relatively cold and dense objects embedded in the hotter corona. When erupting, part of their material appears to escape the solar gravitational field with a strong radial component of motion away from the
solar surface and the remaining part falling back down. In its outward motion, the prominence material becomes fainter in H-$\alpha$ as it is ionized and thus becomes visible in coronal observations in white light, dominated by Thomson scattering of photospheric radiation by prominence electrons, even at distances greater than 5 solar radii
\citep{Chen_CME_review_2011LRSP....8....1C,Webb_CME_review_2012LRSP....9....3W,Bemporad_filament_rotation_2011A&A...531A.147B,Vourlidas_rotating_cme_2011ApJ...733L..23V}.
The associated CMEs often can be interpreted as an erupting flux rope system that displays the typical three-part morphology, where a bright front loop is immediately followed by a dark cavity with an embedded bright core \citep{Illing_classic_cme_structure_1985JGR....90..275I}. The three-part structure is considered to be the standard morphology for CME, although coronagraph observations indicate that only 30$\%$ of CME events possess all the three parts while more than 70$\%$ of large CMEs contain the bright core component \citep{Webb_CME_3part_1987SoPh..108..383W}. This last bright core is often identified as the cold and dense eruptive prominence material, although it is difficult to prove such an association from white-light coronal observations alone. However, recently \cite{Howard_cme_structure_2017ApJ...834...86H} presented the case that the inner core of the three-part coronagraph CME in the most common cases is not a filament, which is rarely observed at large distances from the Sun, but a consequence of flux rope launch or part of the flux rope structure.

Given the present status of our knowledge of CMEs, the study of the eruptive events for the first time observed simultaneously in white light and UV \Lya\ images can be relevant to the general understanding and interpretation of the physics of such phenomena. Our focus on the morphology and kinematics of the UV bright eruptive events can provide constraints or even be the driver for the development of theoretical ideas on CME events.

In this work, Sec.\,\ref{sec:data_calib} is dedicated to the description of the instrument and to the main steps of Metis data processing and calibration. In Sec.\,\ref{sec:observations}, we proceed to describe the morphology and kinematics of each event. In Sec.\,\ref{sec:discussion} we discuss the general features of UV emission in solar eruptions as seen by Metis and compare the results with UVCS observations, finally presenting our conclusions in Sec.\,\ref{sec:summary}.

\section{Metis instrument and data processing}
\label{sec:data_calib}

\subsection{Description of the instrument and of onboard acquisition modes}

Metis is an externally occulted coronagraph, consisting in a single on-axis gregorian telescope which feeds two channels operating in the VL and the UV \citep{metis_instr,Fineschi_Metis_2020ExA....49..239F}. 
The optical paths of the two channels are separated by an interference filter tuned at 121.6 nm that reflects the visible light and transmits the UV light. The UV detector assembly (UVDA) is an intensified unit formed by a micro-channel plate coupled to an active pixel sensor via a fiber optic taper.

The detector assemblies at the focal planes of both channels, the visible light detector assembly (VLDA) and the UVDA, can be operated in parallel. An acquisition session typically involves the acquisition by both detectors of a number of individual images exposed for a given Detector Integration Time (DIT); a programmable number, NDIT, of images can also be averaged on board to create a single image whose effective exposure time is the product of DIT times NDIT. The process is repeated with a given cadence and for a given duration, producing therefore a pre-set number of images that are then transferred to ground for further processing.  The definition of an acquisition session also includes other parameters such as binning and compression factors; once again these parameters can be set independently for the two detectors.
Based on the scientific observing mode, the detector exposure time and relevant parameters are configured to obtain different spatial and time resolutions, depending on the scientific goal.

In addition, the VLDA is designed to measure the linear polarization of the solar corona brightness \citep{DeLeo_VL_2023A&A...676A..45D, metis_instr}. The VLDA acquires a polarimetric sequence $k$ by rotating its linear polarizer over four predefined angles. Thus, an observing sequence consists of a series of $P_{i,k}$ images where for each $k$, $i$ goes from 1 to 4. In general, if averaging over NDIT acquisitions (NDIT=1 meaning no averaging) is required, then images within $k$ and $k+$\,NDIT with the same $i$ are averaged together with a cadence given by the time difference between $P_{i,k}$ and $P_{i,k+\mathrm{NDIT}}$ resulting in an effective exposure time of NDIT\,$\times$\,DIT for each polarization angle, where DIT indicate the acquisition duration of a single image.

\subsection{Data processing}

Once the data have been transmitted to ground, the data processing pipeline is applied, which includes standard operations such as bias and dark subtraction, flat-field and vignetting corrections, and exposure-time normalization \citep[see][]{first_light_2021A&A...656A..32R}.

In the case of the UV channel, a further dark correction is performed. As described in \cite{Andretta_2021_first_CME}, at the time of the observations analyzed here the UV detector was subject to a transient effect at the beginning of each acquisition sequence \footnote{A software update in April 2022 has largely resolved this problem.}. 
Such a transient affects the first frames of each acquisition. If NDIT\,=\,1, only the first images are influenced and can be discarded after being received on ground. For NDIT\,>\,1 (on board averaging) the transient is restarted at the beginning of each group of NDIT frames, thus all the resulting images will be affected by an amount that depends on NDIT\,$\times$\,DIT (the longer NDIT\,$\times$\,DIT the smaller the effect). Since the effect depends on the combination of parameters used (DIT, NDIT and cadence), acquisitions of sets of dark frames have been planned with the same parameters used for science observations. The images have been then dark corrected using the dark frames acquired closer in time, among those with appropriate parameters. However, we noticed variations in the dark level on short time scale, so darks acquired even a few hours later may not properly correct the images. 

Analyzing these variations on dark image sets, we found that they are not consistent with neither offset nor purely gain variations. Instead, assuming a linear combination of offset and gain variation, it is possible to satisfactorily reproduce these fluctuations. We considered the average values in 5 boxes arranged in the darkened areas of the detector (the 4 corners and the occulted center). In fact, due to the uneven distribution of the dark current on the sensor, these boxes provide constraints on a high dynamic range. Thus, it is possible to estimate the function to be applied to the reference dark to minimize residuals for those areas, and the result is very similar with respect to minimizing residuals over the entire frame. Since these boxes do not contain signals during observations, they have been used as reference to compute the optimal dark subtraction for each scientific image.

\subsection{Data calibration}

The in-flight radiometric calibration of the two channels of Metis was performed by means of observations of selected calibration stars. Stars with known and stable fluxes play a crucial role in determining the instrumental response and monitoring its evolution over time. 

In particular, calibration stars passing through the FoV allow inferring the radiometric calibration coefficient, which converts the acquired frames (level 0 or L0 data) in Digital Numbers (DNs) into calibrated frames (level 2, or L2 data) expressed in radiance ($\rm photons\; cm^{-2}\,\; sr^{-1}\; s^{-1}$).

For both channels, after pre-processing the frames as described in the previous section, it is possible to obtain the radiometric calibration coefficient $\epsilon_{ch}$ by using the general inversion formula, as follows:
\begin{equation}
\epsilon_{ch}=\frac{N_*\,{\rm(FoV)}}{\overline{f_*}\cdot\ A_{pup}\cdot\,VF{\rm(FoV)}} 
\label{inv_form}
\end{equation}
where $\rm N_*$ is the star count rate (in DN/s), $\overline{f_*}$ is the averaged flux of the star in the bandpass of the considered channel, $A_{pup}$ is the pupil area of the telescope, and  $\rm VF$ is the vignetting function that depends on the star position across the frame.

Due to its band pass, the VL channel can be assimilated to a non-standard red filter. In order to find the correct stellar flux $\overline{f_*}$ to use in the inversion formula (\ref{inv_form}), it is necessary to determine the conversion factor between the Metis non-standard photometric system and a standard one (i.e. Johnson-Cousins). The resulting radiometric calibration coefficient is in good agreement, within the uncertainties, with the value estimated from the ground calibration campaign, performed before delivering the instrument for integration into the spacecraft. The approach used and the results are described in \cite{DeLeo_VL_2023A&A...676A..45D}.

The stellar observations also proved the uniformity of the spatial response of the VL channel once accounting for the VF (measured during the on-ground Metis calibration activities). However, in the case of the UV channel, the stellar transits reveal that the spatial response is significantly non-uniform even after the correction for the UV VF. Thus, for the UV channel radiometric calibration, the procedure to evaluate the radiometric coefficient requires one more step with respect to the VL channel.\\
Since the Metis heat shield (HS) door is not light-tight, even if closed, a significant amount of both visible and UV light is reflected back into the instrument. By assuming that the ratio of the UV and VL  reflectivities of the HS door is approximately constant over the illuminated area, the ratio of the images obtained with the HS door closed provides an estimate of the spatial variation over the FoV of the relative efficiency of the two channels. 
For instance, the images of the HS door back-reflection indicate a significant difference in response between the east and the west side of the FoV, consistent with the results from the stellar transits. Therefore, it is possible to use the ratio of the UV to VL images obtained with the door closed to correct for the spatial response non-uniformity \citep[see][]{Andretta_2021_first_CME}.

This spatial correction model is applied in addition to the correction by the standard UV vignetting function. After a refinement of the model by using the stellar measurements, it is possible to proceed with the data inversion, using Eq. (\ref{inv_form}). The radiometric calibration of the UV channel, including the correction of the spatial response, is described in detail in \cite{De_Leo_2023b_UV}.

\section{Characterization of the UV-bright eruptions}
\label{sec:observations}

The six events discussed here were observed when Solar Orbiter was at distances greater than 0.5~au from the Sun. While the associated CMEs were rather faint in all but two cases, by contrast, all of them exhibited bright UV structures that were visible well beyond 5 solar radii, in one case up to 9 solar radii in the Metis plane of the sky (PoS), which translates into 23 solar radii when the projection angle is taken into account (see Sec.\,\ref{subsec:morphology_kinematics}). The associated UV features appear to be relatively compact features that allowed us to track their brightest parts through the instrument FoV and, consequently, to estimate their apparent velocity and acceleration on the plane of the sky. 

The events and the associated CMEs were also observed with space-based coronagraphs like LASCO-C2 and/or COR2 respectively onboard SOHO and Solar TErrestrial RElations Observatory \citep[STEREO-A,][]{STEREO_2008SSRv..136....5K}, from different points of view based on their space location. The relative configurations of the spacecraft involved for the events discussed in this work are shown in Fig.~\ref{fig:gse_all_events}. We report on the possible source regions of the eruptions in Appendix\,\ref{sec:appendix_source_region}. 

In all cases, as described in detail below, it was possible to observe features in the images of at least one of the LASCO/C2 and STEREO/COR2 coronagraphs that could be identified as the VL counterparts of the UV features observed by Metis.  It was therefore possible to take advantage of at least two of the three coronagraphs to estimate the 3D position of the selected points in the eruptive features with the triangulation method provided by the routine called \texttt{scc\_measure} available in the Solar Soft library \citep{Thompson_sccmeasure_2009Icar..200..351T}. The routine estimates the Stonyhurst heliographic coordinates of a tie-pointed feature \citep{epipolar_Inhester_2006} when the user locates the same feature (selected with a cursor), in two images acquired almost simultaneously from different views. We accounted for the different light travel times among the various spacecraft and the Sun modifying the observation time of each frame as the UT at Sun's surface. 
The estimated coordinates are then converted to the Carrington coordinate system; the error on the final value was estimated by computing the standard deviation on the average of all the positions selected on the simultaneous images.

In coronagraphy, it is usually convenient to analyze polarized Brightness (from now on pB) images, since they do not include the contribution of F corona, which is mainly due to dust and therefore not polarized, and thus they primarily show the signal due to free electrons in the corona. However, as discussed by \cite{Vourlidas-Howard_2006} and \cite{Howard-DeForest_2012}, at large distances from the Sun and for structures significantly far from the coronagraph's PoS, the polarized signal might not be the most convenient quantity to analyze: total Brightness (\textit{B}) images might provide a better and more contrasted signal. In this work, we will choose to analyze both quantities, expressed in units of mean solar brightness (MSB), that is 4.67$\times$10$^{20}$ ph\,cm$^{-2}$\,s$^{-1}$\,sr$^{-1}$ for the bandpass of the Metis VL channel, choosing to display either or both on a case-by-case basis.  In particular, for the events with angles from the plane of the Metis sky greater than 50\,$^\circ$ (see measured values in Tab.\,\ref{table:events_measured_info} and \ref{table:radial_fit_params}), total brightness images are shown instead because pB weakens as the eruption moves away from the observer's sky plane.

Knowledge of the 3D coordinates of those eruptive features also allowed to ``de-project'' the measured, apparent velocities determined from Metis images, to determine the true radial velocity with respect to the Sun. The same procedure makes it possible to determine the true distance of the observed UV features from the solar surface.

In the following subsections, we describe in detail the kinematical and dynamical characteristics of the individual events. The results of the measurements are summarized in Tab.\,\ref{table:events_measured_info}, and the kinematics is also shown in Fig.\,\ref{fig:polarang_position_all} and will be discussed later in Sec.\,\ref{sec:discussion}. 
An overview of the Metis instrument acquisition settings is reported in Tab.\,\ref{table:events_info}. In addition, all the acquisition times and dates given in the text and figures refer to the mid-time of the acquisition. All images are scaled for different minimum and maximum values to enhance the visibility of the features. Quantitative information about the intensities in the two channels is then reported in Fig.\,\ref{fig:area_br_rsun_all} and in Tab.\,\ref{table:radial_fit_params}.
\begin{figure}
  \centering
   \includegraphics[trim=50 25 275 75,width=\linewidth]{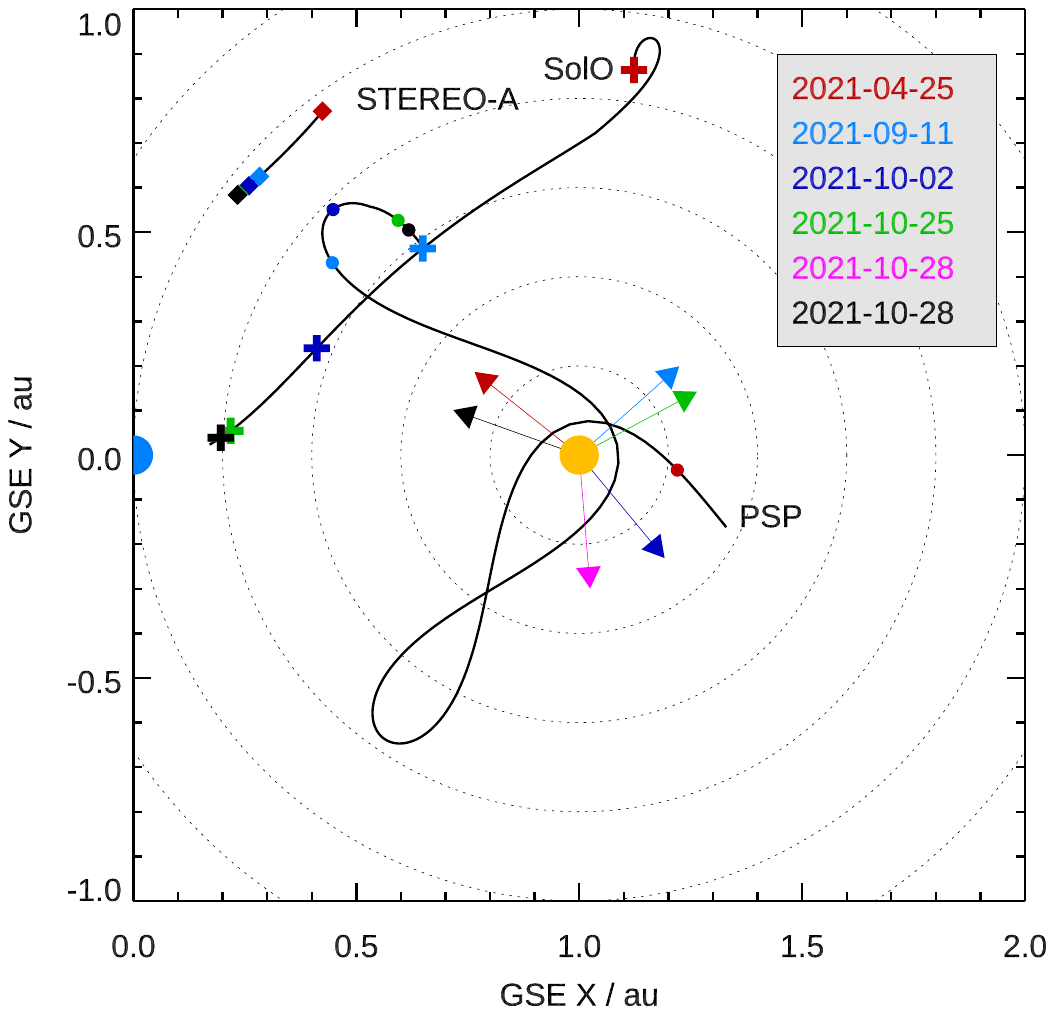}
   \caption{Position of Solar Orbiter, STEREO-A, and near-Earth facilities (including SOHO close to the Earth position at (0,0) coordinates) for all the events considered in this work, in Geocentric Solar Ecliptic (GSE) coordinate systems: In this coordinate system, X is the Earth-Sun line, and Z is aligned with the north pole for the ecliptic. The dots representing Earth (blue) and Sun (yellow) are not in scale.  The arrows show the estimated directions of the eruption as determined by triangulation (see text for more details). The color code used for each event throughout this work is: April 25$^\mathrm{th}$ event in red, September 11$^\mathrm{th}$ in light blue, October 2$^\mathrm{nd}$ in dark blue, October 25$^\mathrm{th}$ in green, October 28$^\mathrm{th}$ (north-west) in magenta, and October 28$^\mathrm{th}$ (south-east) in black.}
   \label{fig:gse_all_events}%
\end{figure}

\begin{table*}
\caption{Observing parameters for each event.}             
\label{table:events_info}      
\centering          
\begin{tabular}{c c c | c c | c c | c c | c c | c c | c c}   %
\hline\hline       
Event & Distance & Annular FoV & \multicolumn{2}{c}{Binning} & \multicolumn{2}{c}{DIT} & \multicolumn{2}{c}{NDIT} &  \multicolumn{2}{c}{T$_{exp}$} & \multicolumn{2}{c}{Cadence} & \multicolumn{2}{c}{Spatial scale}\\ 
date & [au] & range [R$_{\sun}$] & & & \multicolumn{2}{c}{[s]} & & & \multicolumn{2}{c}{[min]} & \multicolumn{2}{c}{[min]} & \multicolumn{2}{c}{[10$^3$ km/px]} \\
\cline{4-15}
 & & & VL & UV & VL & UV & VL & UV & VL & UV & VL & UV & VL & UV \\
\hline                    
   25 Apr. 2021 & 0.87 & 5.3 - 11.2 & 4x4 & 4x4 & 30 & 60 & 15 & 15 & 7.5 & 15 & 30.5 & 16 & 29 & 59 \\  
   11 Sep. 2021 & 0.59 & 3.5 - 7.6 & 2x2 & 4x4 & 30 & 60 & 10 & 1 & 5 & 1 & 24 & 3 & 8.6 & 34 \\
   2 Oct. 2021 & 0.64 & 3.8 - 8.2 & 2x2 & 4x4 & 30 & 60 & 14 & 1 & 5 & 1 & 30 & 2 & 9.4 & 37.7 \\
   25 Oct. 2021 & 0.78 & 4.7 - 10.0 & 2x2 & 4x4 & 30 & 60 & 14 & 1 & 7 & 1 & 30 & 2 & 11 & 46 \\
   28 Oct. 2021 & 0.8 & 4.8 - 10.3 & 2x2 & 4x4 & 30 & 60 & 14 & 1 & 7 & 1 & 30 & 2 & 11.7 & 47.1 \\
\hline                  
\end{tabular}
\tablefoot{For each event we show: the event date; SolO distance from the Sun; annular FoV range of Metis; data acquisition settings for the two Metis channels in terms of binning, detector integration time (DIT), number of images averaged on board (NDIT), exposure time (T$_{exp}$) and cadence, and spatial scale of the images after binning.}
\end{table*}
\begin{table*}
\begin{threeparttable}
\caption{Measured parameters for all the events.}             
\label{table:events_measured_info}      
\centering          
\begin{tabular}{c | c c | c c c | c c }   
\hline\hline       
\multirow{6}{*}{\centering Event date} & & &  & De-projected & & & \\ 
 & \multicolumn{2}{c}{Event} & PoS & radial & Residual & \multicolumn{2}{c}{Direction}  \\
  & \multicolumn{2}{c}{range} & velocity & velocity & acceleration & \multicolumn{2}{c}{Carr. coord.} \\
  & \multicolumn{2}{c}{[UT]}  & [km/s] & [km/s] & [m/s$^2$] & \multicolumn{2}{c}{[deg]}  \\
   &  &  &  & (Doppler & & & \\
   &  &  &  & dimming & & & \\
\cline{2-3} \cline{7-8}
 & pB & UV &  & coefficients)  &  & Lon. & Lat. \\
\hline           
   \multirow{4}{*}{\centering 25\,-\,26 Apr. 2021} & from & from & \multirow{4}{*}{\centering 152 $\pm$ 6 } &  & \multirow{4}{*}{\centering 1.7 $\pm$ 0.4} & \multirow{4}{*}{\centering 179.5\,$\pm$0.3} & \multirow{4}{*}{\centering $-$6\,$\pm$\,1} \\
    & 15:20 on 25/04  & 22:44 on 25/04 &  & 177\,$\pm$\,7 &  &  &   \\  
    & to  & to &  & (0.17-0.29) & & & \\
   & 04:20 on 26/04 & 4:44 on 26/04 & &  & & & \\
   \hline 
   \multirow{2}{*}{\centering 11 Sep. 2021} & \multirow{2}{*}{\centering 07:03\,-\,11:07} & \multirow{2}{*}{\centering 10:05\,-\,11:17} & \multirow{2}{*}{\centering 375 $\pm$ 5}  &  377\,$\pm$\,5 & 
   \multirow{2}{*}{\centering 7 $\pm$ 1} & \multirow{2}{*}{\centering 44.1\,$\pm$\,0.6} & \multirow{2}{*}{\centering $-$34\,$\pm$\,1} \\
    & &  & & (0.001-0.006) &  & & \\
   \hline 
   \multirow{2}{*}{\centering 2 Oct. 2021} & 14:19\,-\,15:19 & \multirow{2}{*}{\centering 17:05\,-\,19:49} & \multirow{2}{*}{\centering 228 $\pm$ 13 } & 367\,$\pm$\,22 & compatible & \multirow{2}{*}{\centering 30\,$\pm$\,1} & \multirow{2}{*}{\centering $-$28\,$\pm$\,1} \\
    & 17:33 - 19:49 &  & & (0.001-0.004) & with zero & & \\
    \hline 
   \multirow{2}{*}{\centering 25 Oct. 2021}  & 11:49\,-\,16:19 & 14:13\,-\,16:33 & \multirow{2}{*}{\centering 150 $\pm$ 1 } & 265\,$\pm$\,3 & \multirow{2}{*}{\centering 0.7 $\pm$ 0.1} & \multirow{2}{*}{\centering 169.6\,$\pm$\,0.5} & \multirow{2}{*}{\centering $-$11\,$\pm$\,1} \\
    & 18:19\,-\,20:49 & 18:05\,-\,20:13 & & (0.009-0.04) & & & \\
    \hline 
   28 Oct. 2021  & 15:49 - 16:19 & \multirow{2}{*}{\centering 18:07 - 19:11} & \multirow{2}{*}{\centering 244 $\pm$ 10 } & 246 $\pm$ 10 & \multirow{2}{*}{\centering 15 $\pm$ 3}  & \multirow{2}{*}{\centering 15.6\,$\pm$\,0.5} & \multirow{2}{*}{\centering 22.9\,$\pm$\,0.8}  \\
   North-West & 18:19 - 20:49 & & & (0.02-0.08) & & & \\
   \hline 
   28 Oct. 2021  & \multirow{2}{*}{\centering 18:19 - 20:49} & \multirow{2}{*}{\centering 18:05 - 22:57} & \multirow{2}{*}{\centering 152 $\pm$ 1 } & 437 $\pm$ 4 & \multirow{2}{*}{\centering $-$8.5 $\pm$ 0.1}  & \multirow{2}{*}{\centering 258.5\,$\pm$\,0.4} & \multirow{2}{*}{\centering $-$8.3\,$\pm$\,0.6}  \\
   South-East & & & & (2.7-5.2)$\cdot$10$^{-4}$ & & &\\
\hline                  
\end{tabular}
  \end{threeparttable}
  \tablefoot{For each event we show: the event date, the time range of the events in the two channels, the apparent PoS velocity, radial velocity, and, in brackets, Doppler dimming coefficients calculated considering the electron temperatures distributions of \cite{Gibson_1999JGR...104.9691G} and \cite{Vasquez_2003ApJ...598.1361V} extrapolated at the radial position of bright structures, the de-projected residual acceleration, and Carrington coordinates of the eruption directions estimated by the triangulation method as explained in the text.}
\end{table*}

\subsection{25$^{\mathrm{th}}$ April 2021}
\label{subsec:25_apr}

Between April 25$^\mathrm{th}$ and 26$^\mathrm{th}$ 2021, Metis observed a CME above the west limb of the Sun while running a low-cadence synoptic observation program from a distance of 0.87\,au.

\begin{figure*}
   \centering
   \includegraphics[clip, trim=0.5cm 9.2cm 1cm 7.3cm,width=18.4cm]{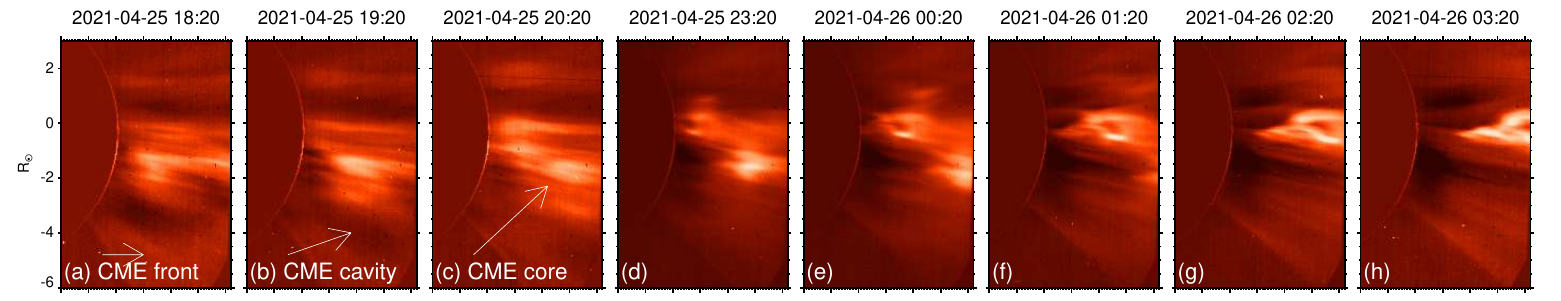}
   \includegraphics[clip, trim=0.5cm 7.8cm 1cm 8.2cm,width=18.4cm]{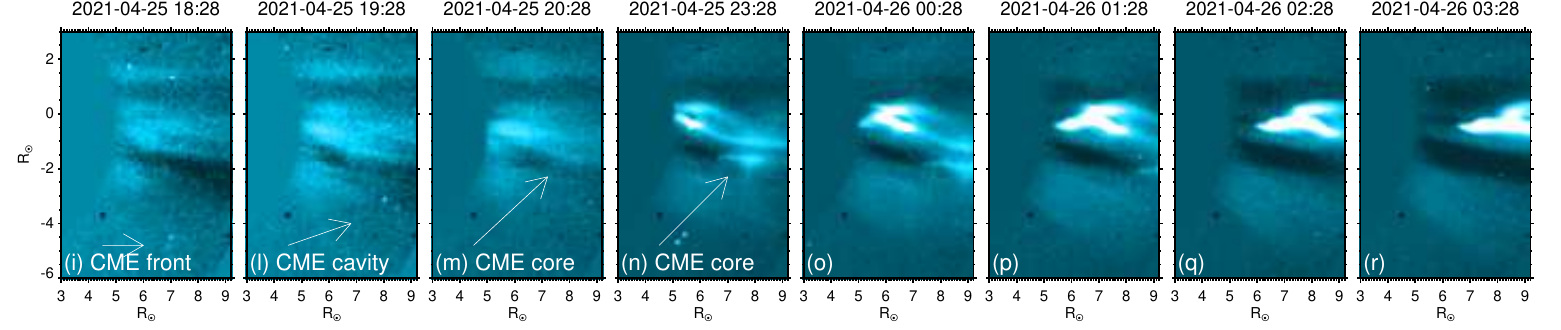}
   \caption{Main frames which show the event on April 25$^\mathrm{th}$ and 26$^\mathrm{th}$ 2021. At the top, a sequence of eight frames of the Metis pB base difference. The base frame is at 15:20\,UT. At the bottom, the bright event observed in \ion{H}{i}~\,\Lya\ is visible in nearly simultaneous base differences, using the base frame at 15:28\,UT. The CME structure is highlighted with arrows in frames (a), (b) and (c), which contain respectively front, cavity and bright core. In UV frames (i), (l) and (m), the same structures are barely visible. In frame (n), the CME bright core and features are both very intense and easily visible in both channels and appear separate. The temporal evolution in both channels is available as an online movie (pb\_20210425.mp4 and UV\_20210425.mp4).}
              \label{fig:pB_UV_prominence_20210425-26}%
    \end{figure*}

The so-called three-part structure of a CME is evident in the polarized brightness base differences as shown in the red-colored (a), (b) and (c) frames at the top of Fig.\,\ref{fig:pB_UV_prominence_20210425-26} and in the associated movie (\textcolor{red}{UV\_20210425.mp4 and pb\_20210425.mp4}). The base frame is shown in Fig.\,\ref{fig:pB_base} and is subtracted to remove the F-corona contribution that dominates at the heights we observe. The bright frontal loop, the dark cavity and the embedded bright core are indicated with white arrows. The CME front appeared at 15:20\,UT on April 25$^\mathrm{th}$ and spanned a distance of 5\,R$_{\sun}$ on the plane of the sky in almost 4 hours. 
The three-part structure is barely visible in the UV channel (i), (l) and (m) base difference frames, but the bright core is well recognizable in frame (n) as indicated by the white arrow.

The Metis UV channel also shows a very bright feature from 22:44\,UT on April 25$^\mathrm{th}$, until 4:44\,UT on April 26$^\mathrm{th}$ that moves horizontally to the west, just above the CME bright core and appears to be separated from it, as shown in the bottom panels of Fig.\,\ref{fig:pB_UV_prominence_20210425-26}. The apparent shape of the \Lya\ feature resembles a ''fork'' with an evident concavity, which might also be consistent with a helical structure \citep{Ciaravella_helical_CME_2000ApJ...529..575C,Suleiman_UVCS_helical_CME_2005IAUS..226...71S}. The fork-like shape is also well recognizable in the pB base difference frames it crossed in almost 6 hours, starting from panel (d) at the top of Fig.\,\ref{fig:pB_UV_prominence_20210425-26}.

The CME and the eruption were also observed with the space-based coronagraphs SOHO/LASCO-C2 and STEREO-A/COR2. As visible from Fig.\,\ref{fig:gse_all_events}, the SolO spacecraft (S/C) on that day formed an angle of almost 100\,$^\circ$ with the Earth and SOHO on the plane of orbit, while forming almost an angle of 45\,$^\circ$ with STEREO-A.

\begin{figure}
  \centering
   \includegraphics[clip, trim=7.5cm 3cm 7cm 3.8cm,width=4.9cm]{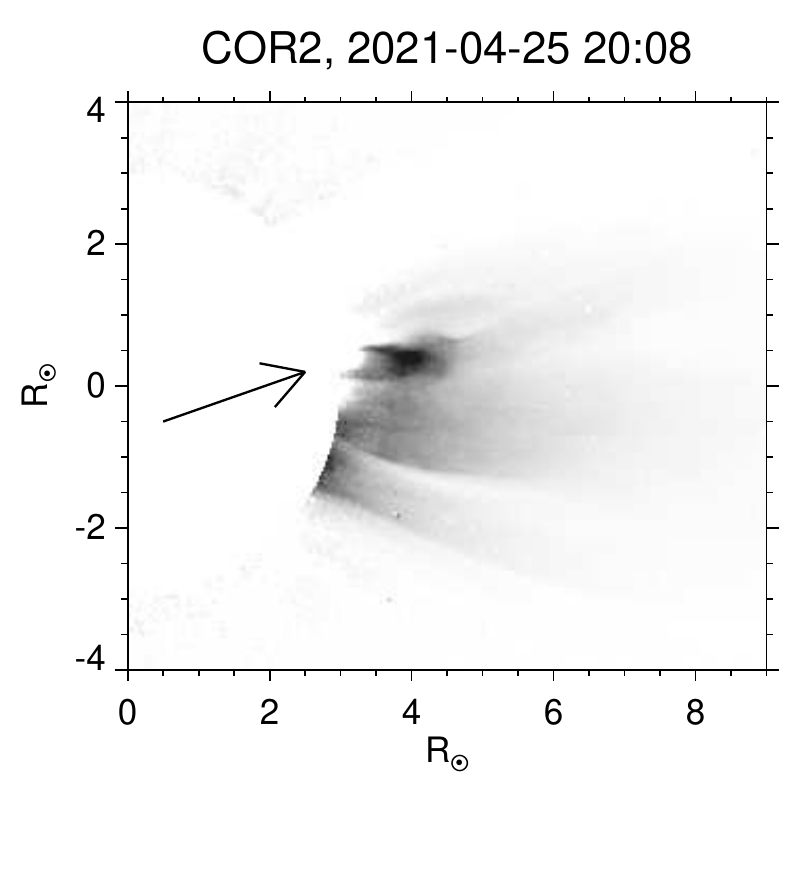}
   \includegraphics[clip, trim=9.8cm 4cm 9cm 4cm,width=4cm]{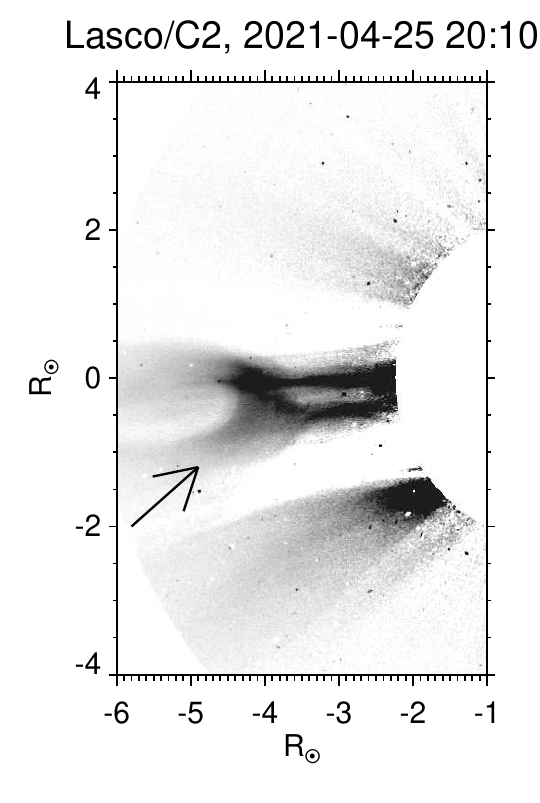}
   \caption{Eruption event on April 25$^\mathrm{th}$, as seen by SOHO/LASCO-C2 on the right, and STEREO-A/COR2 on the left in almost simultaneous frames and in reverse colors. Note that the event is moving eastward in SOHO/LASCO-C2 frame and westward in STEREO-A/COR2. These frames were used to estimate the structure direction using triangulation, with the fork shape indicated by the black arrow as the leading point.}
              \label{fig:Lasco_COR2_prominence_20210425-26}%
\end{figure}

The CME front appeared first in the SOHO/LASCO-C2 coronagraph FoV at 13:46\,UT on April 25$^\mathrm{th}$, in the eastern limb, while the fork-like structure was later observed at 17:22\,UT, almost overlapping with the CME core towards the north. The eruption has almost an hour difference with Metis because of the smaller FoV of LASCO-C2. This means that the event goes through the entire FoV of LASCO-C2 before entering the frames of Metis. In Fig.\,\ref{fig:Lasco_COR2_prominence_20210425-26} the fork shape of the eruptive prominence is also well recognizable from the LASCO-C2 frame.

The STEREO-A/COR2 coronagraph showed an eruption starting around 18:08\,UT at the west limb, as in Metis images, with a peculiar shape as in Fig.\,\ref{fig:Lasco_COR2_prominence_20210425-26} slowly expanding with time and similar to a streamer blowout. 

For this event it is possible to select simultaneous frames of the three coronagraphs to estimate the 3D position of the bright structure. The routine \texttt{scc\_measure} has been applied independently to pairs of images of the three coronagraphs. The leading points selected by hand, in this case, are the structures in the simultaneous SOHO/LASCO-C2 and STEREO-A/COR2 frames shown with black arrows in Fig.\,\ref{fig:Lasco_COR2_prominence_20210425-26}. A comparable result is obtained with the same approach by using a simultaneous couple of Metis UV and STEREO-A/COR2 frames.

The direction of the prominence thus measured is indicated with the yellow star in the first synoptic map of Fig.\,\ref{fig:magnetograms} in Appendix\,\ref{sec:appendix_source_region} at Carrington longitude of 179\,$^\circ$\,$\pm$\,3\,$^\circ$ and latitude of $-$6\,$^\circ$\,$\pm$\,1\,$^\circ$ as also reported with the dark red arrow in the orbit plot of Fig.\,\ref{fig:gse_all_events}. The position of the yellow star in Fig.\,\ref{fig:magnetograms} is in between the active region (AR) NOAA AR.12820 located at S22E08 at the western limb of the Sun with respect to SolO (also indicated by a cyan star in the first panel of Fig.\,\ref{fig:magnetograms}) and along a filament structure on the disk to the east, as visible in Fig.\,\ref{fig:EUVI_20210425-26}, taken with STEREO-A Extreme-Ultraviolet Imager \citep[EUVI,][]{EUVI_2008SSRv..136...67H} in the wavelength 304\,\r{A}.

The apparent velocity on the PoS of the bright part of the eruption has been calculated following the path of the central part of the fork feature along the Metis FoV, used as a leading point (green dashed line in Fig.\,\ref{fig:UV_velocity_track_all}). The estimated value is 152\,$\pm$\,6\,km/s, while the radial velocity is estimated at 177\,$\pm$\,7\,km/s. By extrapolating the time-distance relation and assuming a constant velocity on the plane of the sky, the eruption on the solar disk is estimated to have occurred around 17:00\,UT. This time coincides with the activation of AR.12820 as can be seen in Fig.\,\ref{fig:EUI_FSI_20210425-26}, taken with Extreme-Ultraviolet Imager \citep[EUI,][]{EUI_2020A&A...642A...8R} instrument and its Full-Sun Imager (FSI) telescope onboard SolO, working at wavelength 304\,\r{A}, making this region a good source candidate on disk. 

\subsection{11$^{th}$ September 2021}
\label{subsec:11_sep}

On September 11$^\mathrm{th}$ 2021, from a distance of 0.59\,au, 
Metis observed a second, more structured CME in both UV and VL channels, occurring above the southeast limb.

\begin{figure*}
  \centering
  \includegraphics[clip, trim=0.8cm 8.7cm 1.5cm 8.3cm,width=18.5cm]{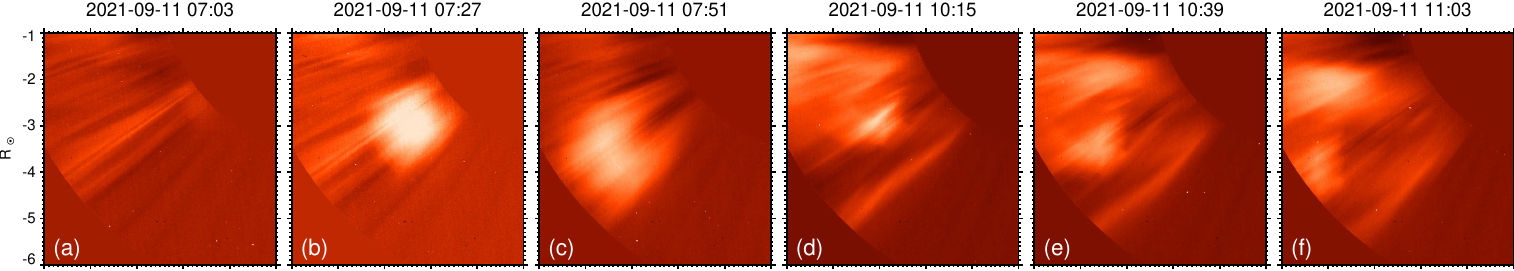}
   \includegraphics[clip, trim=0.8cm 8cm 1.5cm 8.3cm,width=18.5cm]{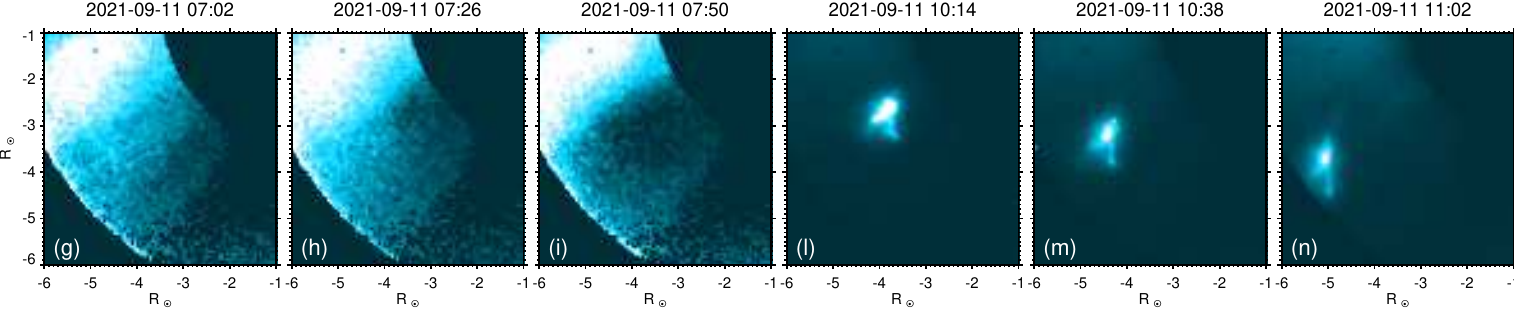}
   \caption{The CME front of the event on September 11$^\mathrm{th}$ 2021 can be seen in the VL pB base differences in the red-colored frames (a), (b) and (c) and in the three simultaneous blue-colored (g), (h), and (i) UV frames below, processed via SiRGraF algorithm \citep{sirgraf}. The base frame is at 01:03\,UT. The eruptive prominence is visible in both channels starting from 10:15\,UT. The temporal evolution in both channels is available as an online movie (pb\_20210911.mp4 and UV\_20210911.mp4).}
              \label{fig:pB_UV_20210911}%
   \end{figure*}

The CME front is visible in the pB base differences starting at 07:03\,UT as shown in Fig.\,\ref{fig:pB_UV_20210911}, extending only into the first three frames in almost 50\,min, before a data gap of a couple of hours. The dark cavity is distinguishable only in the base difference in panel (c). 
The CME structure appears as a dark absorption cavity through the FoV of the Metis UV channel in the simultaneous blue-colored frames (g), (h) and (i) in Fig.\,\ref{fig:pB_UV_20210911}, and has been enhanced by applying the Simple Radial Gradient Filter \citep[SiRGraF,][]{sirgraf} algorithm. This consists in subtracting the minimum background image from each frame and normalizing it with a uniform-background image with a circularly symmetric intensity gradient.

The bright eruptive prominence entered the UV frames of Metis at 10:05\,UT, after a couple of hour data gap, crossing nearly 3\,R$_{\sun}$ on the PoS in just over an hour, and it is shown in panels (l), (m) and (n) of Fig.\,\ref{fig:pB_UV_20210911}. The structure corresponds to the bright features in the pB base differences of panels (d), (e) and (f), which could coincide with the bright core of the three-part CME. 

\begin{figure}
  \centering
   \includegraphics[clip, trim=2cm 4cm 2cm 4cm,width=6cm]{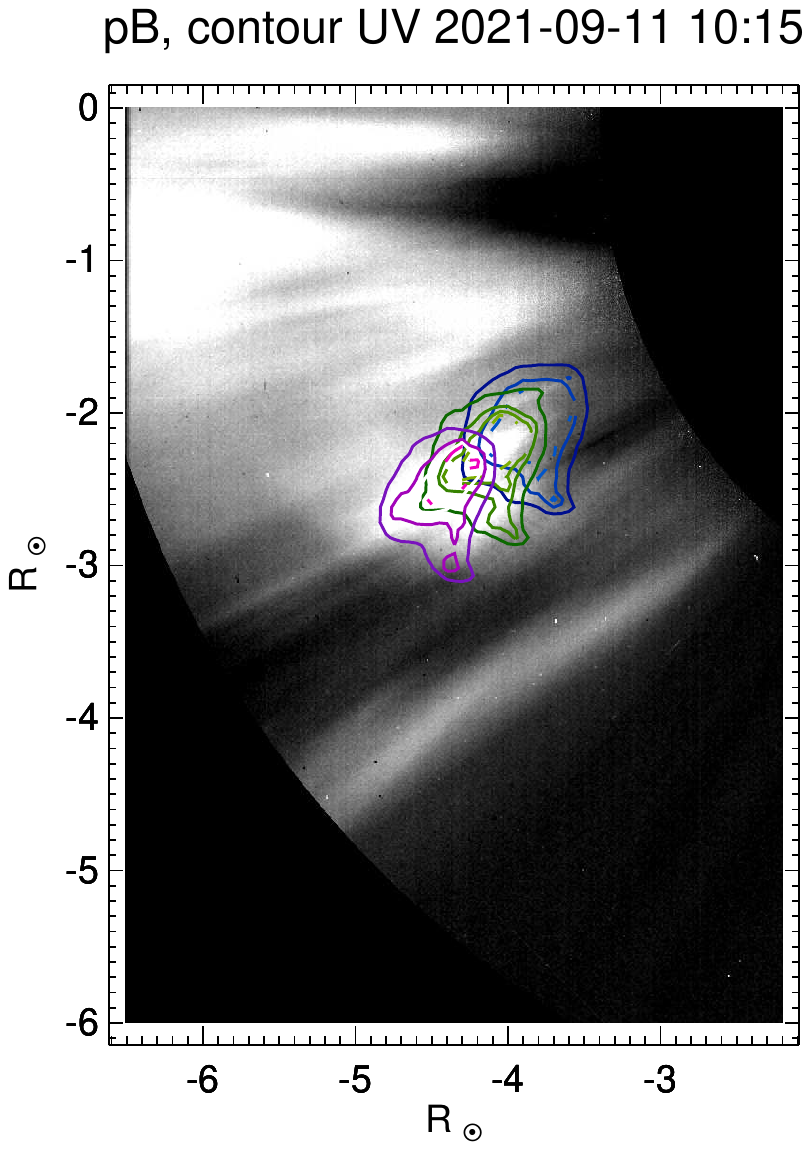}
   \caption{Eruptive prominence observable in the pB base difference frame of 10:39\,UT on September 11$^\mathrm{th}$ 2021. The contour surfaces of the structure containing between 10 and 100\,$\%$ of the maximum intensity from the UV frames at 10:29\,UT (blue curves), 10:38\,UT (green curves) and 10:50\,UT (magenta curves) are shown superimposed on the pB image.}
              \label{fig:UV_contour_20210911}%
   \end{figure}

Moreover, since the pB frames take much longer time to be acquired than the UV frames, it makes sense to overlay the contour surfaces of the prominence eruption from multiple UV frames on a pB frame. These contours contain between 10 and 100\,$\%$ of the highest intensity pixel on the UV frame, as shown in Fig.\,\ref{fig:UV_contour_20210911}.
For this acquisition, the UV channel has a cadence eight times higher than the VL channel, so that during the single pB acquisition of 24\,min duration, the UV channel acquires 8 frames (see numbers in Tab.\,\ref{table:events_info}). Greater integration time in the VL channel results in a smearing effect of the captured structure as observable, for example, in the visible light pB frame at 10:39\,UT, in Fig.\,\ref{fig:UV_contour_20210911} and in the frame (e) of Fig.\,\ref{fig:pB_UV_20210911}.
In Fig.\,\ref{fig:UV_contour_20210911}, the three surface contours of UV frames are taken simultaneously with the pB below, one coinciding with the beginning of the pB acquisition, one at the center of acquisition, and the last at the end. This highlights how the high temporal and spatial cadence of the UV channel allows a far better analysis of the morphology and kinematics of these fast-moving structures.

The smearing effect appears in the pB frames of all six events because the cadence is always around 30\,min (see Tab.\,\ref{table:events_info}). In a minor way, the effect is visible also for the April 25$^\mathrm{th}$ where the cadence of the UV channel is half that of the VL. 

\begin{figure}[!ht]
  \centering
   \includegraphics[clip, trim=2.5cm 5cm 2cm 3.5cm,width=4.5cm]{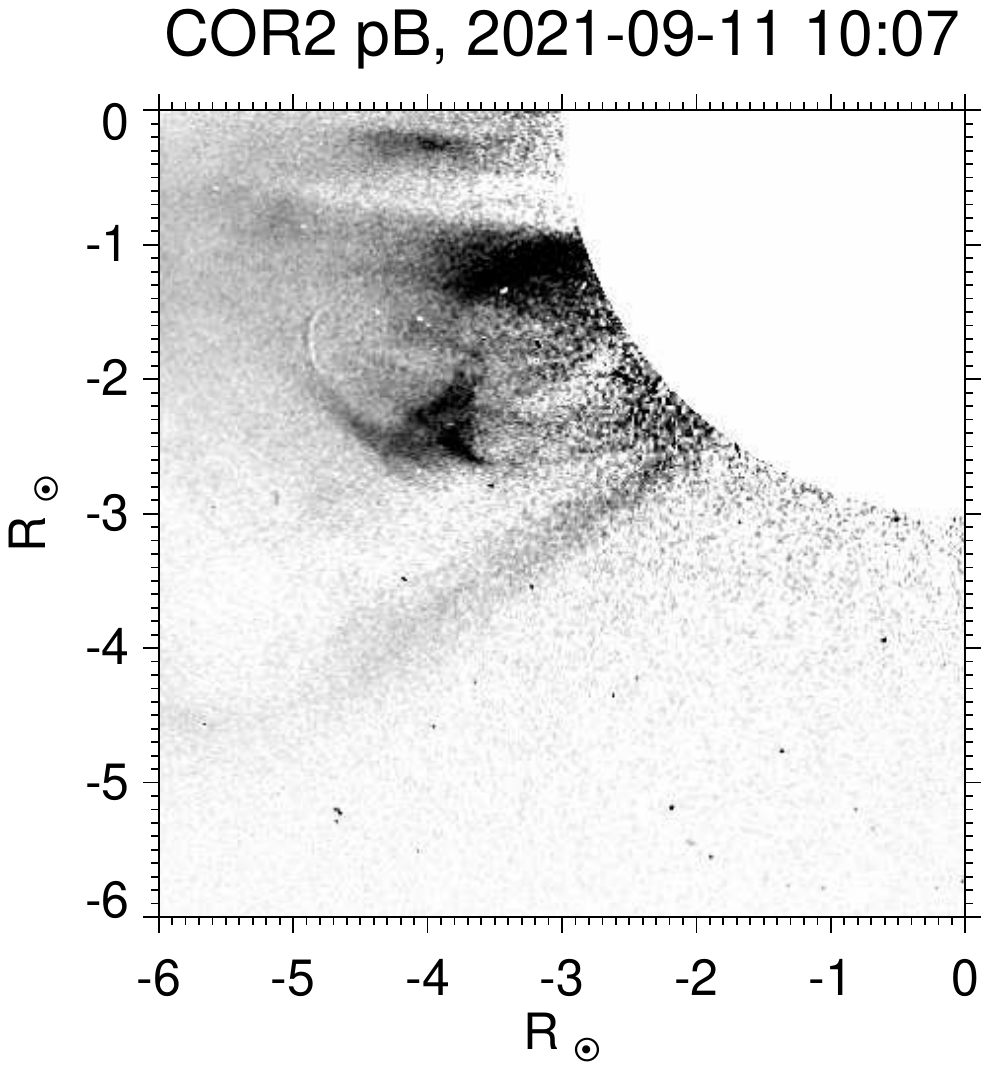}
   \includegraphics[clip, trim=6cm 0.5cm 4.7cm 0.5cm,width=4.1cm]{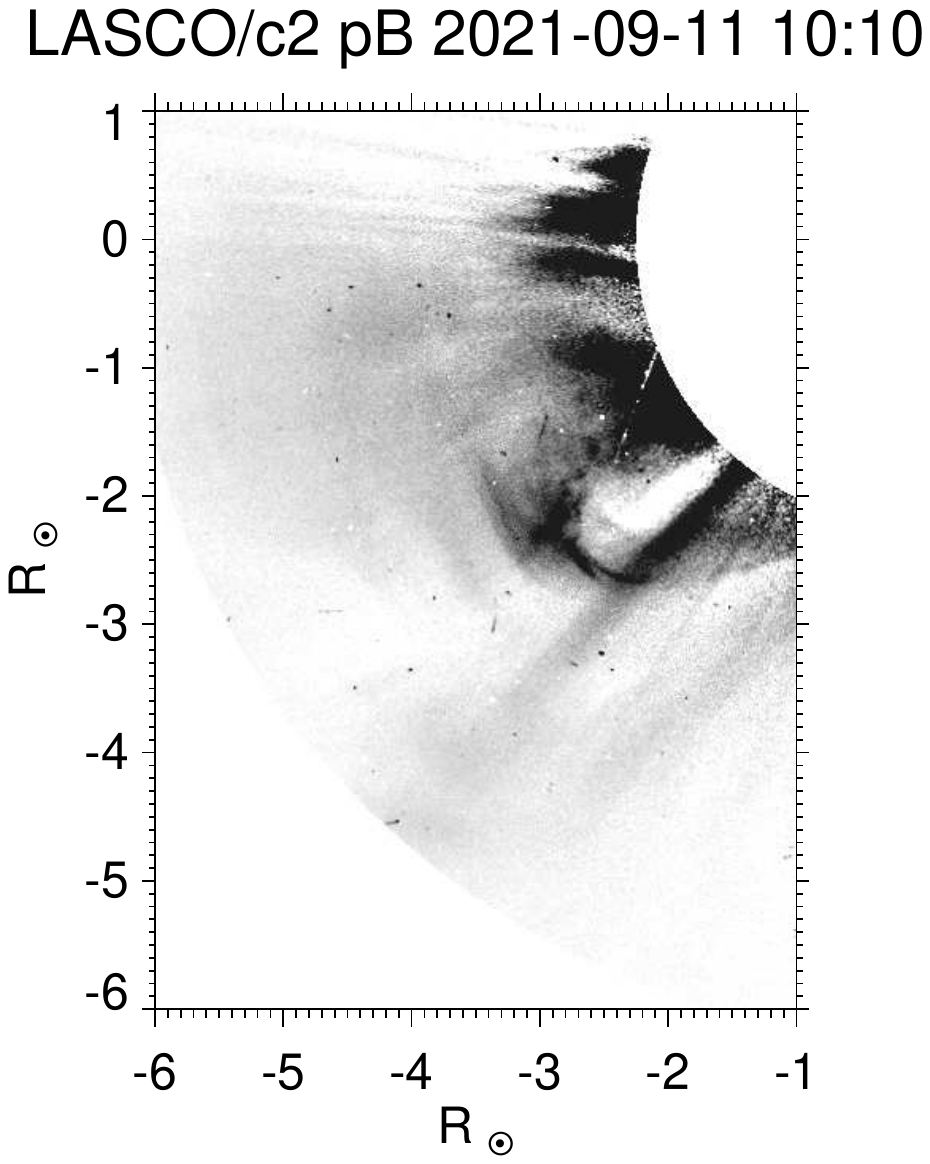}
   \caption{Eruptive prominence on September 11$^\mathrm{th}$ 2021, as seen by SOHO/LASCO-C2 on the right, and STEREO-A/COR2 on the left in almost simultaneous frames and shown here in reverse colors. These frames have been used for the structure direction estimation with the triangulation method.}
              \label{fig:Lasco_COR2_prominence_20210911}%
\end{figure}

Also in the case of the September 11$^\mathrm{th}$ event, the eruption was observed by the other two coronagraphs in space SOHO/LASCO-C2 and STEREO-A/COR2, which formed an angle of respectively $\sim$\,50\,$^{\circ}$ and $\sim$\,10\,$^{\circ}$ with SolO in the orbit plane. In both white-light imagers, the distinctive triangular shape of the structure is recognizable, and its features can easily be used to estimate the direction of the eruption using the triangulation method (see Fig.\,\ref{fig:Lasco_COR2_prominence_20210911}). We estimate a Carrington long. of 44.1\,$^{\circ}$\,$\pm$\,0.6\,$^{\circ}$ and a lat. of $-$34\,$^{\circ}$\,$\pm$\,1\,$^{\circ}$, just behind the solar limb as seen by Metis and depicted with the light blue arrow in Fig.\,\ref{fig:gse_all_events}. These coordinates are confirmed if we use a different couple of images of the three instruments for the calculation, and the position is also shown with the yellow star in the magnetogram of Fig.\,\ref{fig:magnetograms}.

The PoS velocity can be estimated following the bright peak of the prominence through the Metis UV FoV (green dashed line in Fig.\,\ref{fig:UV_velocity_track_all}), with a value of 375\,$\pm$\,5\,km/s. The velocity can be compared to the estimate of the online CACTus catalog \citep{cactus_2004} for the CMEs observed by STEREO-A/COR2, with a value of 337\,$\pm$\,79\,km/s. However, it is important to take into account that the two coronagraphs have different PoS in this particular case. The velocity estimated by Metis is within three sigmas of that estimated from the catalog of CMEs observed by SOHO/LASCO-C2, which reports a value of \,508\,$\pm$\,48\,km/s. It is worth noting that the CACTus catalog is based on an automatic algorithm for the detection of CMEs in LASCO images, and it measures a linear speed profile as a function of the angle around the occulter and lists the median value \citep{Cactus_2_2009ApJ...691.1222R}, so, there is no distinction between front or core speed of the CME. 

The Metis radial velocity is instead 377\,$\pm$\,5\,km/s. 
By extrapolating the time of the eruption at the solar disk and at a constant velocity on the PoS, the low corona counterpart of the eruption leaving the Sun's surface is recognizable in the frame of STEREO-A/EUVI in the wavelength 304\,\r{A} around 7:45\,UT, in Fig.\,\ref{fig:EUVI_20210911}. No active regions are reported corresponding on the solar disk, but a large erupting prominence is seen leaving the solar surface, as in Fig.\,\ref{fig:EUVI_20210911}.

\subsection{2$^\mathrm{nd}$ October 2021}
\label{subsec:2_oct}

   \begin{figure*}
   \centering
   \includegraphics[clip, trim=0.5cm 7.6cm 1cm 7cm,width=18cm]{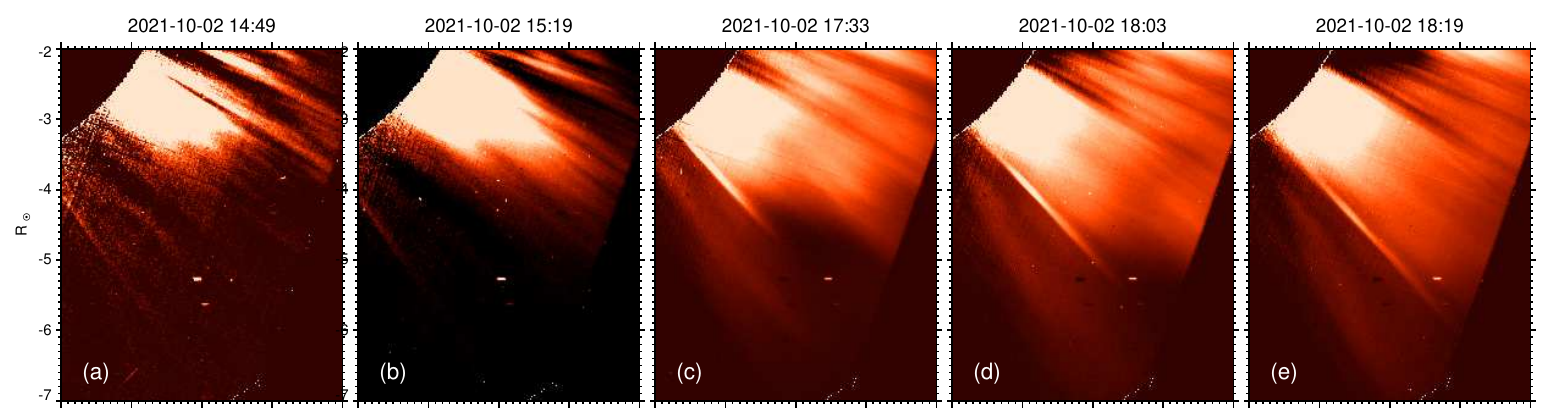}
   \includegraphics[clip, trim=0.5cm 7cm 1cm 7.6cm,width=18cm]{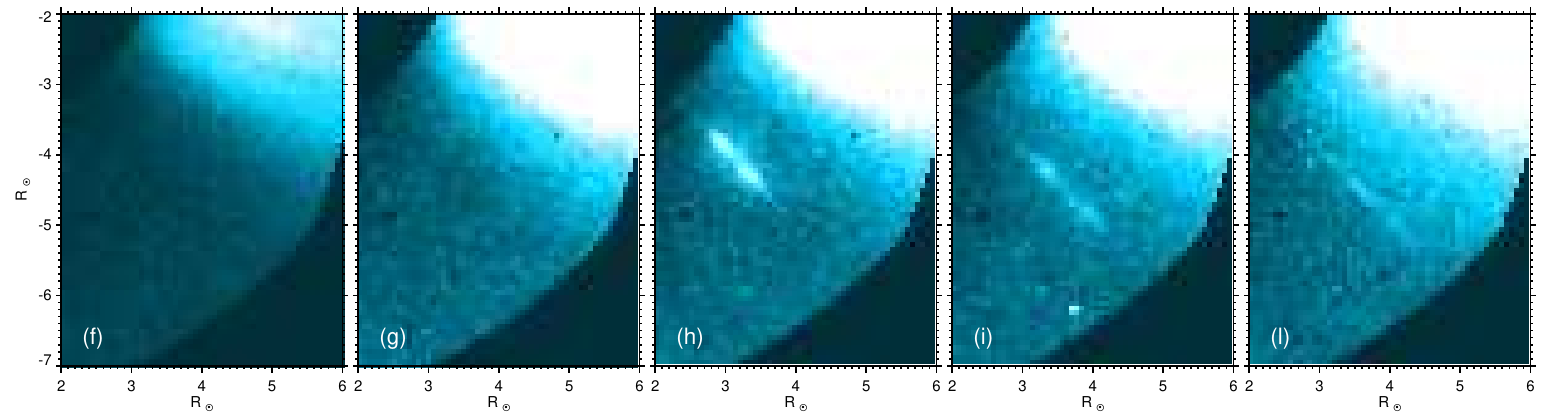}
   \caption{Event on October 2$^\mathrm{nd}$ as seen by Metis VL channel in the \textit{B} base difference (base frame at 14:19\,UT) on top and by Metis UV channel in the blue-colored frames at the bottom. The CME front is detectable in panel (a) and (b) while the bright structure is visible in both channels starting from 17:33\,UT. The temporal evolution in both channels is available as an online movie (tb\_20211002.mp4 and UV\_20211002.mp4).}
              \label{fig:pB_UV_prominence_20211002}%
    \end{figure*}
On October the 2$^\mathrm{nd}$ Metis observed another event at the southwest limb of the Sun, from a distance of 0.64\,au. 

The front of the event is barely visible in Metis pB but appears in \textit{B} frames at 14:49\,UT and 15:19\,UT as in Fig.\,\ref{fig:pB_UV_prominence_20211002}, panels (a) and (b). After a data gap of a couple of hours, an elongated feature was well observable stretching in both Metis channels, between 17:33 and 19:49\,UT as in Fig.\,\ref{fig:pB_UV_prominence_20211002}. It became very feeble after 19:19\,UT. 

\begin{figure}
  \centering
   \includegraphics[clip, trim=2cm 3cm 2cm 2.5cm,width=8cm]{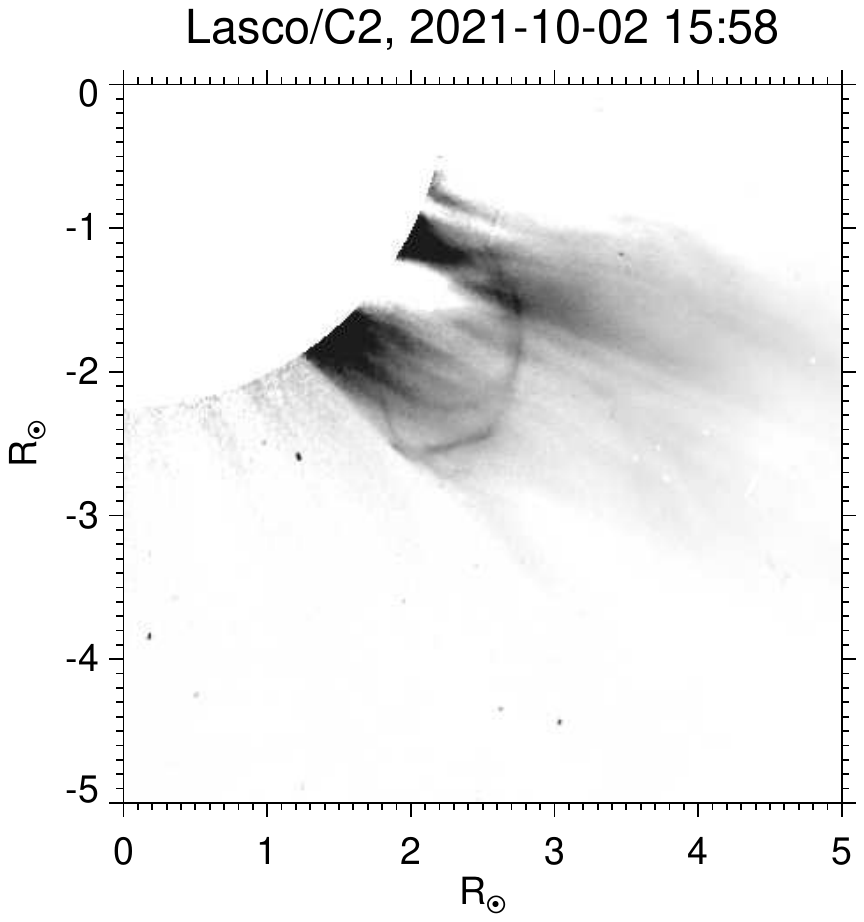}
   \caption{October 2$^\mathrm{nd}$ eruption event as seen by SOHO/LASCO-C2 coronagraph at 15:58\,UT, a time when Metis was not yet observing.}
              \label{fig:Lasco_pB_20211002}%
\end{figure}

The eruption appears more structured in the SOHO/LASCO-C2 coronagraph images between 14:24\,UT and 17:10\,UT, mostly during the Metis data gap. The feature appears as an arc-like shape front that crosses the LASCO-C2 FoV before entering the Metis frames, with only one overlapping solar radius between the two (Fig.\,\ref{fig:Lasco_pB_20211002} displays the appearance of the structure in the LASCO-C2 FoV during Metis data gap). The CME three-part structure is not well recognizable in SOHO/LASCO-C2 frames and no data are available for STEREO-A/COR2 coronagraph for this event. Moreover, SOHO and SolO were separated by only $\sim$\,20\,$^\circ$ on the orbit plane.

This means that it is possible to use only one simultaneous image of SOHO/LASCO-C2 at 17:10\,UT to apply the triangulation method and estimate the direction of the event with respect to the bright features in the Metis UV frame taken at the same time. We estimate a Carrington longitude of 30\,$^\circ$\,$\pm$\,1\,$^\circ$ and a latitude of $-$28\,$^\circ$\,$\pm$\,1\,$^\circ$ selecting poorly visible features in LASCO-C2 FoV (see also the position shown in Fig.\,\ref{fig:magnetograms}). 

The bright part of the event in the Metis UV channel can be used to also estimate the PoS velocity, following the bright peak of the elongated feature. The velocity estimate is 228\,$\pm$\,13\,km/s while the de-projected value is 367\,$\pm$\,22\,km/s. 
The estimated velocity can be compared with the evaluation of the CACTus catalog from SOHO/LASCO-C2 data \citep{cactus_2004}, which is 440\,$\pm$\,203\,km/s. This value can also be de-projected, determining an estimate of 546\,$\pm$\,342\,km/s. 

\subsection{25$^\mathrm{th}$ October 2021}
\label{subsec:25_oct}

On October 25$^\mathrm{th}$ 2021, Metis observed a CME with both UV and VL channels occurring above the southeast limb, at a distance of 0.78\,au. 

\begin{figure*}
  \centering
  \includegraphics[clip, trim=0.5cm 8.8cm 1cm 7cm,width=18cm]{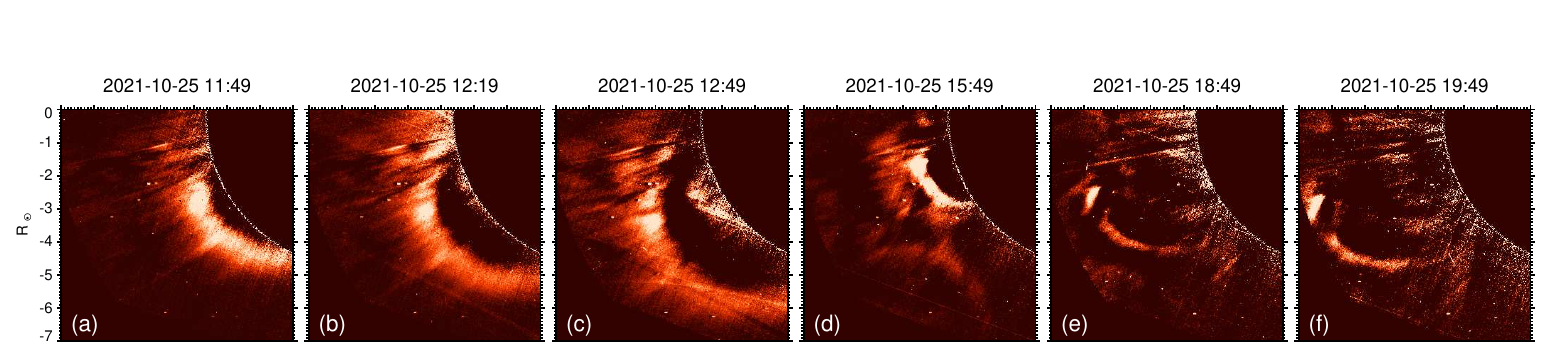}
   \includegraphics[clip, trim=0.5cm 7.8cm 1cm 8.5cm,width=18cm]{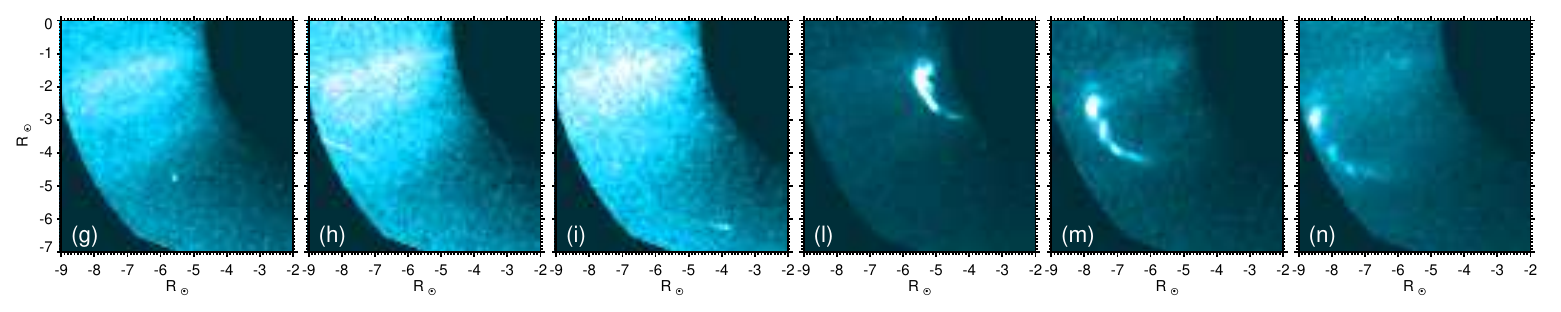}
   \caption{Development of the event on September 25$^\mathrm{th}$. The CME structure is observable in the first three frames of total brightness running differences on top but is not detectable in UV frames in (g), (h) and (i) panels. The line-shape features which appear in random directions on (b) and (c) frames are artifacts due to the particle shower on the detector and cannot be eliminated in post-production. The eruptive prominence is visible in the total brightness running differences from 15:49\,UT in (d), (e) and (f) panels while the bright UV \ion{H}{i}~\Lya\ bright structure can be seen in the (l), (m) and (n) UV frames at the bottom. The temporal evolution in both channels is available as an online movie (tb\_20211025.mp4 and UV\_20211025.mp4).}
              \label{fig:tB_UV_eruption_20211025}%
\end{figure*}

The front is only clear in the total brightness starting from the frames around 11:49\,UT, as shown in panels (a), (b) and (c) of Fig.\,\ref{fig:tB_UV_eruption_20211025}, in almost 1\,h of acquisition. This is completely invisible to the UV channel as in panels (g), (h) and (i). Then, a UV bright eruptive prominence went through the Metis FoV starting from 14:09\,UT, as can be seen in Fig.\,\ref{fig:tB_UV_eruption_20211025} in the bottom frames. It appears as an arc-shaped feature progressively expanding in the radial direction. The eruption is also evident in the simultaneous total brightness running differences (top frames of Fig.\,\ref{fig:tB_UV_eruption_20211025}) for almost 8\,h with a data gap between 16:33\,UT and 18:05\,UT. After the data gap, the prominence appears segmented into three parts, and its angular width widens through the UV frames as shown up in Fig.\,\ref{fig:UV_velocity_track_all} with the colored dashed lines which indicate the path of each bright point through the FoV.  
   
The eruption was also observed with the space-based coronagraphs SOHO/LASCO-C2 and STEREO-A/COR2. As visible from Fig.\,\ref{fig:gse_all_events}, the position of SolO and SOHO was almost on the same line of sight (LoS), while STEREO-A formed an angle of almost 34$^\circ$ with SolO. 

\begin{figure}
  \centering
   \includegraphics[clip, trim=6cm 8cm 6cm 8cm,width=4cm]{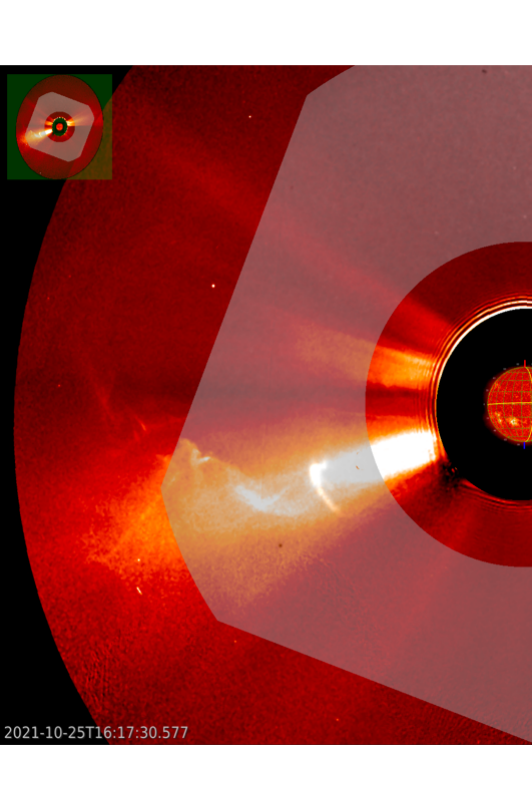}
   \includegraphics[clip, trim=6cm 8cm 6cm 8cm,width=4cm]{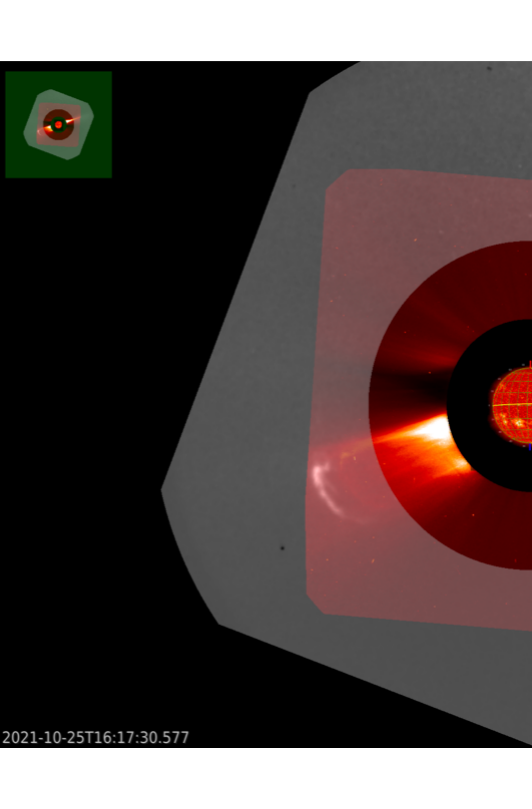}
   \caption{Comparison between the FoVs of Metis (4.7\,-\,10.0\,R$_{\sun}$) and SOHO/LASCO-C2 FoV (2.2\,-\,6\,R$_{\sun}$) on the right, and STEREO-A/COR2 FoV (3\,-\,15\,R$_{\sun}$) on the left on October 25$^\mathrm{th}$. Despite the 34-degree angle between the two instruments PoS, the structure visible in the UV channel of Metis almost matches the bright central core of the CME visible from COR2, as shown in the picture on the left. The images are produced with the open-source software JHelioviewer \citep{JHV}.}
              \label{fig:FOV_Metis_Lasco_COR2_20211025}%
\end{figure}

The CME front appeared in the SOHO/LASCO-C2 instrument FoV at 9:31\,UT, while the arc shaped feature appeared at 11:19\,UT, a couple of hours later than Metis due to the smaller FoV. The STEREO-A/COR2 coronagraph only captured the eruption in a few frames due to longer cadence, with the front appearing at 9:07\,UT and a bright core at 11:07\,UT. This time, the FoV of SOHO/LASCO-C2 almost overlapped with that of Metis, as shown in Fig.\,\ref{fig:FOV_Metis_Lasco_COR2_20211025}. The left panel shows the bright Metis UV structure appearing close to the bright CME core seen by STEREO/COR2, despite the 34-degree angle between the two fields of view. 
This allows findings of multiple simultaneous images in the three instruments to apply the triangulation method with the routine \texttt{scc\_measure}. 

\begin{figure}[!ht]
  \centering
   \includegraphics[clip, trim=5cm 1cm 5cm 1cm,width=6cm]{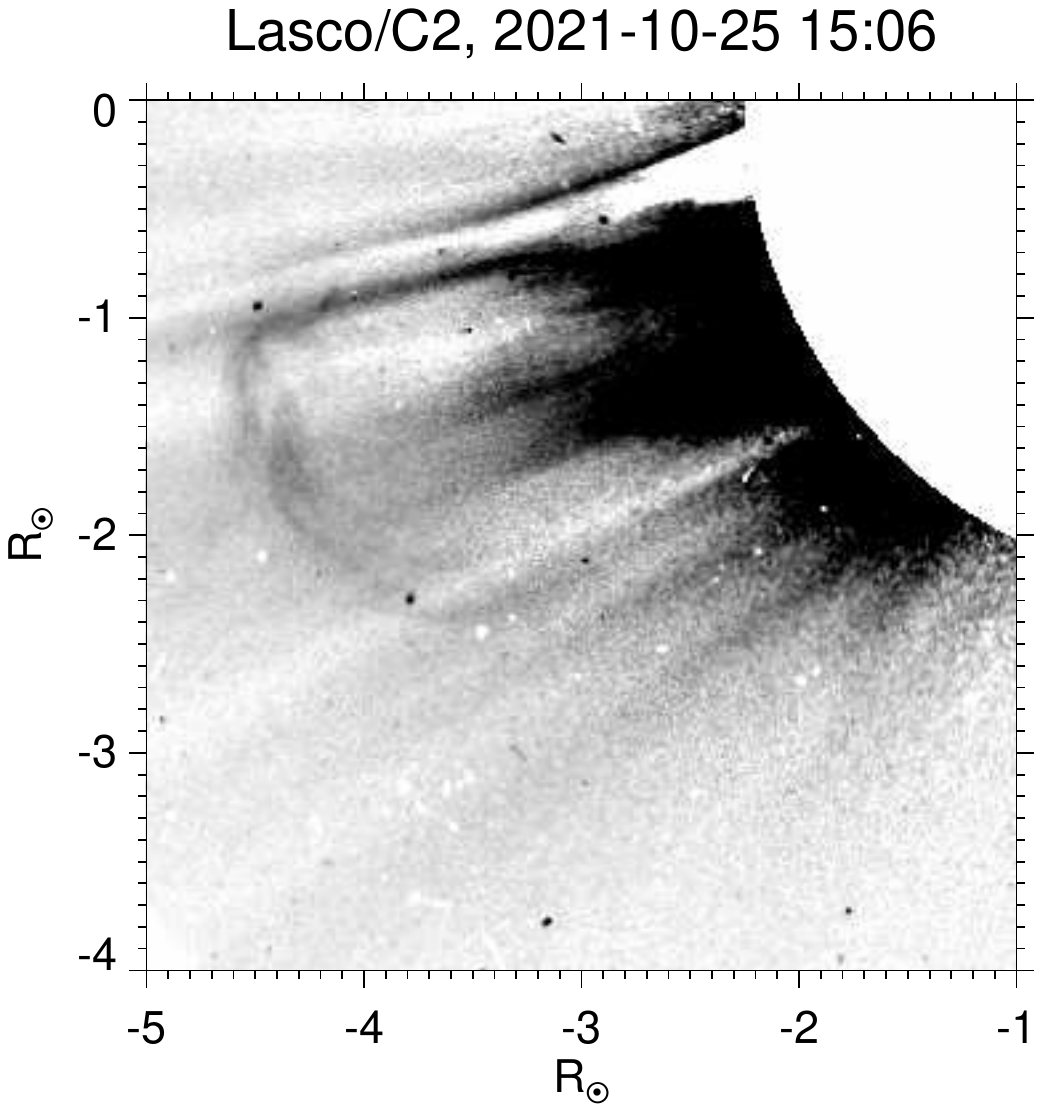}
   \includegraphics[clip, trim=5cm 5cm 5cm 4.8cm,width=6cm]{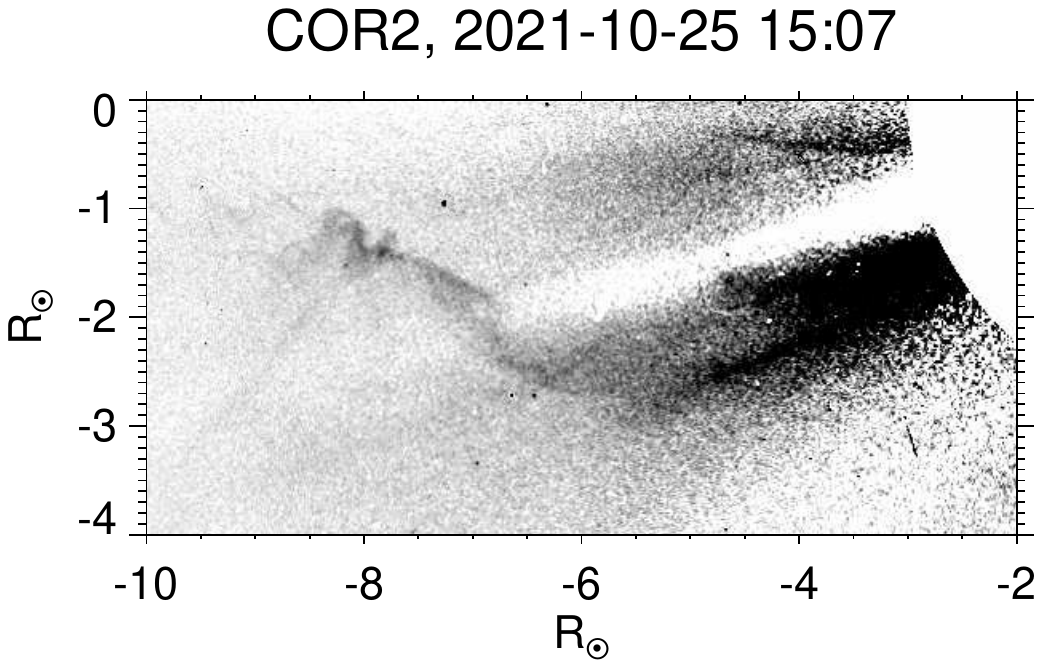}
   \caption{Eruptive prominence on October 25$^\mathrm{th}$ as seen by SOHO/LASCO-C2 on the top, and STEREO-A/COR2 on the bottom. These frames have been used for the eruption direction estimation with the triangulation method, together with the Metis UV simultaneous frame.}
              \label{fig:Lasco_COR2_prominence_20211025}%
\end{figure}

The leading point, the brightest part of the prominence, is visible in the UV channel of Metis and can be recognized in SOHO/LASCO-C2 as similar arc-shaped features, as well as in the smoking-like part of the prominence seen by STEREO-A/COR2 (see Fig.\,\ref{fig:Lasco_COR2_prominence_20211025}). The direction of the eruption is indicated with the green arrow in Fig.\,\ref{fig:gse_all_events} at Carrington long. of 169.6\,$^\circ$\,$\pm$\,0.5\,$^\circ$ and lat. of $-$11\,$^\circ$\,$\pm$\,1\,$^\circ$. This means that the eruption happened behind the solar limb with respect to the S/C facing the Sun and no direct observation of the corresponding active region is available for this event (see also the yellow and cyan stars on the magnetogram in the last panel of Fig.\,\ref{fig:EUVI_20211025}). 

Since there are no direct observations of the eruption on the disk and therefore of the magnetic field structure evolution, we can only qualitatively interpret the structure as an expanding flux rope that seems to have a helical structure with the cold plasma material mostly condensed into the dips. This scenario is similar to that proposed in Fig.\,7 of \cite{Sasso_2014A&A...561A..98S}, where the authors plot, schematically, two magnetic field lines with different degrees of twist belonging to a flux rope, to explain the behavior of an activated filament observed through spectropolarimetric observations. This flux rope, indeed, is entirely expanding upwards, lifting also the plasma trapped into the dips (in analogy with the prominence studied in this paper), while some other plasma material is falling down, mostly close to the filament footpoints, following the less twisted magnetic field lines.

A similar structure with bright UV knots was detected on 1996 December 23 by UVCS and SOHO/LASCO as reported in \cite{Antonucci_CME_3knot_1997ApJ...490L.183A} and \cite{Dere_CME_3knots_1997SoPh..175..601D}. Through the UVCS slit, a bright region with three bright knots spreads with time, but it is closer to the Sun disk at a heliocentric distance of 1.5\,R$_\sun$. In this case the \Lya\ bright material corresponded to the cool eruptive prominence gas, with a temperature below 10$^5$\,K and line of sight velocity of 200\,km/s. The bright knots showed red and blueshifts which suggests that the slit crosses a large expanding loop and the knots could be ripples in the sheet or density enhancements.

Following the maximum intensity pixel of the bright structure in the Metis UV frames (green dashed line in Fig.\,\ref{fig:UV_velocity_track_all}), it is also possible to assess the plane of sky velocity of the eruption, which is around 150\,$\pm$\,1\,km/s, while its de-projected value is 265\,$\pm$\,3\,km/s. 
The estimated radial velocity is comparable with the de-projected value provided by the CME catalog CACTus which is $\sim$\,303\,$\pm$\,66\,km/s as estimated from SOHO/LASCO-C2 data.

By extrapolating the time of the eruption at solar disk from the height-time fit at a constant velocity on the PoS, we recognize a low corona feature that could be associated with the prominence leaving the solar surface in the frames of STEREO-A/EUVI instrument in the wavelength 304\,\r{A}, starting from 08:05\,UT as shown in Appendix\,\ref{sec:appendix_source_region} in Fig.\,\ref{fig:EUVI_20211025}.

\subsection{28$^\mathrm{th}$ October 2021}
\label{subsec:28_oct}

On October 28$^\mathrm{th}$ 2021 Metis was also observing during the first X-class flare of the Solar Cycle 25 that occurred in the solar active region NOAA AR.12887 around the disk center with a peak at 15:35\,UT producing the rare event of ground-level enhancement of the solar relativistic proton flux and a global extreme-ultraviolet (EUV) wave, along with a fast halo CME \citep{Hou_xflare_2022ApJ...928...98H}. 
A few hours before the flare a narrower and slower coronal mass ejection emerged starting from an AR just behind the northwestern solar limb as seen by Earth \citep{Papaioannou_xflare_2022A&A...660L...5P}. The flare event was observed by the AIA imagers on board SDO at various wavebands, as well as STEREO–A disk imager EUVI. The associated CME was detected by SOHO/LASCO-C2, and STEREO–A/COR1 and COR2 coronagraphs with a different view angle than SDO, SOHO and SolO. As reported in Fig.\,\ref{fig:gse_all_events}, SolO was almost on the same LoS of the Earth and SOHO, while STEREO-A had a longitudinal separation with Earth of $\sim$\,40$^{\circ}$. The north-west CME is also distinguishable in SOHO/LASCO-C2 and STEREO-A/COR2 coronagraphs.

Metis detected both events; the earlier CME occurred in the north-west (NW) with respect to the solar disk, while the flare-associated event occurred in the south-east (SE), when SolO was at a distance of 0.8\,au. 

\begin{figure*}
  \centering
   \includegraphics[clip, trim=0.5cm 8.2cm 1cm 7.cm,width=18cm]{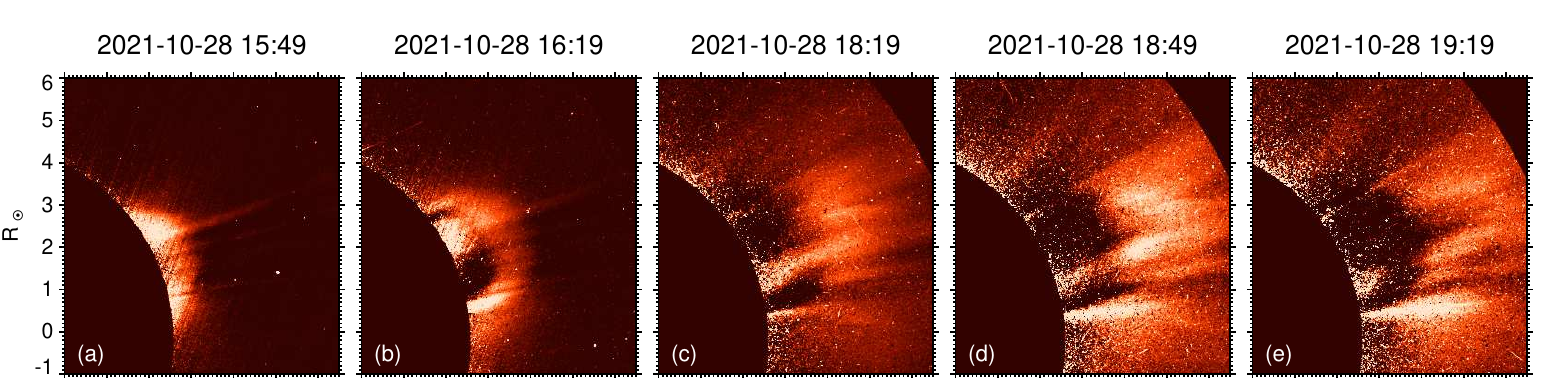}
   \includegraphics[clip, trim=0.5cm 7.2cm 1cm 8cm,width=18cm]{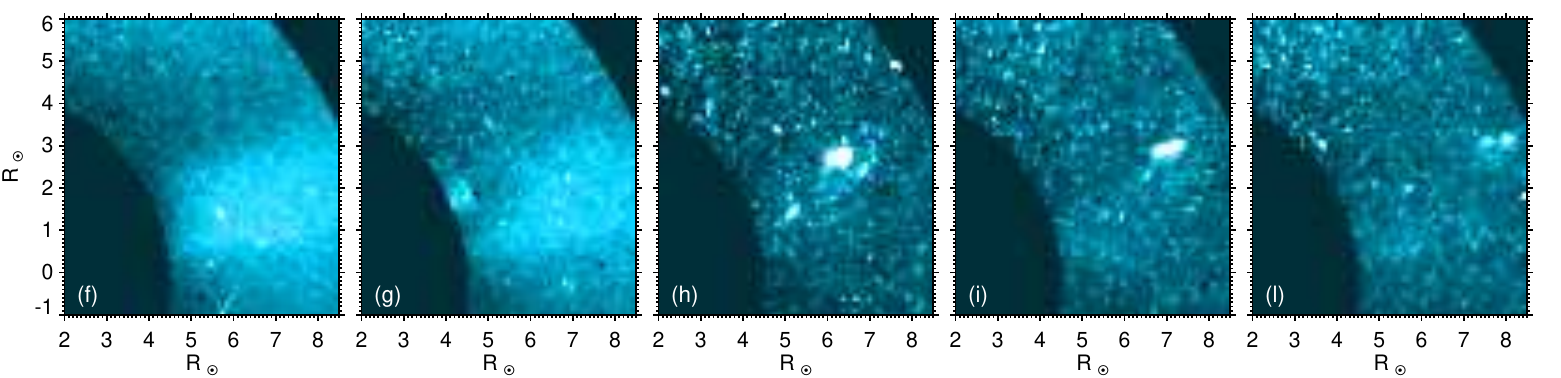}
   \caption{The front of the NW CME on October 28$^\mathrm{th}$ 2021 is observable in Metis pB base difference in the first two panels on 15:49\,UT and 16:19\,UT. In the simultaneous (f) and (g) UV frames, only a slight change in the intensity of the corona is visible. The ellipsoid-like feature of the event is bright in blue-colored (h), (i), (l) frames in the Metis UV channel at the bottom. The CME assumes a fuzzy shape in the pB base differences in (c), (d) and (e) panels at the top. The VL base frame is at 15:19\,UT. Noisy bright points appear after 18:19\,UT because of the particle shower on the detector. The temporal evolution in both channels is available as an online movie (pb\_20211028.mp4 and UV\_20211028.mp4).}
        \label{fig:pb_UV_eruption_20211028}%
\end{figure*}

\begin{figure}
  \centering
  \includegraphics[clip, trim=3cm 6cm 3cm 6cm,width=2.9cm]{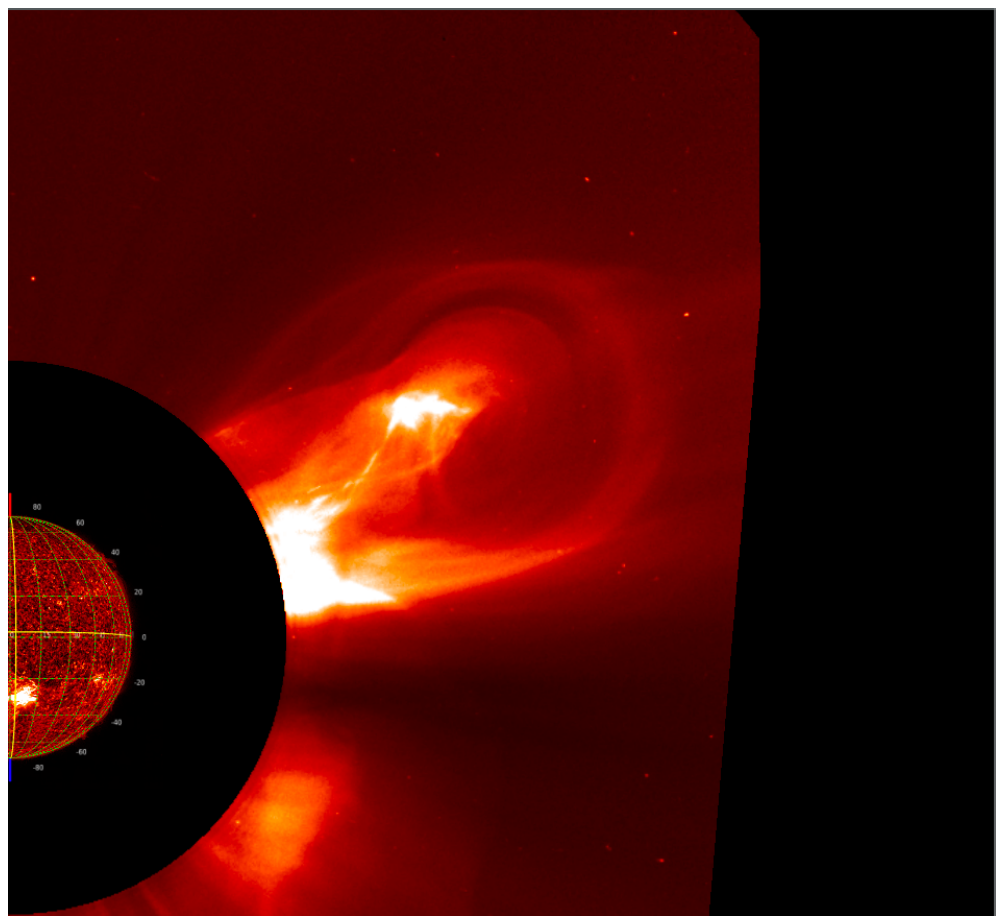}
  \includegraphics[clip, trim=3cm 6cm 3cm 6cm,width=2.9cm]{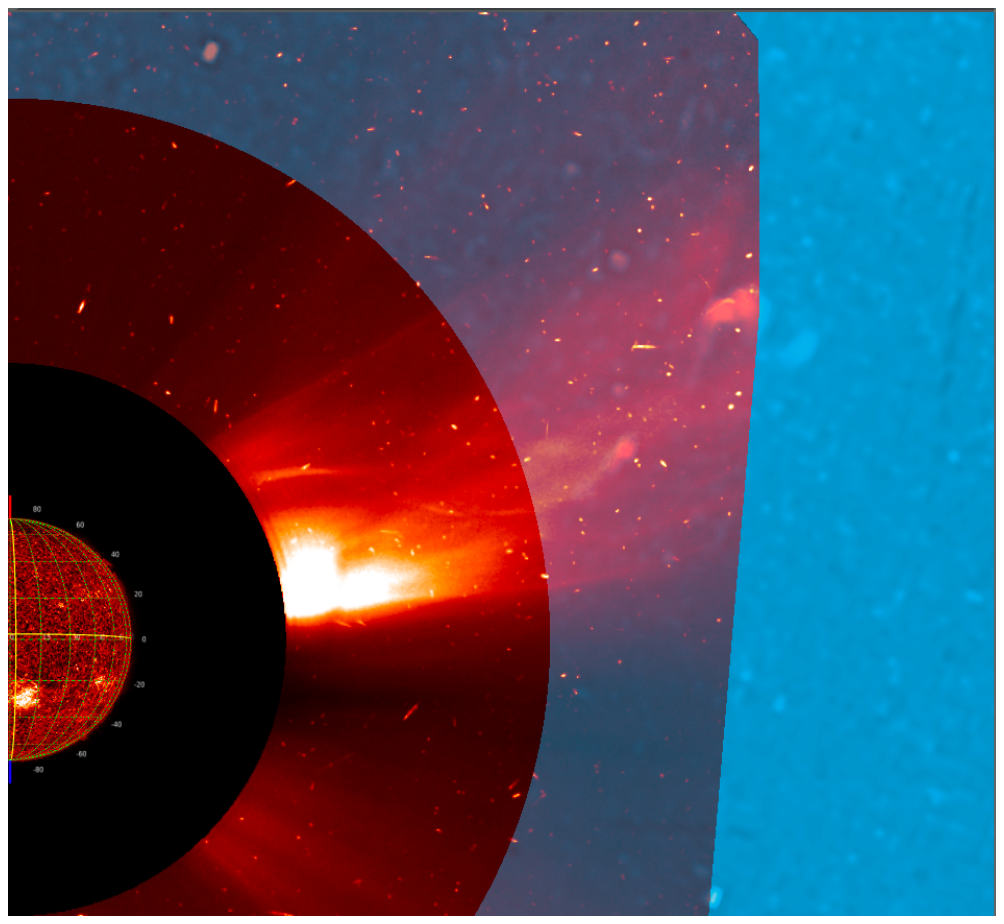}
  \includegraphics[clip, trim=3cm 6cm 3cm 6cm,width=2.9cm]{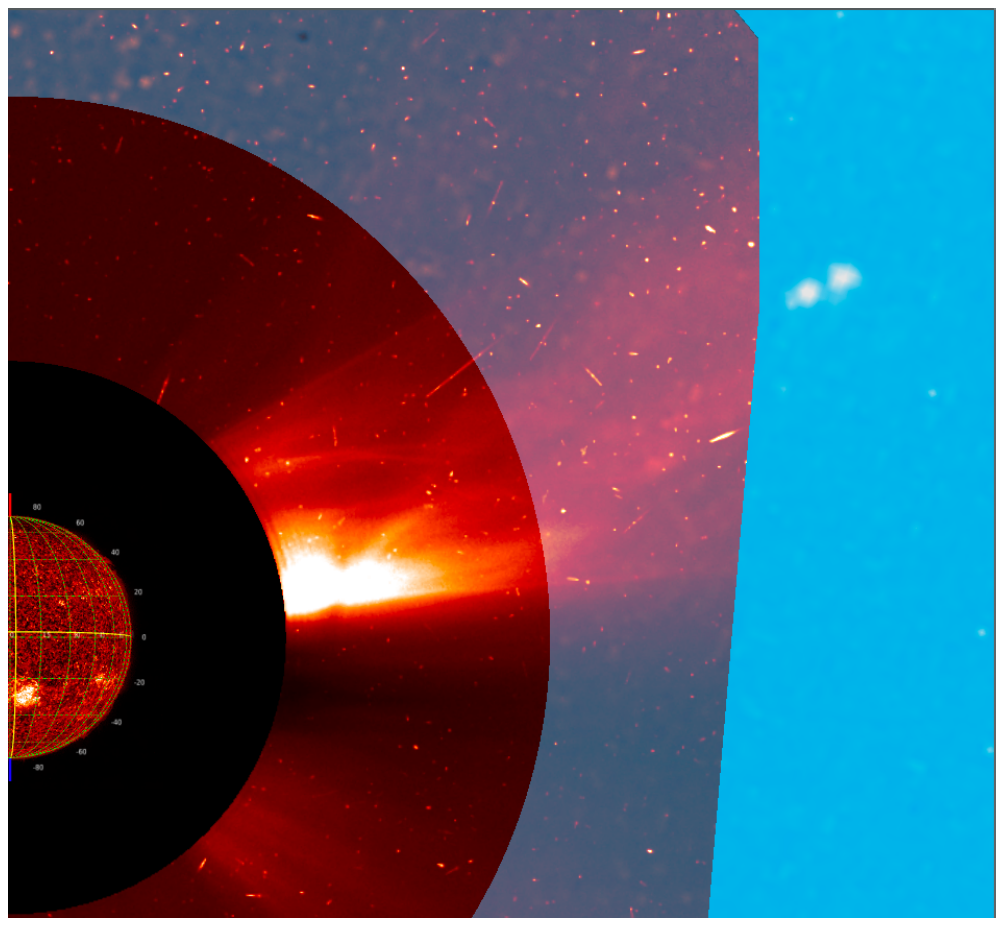}
   \caption{VL frames of SOHO/LASCO-C2 and the UV channel of Metis for the NW CME event on October 28$^\mathrm{th}$ 2021 are shown at 16:13\,UT, 18:15\,UT and 18:39\,UT. Selected images from the different instruments, which observed the CME almost simultaneously, were overlaid with the open-source software JHelioviewer \citep{JHV}.}
              \label{fig:JHV_28oct2021}%
\end{figure}

The north-west event can be seen in Metis pB frames around 15:49\,UT only in two running differences before the two-hour data gap between 16:19 and 18:19\,UT (see panel (a) and (b) in  Fig.\,\ref{fig:pb_UV_eruption_20211028}). The front also appears in the corresponding UV frames, although weaker (panels (f) and (g)). LASCO-C2, on the same LoS as Metis, was observing during the Metis data gap and clearly detected the CME three-part structure as in the first panel on the left in Fig.\,\ref{fig:JHV_28oct2021}. The CME moved out of LASCO's field of view at 17:10\,UT (second panel of Fig.\,\ref{fig:JHV_28oct2021}) and entered into Metis' FoV with a bright ellipsoid-like feature visible in Metis UV frames just after the data gap, from 18:07\,UT to 19:11\,UT, becoming progressively weaker as time passed and revealing two circular formations moving together (see (h), (i) and (l) frames in blue in Fig.\,\ref{fig:pb_UV_eruption_20211028}). Looking at the overlayed images produced with the software JHelioviewer \citep{JHV} in Fig.\,\ref{fig:JHV_28oct2021}, the bright UV structure can be associated with the bright core of the CME. The strong X-class flare is responsible for the production of high-energy solar particles and both Metis channels appear much noisier because of the particle (protons) shower on the detectors. 

\begin{figure}
  \centering
   \includegraphics[clip, trim=2cm 3cm 2cm 2.8cm,width=8cm]{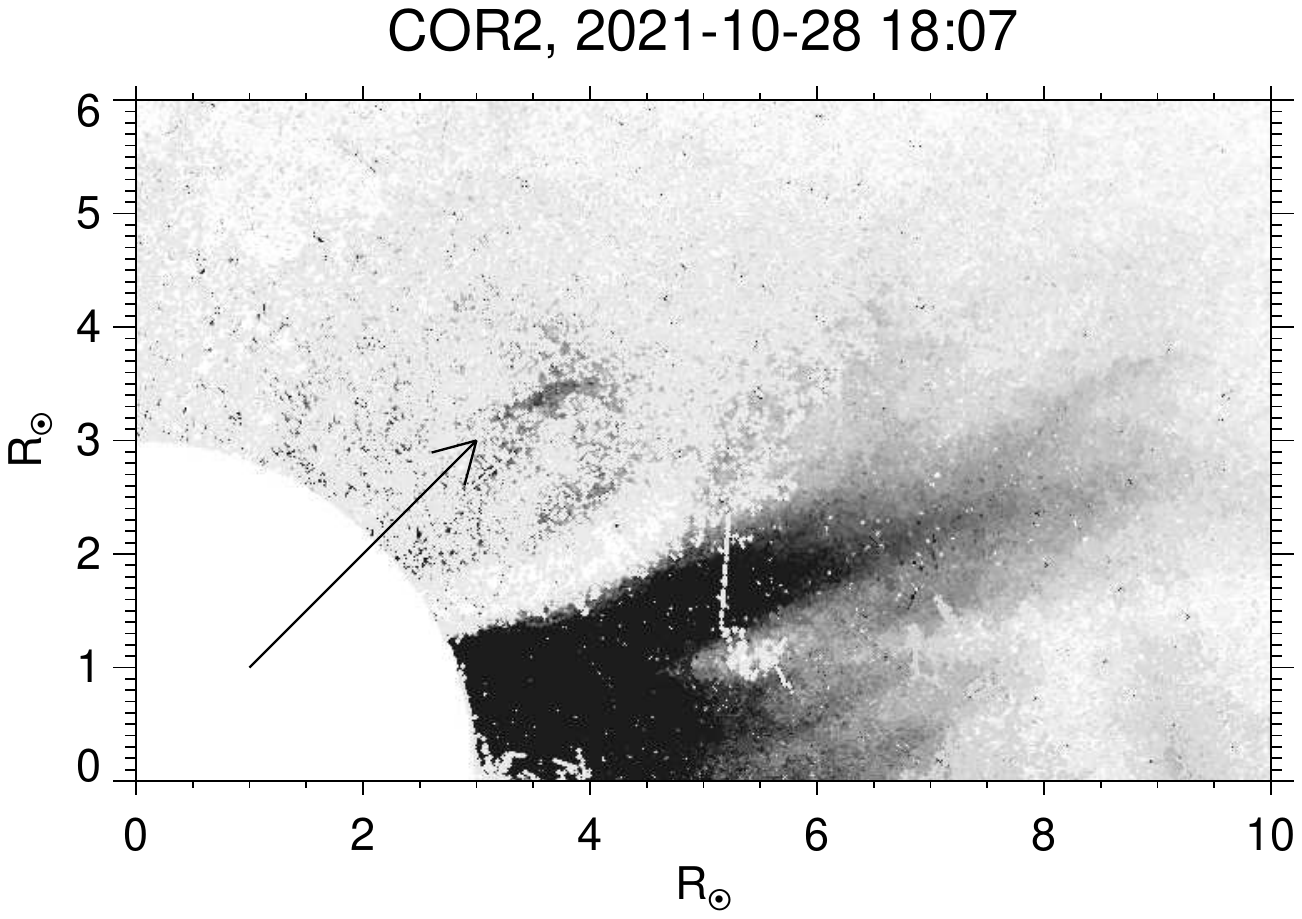}
   \caption{October 28$^\mathrm{th}$ eruption event as seen by STEREO/COR2 coronagraph at 18:07\,UT in reverse colors and used for the triangulation method. The CME core feature corresponding to the Metis bright structure is indicated with the black arrow.}
              \label{fig:COR2_pB_20211028}%
\end{figure}

The NW CME entered the STEREO-A/COR2 FoV at 15:00 UT, and its frames can be used to estimate the direction of the eruption with an image simultaneous with Metis UV channel at 18:08 UT (see Fig. \ref{fig:COR2_pB_20211028}). The Carrington coordinates are 15.6 $\pm$ 0.5$^\circ$ for longitude and 22.9 $\pm$ 0.8$^{\circ}$ for latitude.

It is easy to follow one of the two bright peaks in the ellipsoid-like feature as visible in Metis UV frames to estimate the PoS velocity. This is 244\,$\pm$\,10\,km/s, while the radial value is 246\,$\pm$\,10\,km/s because the direction of the eruption is almost along the Metis PoS.

\begin{figure*}
  \centering
   \includegraphics[clip, trim=0.5cm 7.5cm 1cm 7cm,width=16cm]{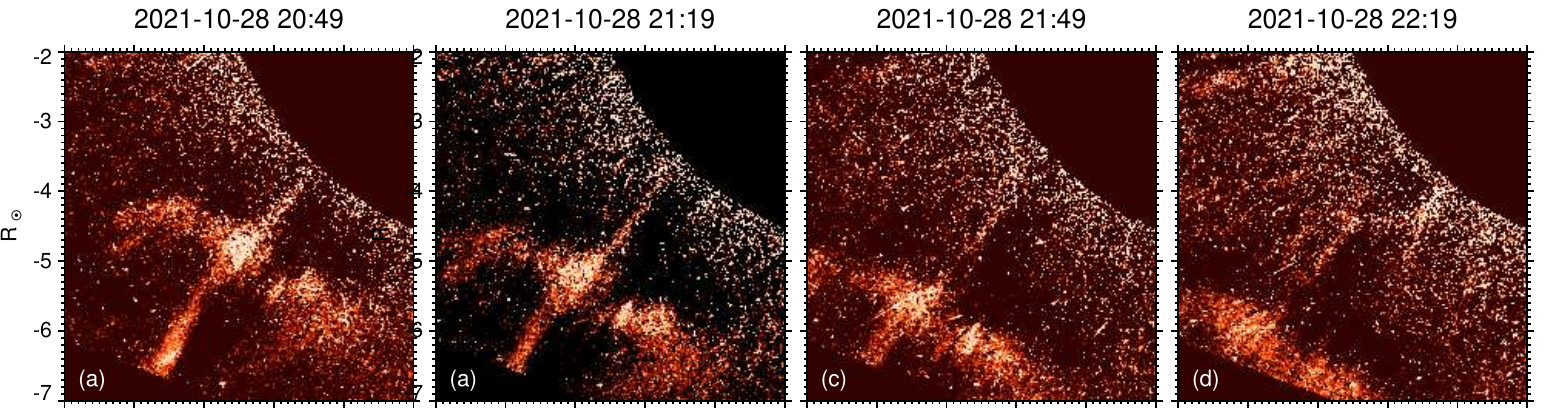}
   \includegraphics[clip, trim=0.5cm 6.7cm 1cm 7.8cm,width=16cm]{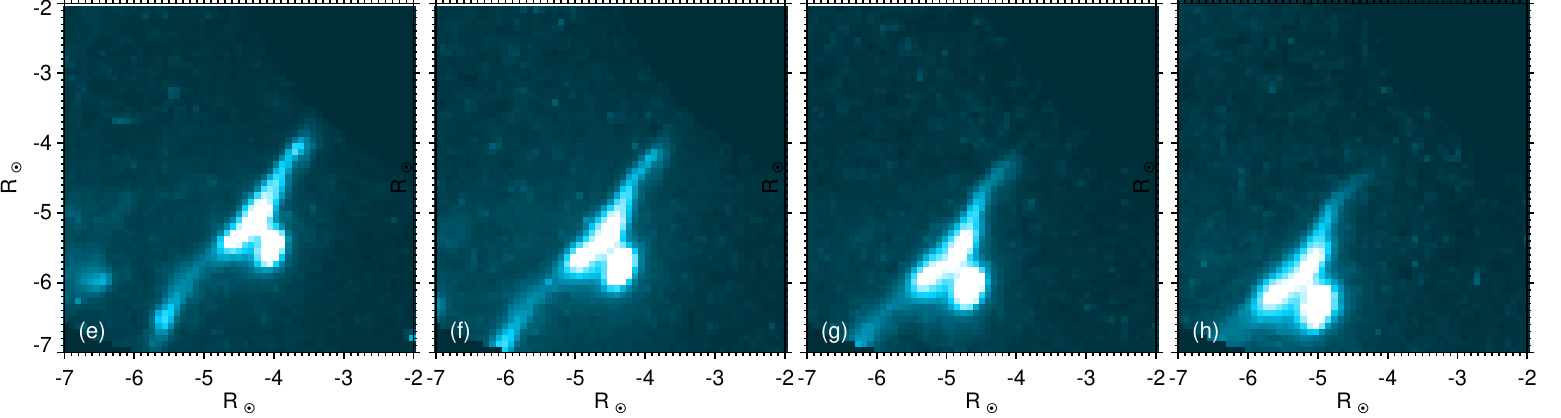}
   \caption{The eruptive phase of the southeastern event on October 28$^\mathrm{th}$. In red, the total \textit{B} running difference, where a long cross-like feature is observable. In blue, the eruption as seen in Metis UV frames.
   The temporal evolution in both channels is available as an online movie (tb\_flare\_20211028.mp4 and UV\_flare\_20211028.mp4).}
    \label{fig:pb_UV_eruption_flare_20211028}%
\end{figure*}

The south-east event entered the Metis FoV after the data gap at 18:05\,UT. In the pB or \textit{B} frames, no clear front phase of the halo CME can be seen because of the gap in the data, but in the UV frames, a very bright fragmented eruption appeared, expanding to 22:57\,UT, as in Fig.\,\ref{fig:pb_UV_eruption_flare_20211028}. 

Thanks again to the STEREO/COR2 frame at 19:07\,UT, it is possible to estimate the direction of the eruption with Carrington coordinates of lon. 258.5\,$\pm$\,0.4\,$^\circ$ and lat. $-$8.3\,$\pm$\,0.6\,$^\circ$, almost on the SolO-Earth LoS (see also the black arrow in Fig.\,\ref{fig:gse_all_events}).

The velocity on the PoS of 152\,$\pm$\,1\,km/s is estimated following the brightest fragment on the left of the UV structure (green dashed line in Fig.\,\ref{fig:UV_velocity_track_all}). The radial velocity, in this case, is 437\,$\pm$\,4\,km/s, the highest value in this series of events and the only event showing deceleration and observed far from the Sun disk above 20\,R$_\sun$. 
The full velocity of the bulk motion of the ejecta in the first hour after the flare is $\sim$\,596\,km/s and was computed in \cite{Xu_flare_celocity_2022ApJ...931...76X} by combining the imaging of the 304\,\r{A} image series taken by either EUI/FSI on SolO or EUVI on STEREO-A and spectroscopic observations from the Extreme ultraviolet Variability Experiment onboard SDO \citep[EVE;][]{EVE_2012SoPh..275..115W}. This value is $\sim$\,35\,$\%$ above the Metis estimate, which detected it many hours after the flare. Far higher are the estimated velocities on the PoS of the halo CME leading edge and of the white light shock, obtained by LASCO-C2, which are respectively $\sim$1240\,$\pm$\,40\,km/s and $\sim$1640\,$\pm$\,40\,km/s \citep{Papaioannou_xflare_2022A&A...660L...5P}. 

\section{Discussion}
\label{sec:discussion}

\subsection{Morphology and kinematics}
\label{subsec:morphology_kinematics}

\begin{figure}
  \centering
  \includegraphics[clip, trim=4.8cm 1cm 2cm 1cm,width=9cm]{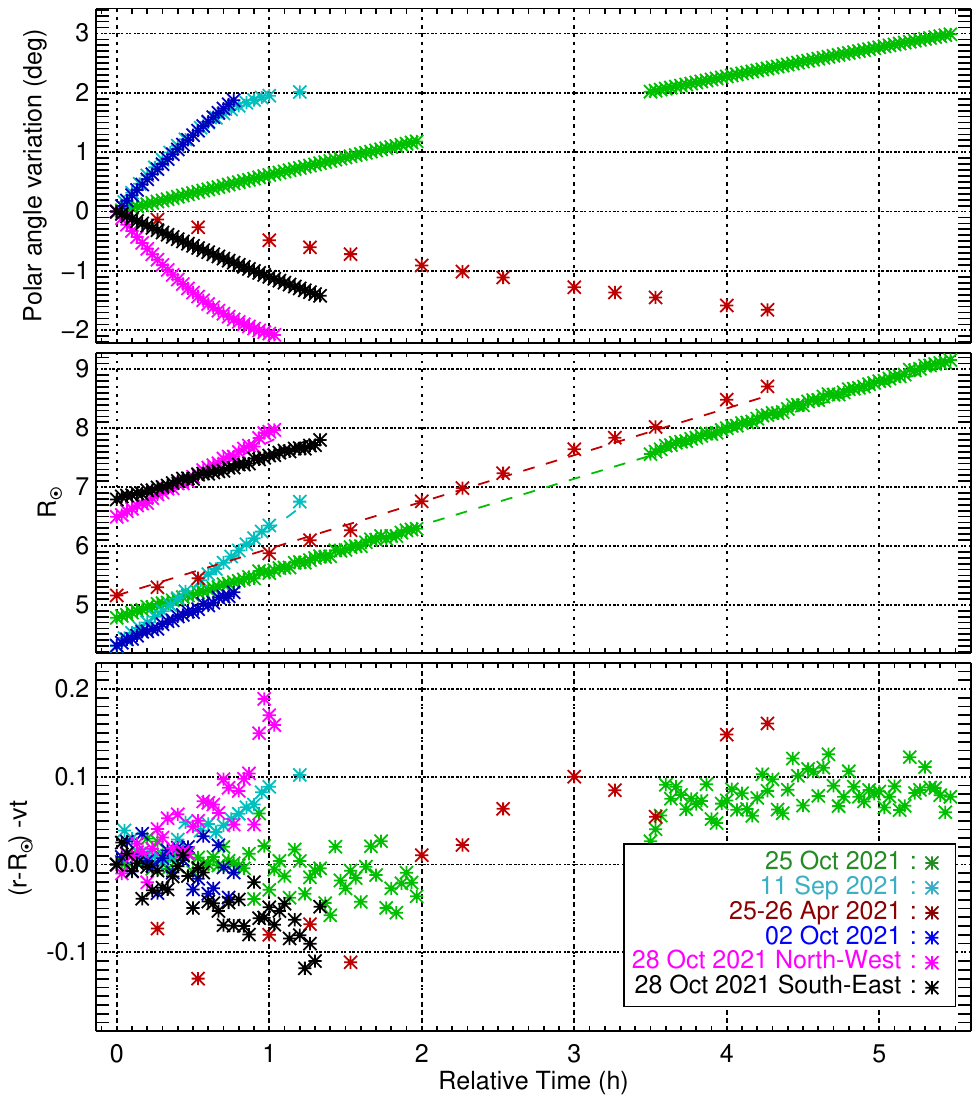}
   \caption{Kinematics of the events. Relative positions of the trailing features of the UV bright structures, frame by frame, expressed as polar angle variation (top panel) and in distance on the PoS in solar radii (central panel) with time. The reference values for the polar angle and time are listed in Tab.\,\ref{table:events_info}. The residual of the position $r$ estimated from the PoS velocity fit, $v$, is shown in the lower panel, providing the acceleration trend of the structures, whose values are also displayed in Tab.\,\ref{table:events_measured_info}.}
              \label{fig:polarang_position_all}%
\end{figure}

\begin{figure*}[!t]
   \centering 
       \includegraphics[clip, trim=5cm 1cm 5cm 1cm,width=6cm]{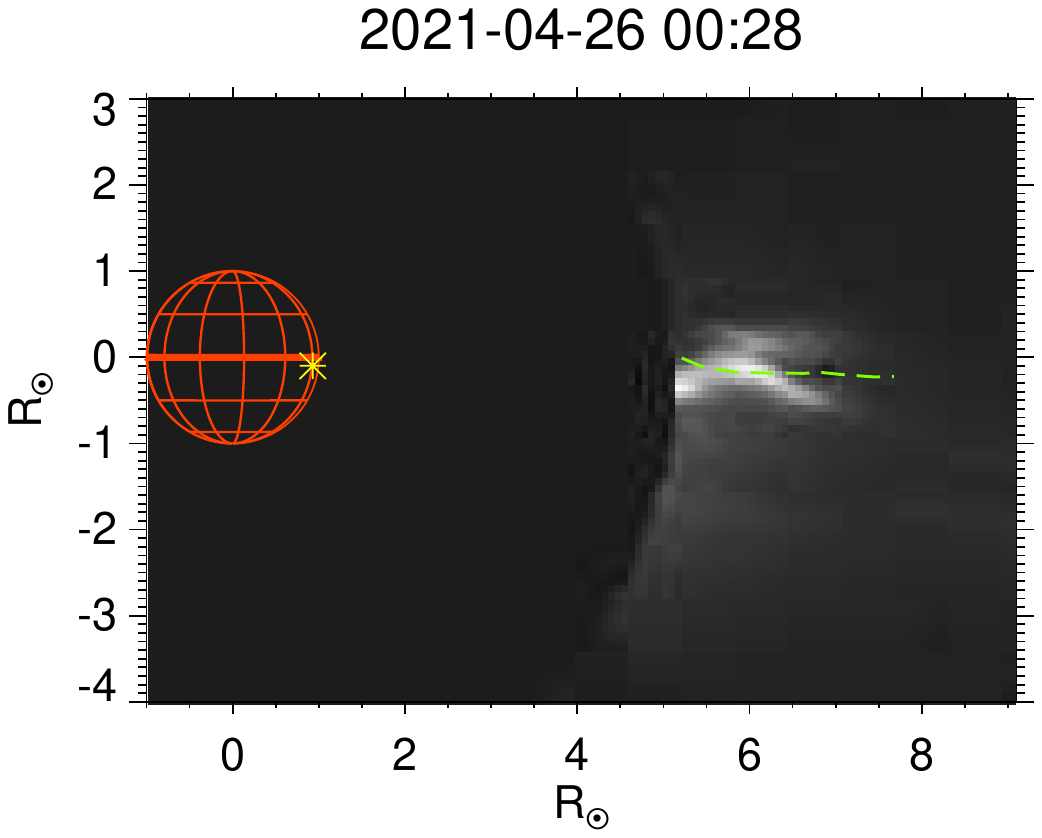}
       \includegraphics[clip, trim=5cm 1cm 5cm 1cm,width=6cm]{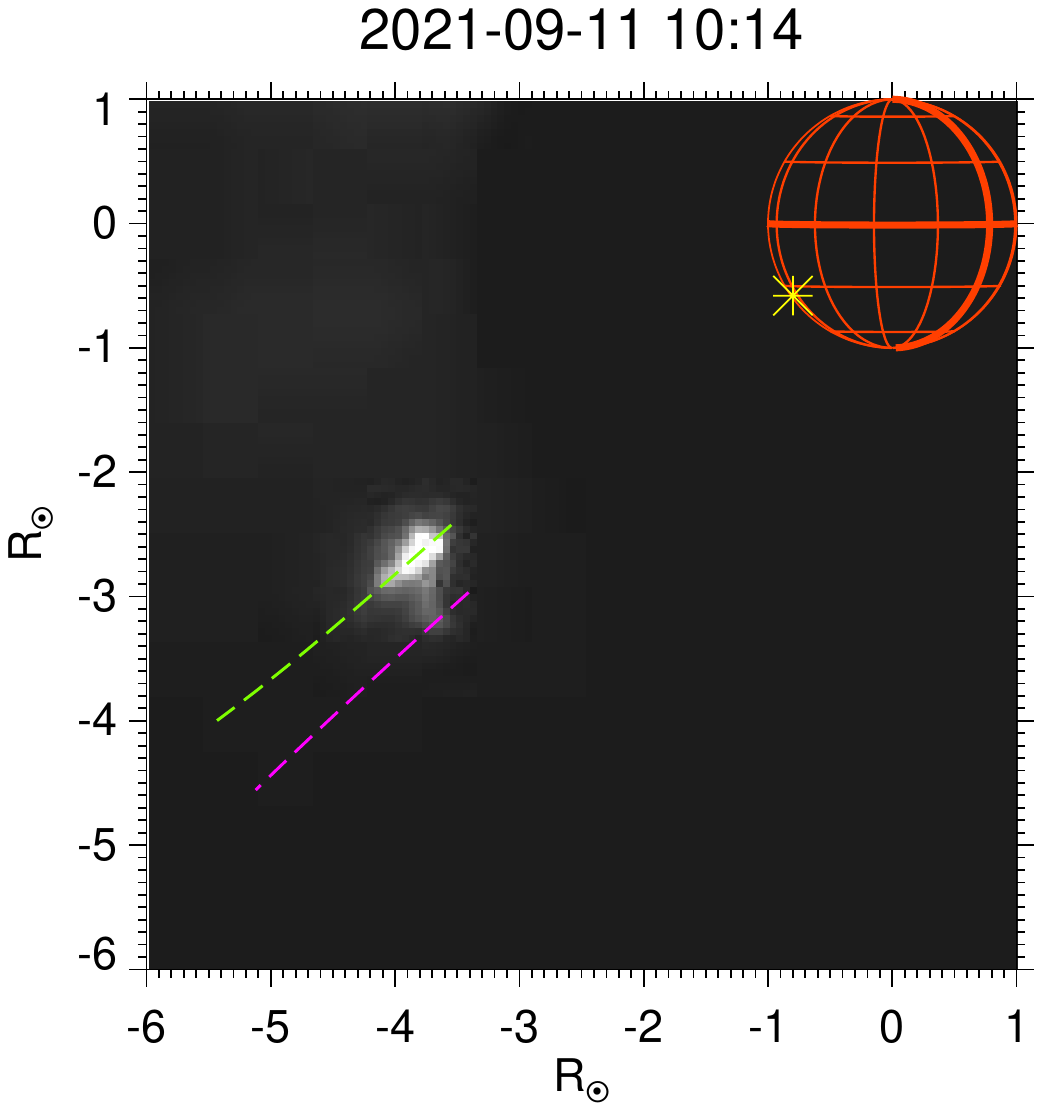}
       \includegraphics[clip, trim=5cm 1cm 5cm 1cm,width=6cm]{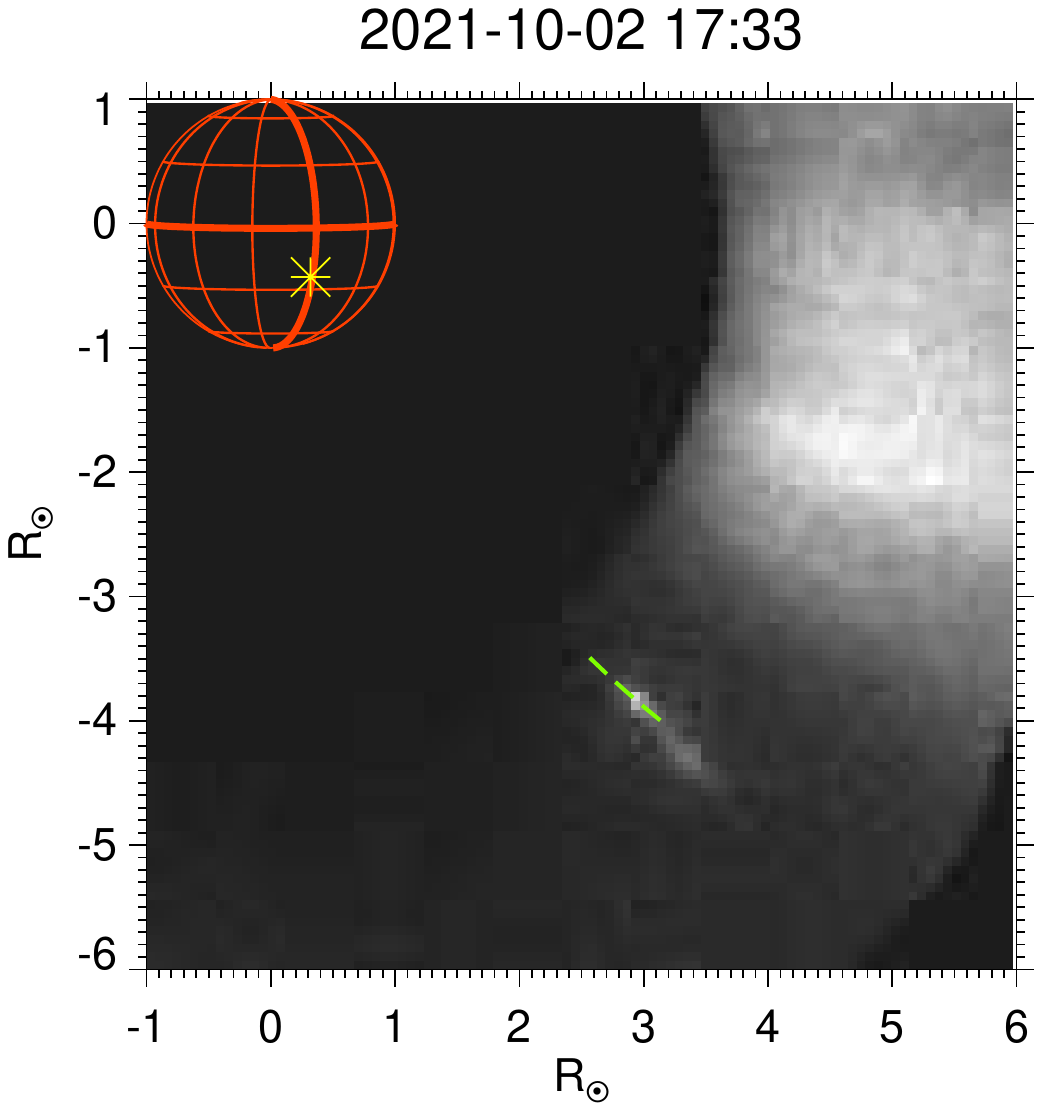}
       \includegraphics[clip, trim=5cm 1cm 5cm 1cm,width=6cm]{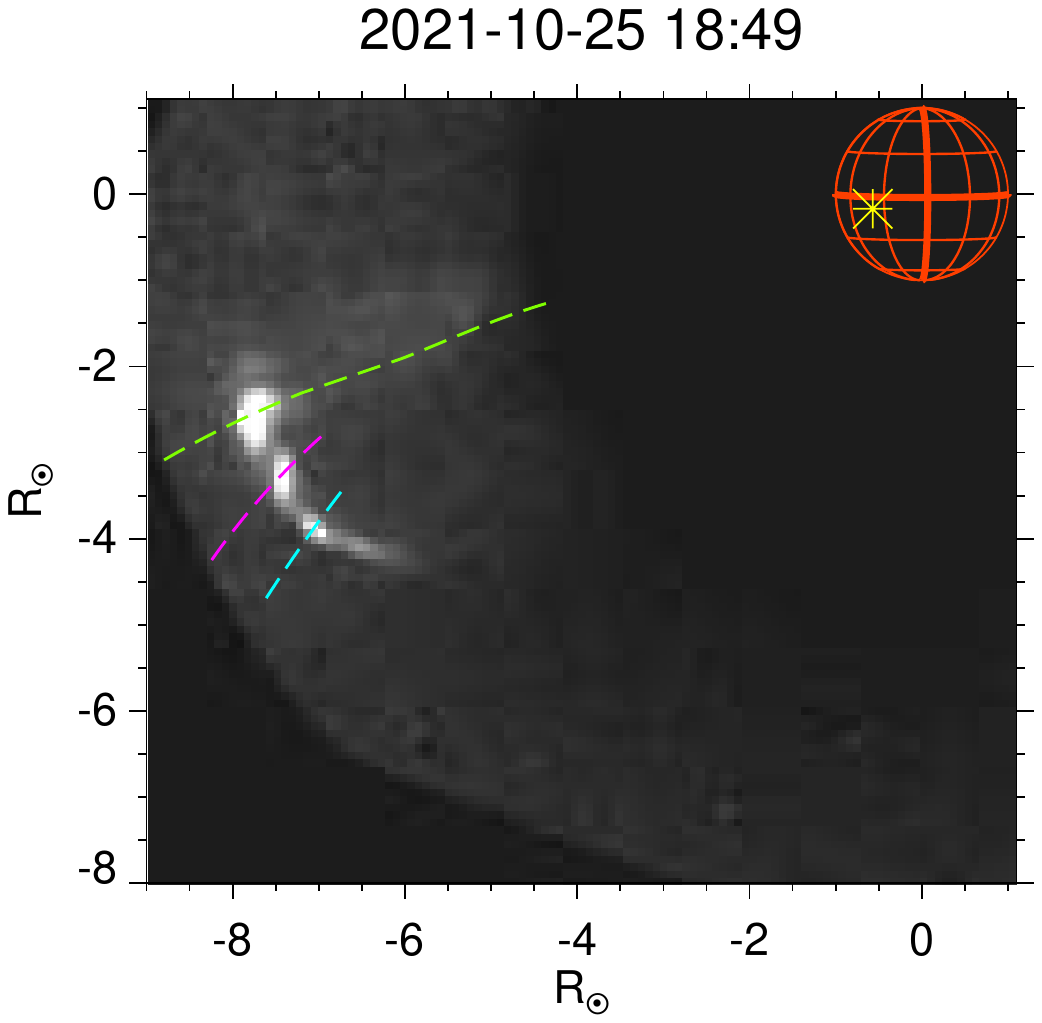}
       \includegraphics[clip, trim=5cm 1cm 5cm 1cm,width=6cm]{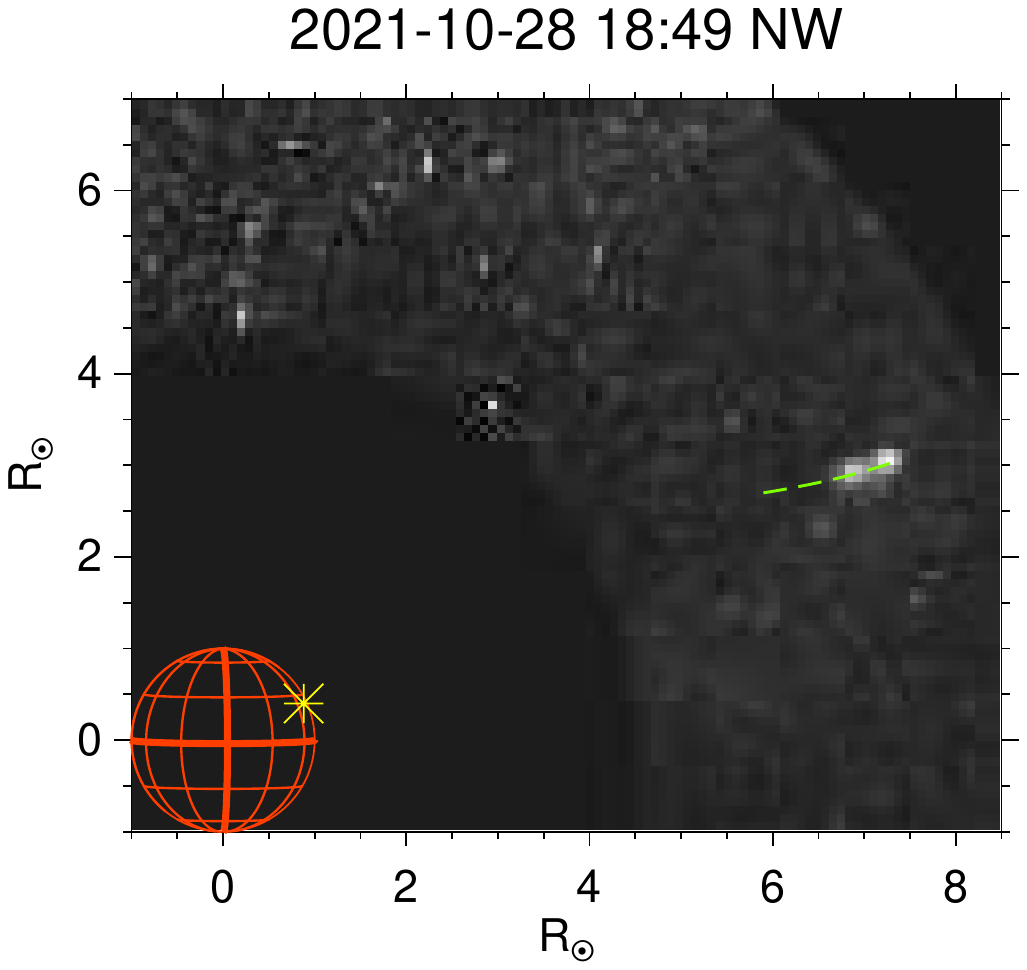}
       \includegraphics[clip, trim=5cm 1cm 5cm 0.8cm,width=6cm]{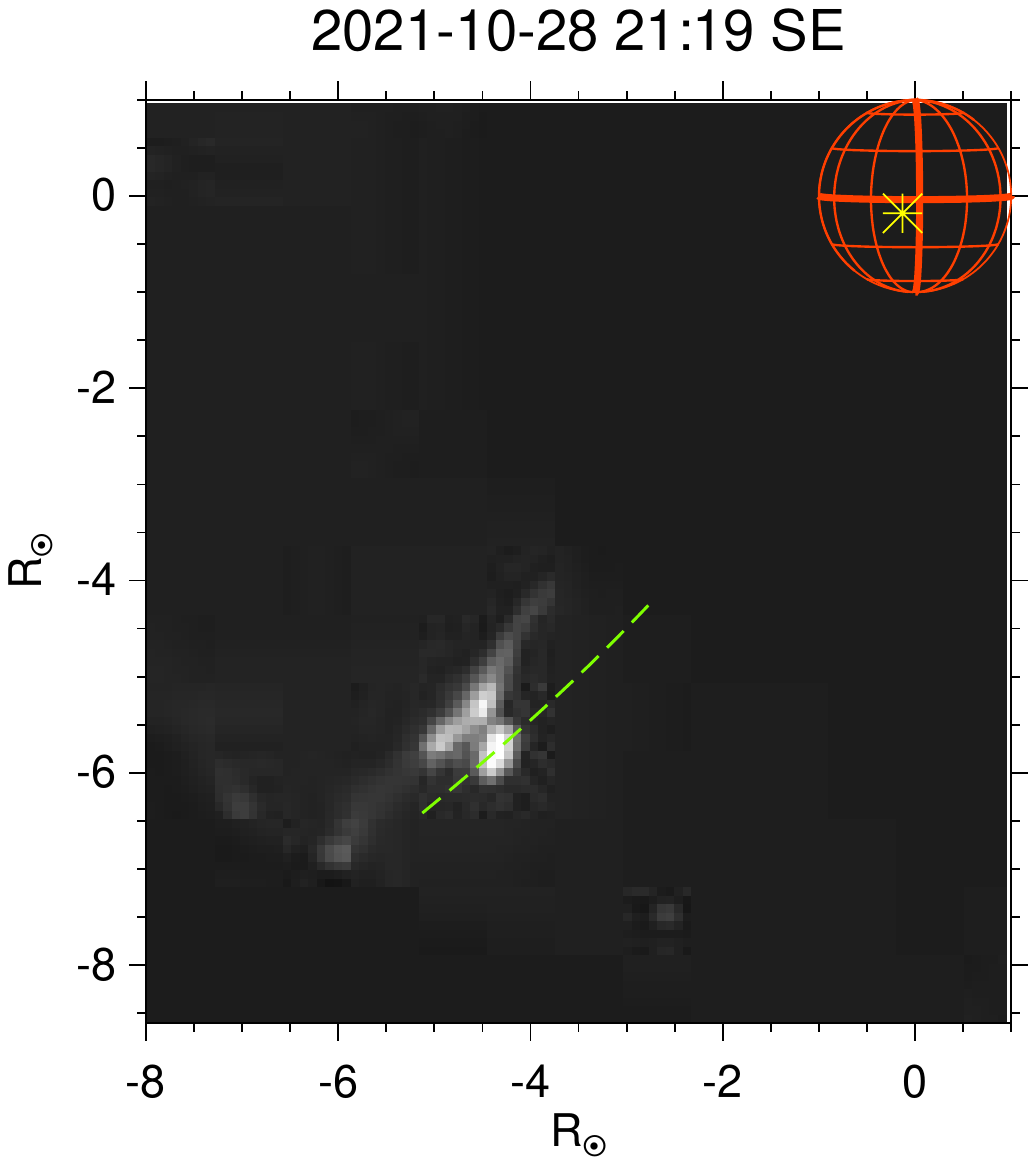}
   \caption{For each event, we show one Metis UV frame identified by the date in the title, where the bright feature is visible. All the frames are taken at the reference time in column two of Tab.\,\ref{tab:electron_densities}, except for October 25$^\mathrm{th}$, for which we chose a frame that shows the evolution of the three bright components of the structure. The green dashed lines are the trajectory of the trailing features in all the available UV frames used to trace the position as a function of time. Other colored dashed lines trace different features of the structures, helping in visualizing the evolution of their angular width. The yellow stars derive from the projection of the Carrington coordinates in columns seven and eight of Tab.\,\ref{table:events_measured_info} on the Metis plane of sky and extrapolated at the Sun's surface.}
              \label{fig:UV_velocity_track_all}%
    \end{figure*}

The spatial and temporal scales of the six events described so far can be summarized in Fig.\,\ref{fig:polarang_position_all} by tracking the trailing features of each prominence in the Metis FoV, which basically correspond to the bright peaks observed in UV frames, as also outlined with green dashed lines in Fig.\ref{fig:UV_velocity_track_all}. 

The structures observed on September 11$^{\mathrm{th}}$ and October 28$^{\mathrm{th}}$-NW show larger deflection in the radial path in time, as shown in the top panel of Fig.\,\ref{fig:polarang_position_all}. As discussed in several recent works \citep[see review by][]{Shen_propagation_CME_2022RvMPP...6....8S}, the deflection of CMEs in corona is mainly driven by the ambient magnetic structure (e.g. coronal holes) and quantitative analyses suggest that CMEs tend to have higher deflection rates below 4\,R\,$_\sun$ \citep{Gui_CMEdeflection_2011SoPh..271..111G}. The magnitude and direction of the deflection are determined by CME parameters such as mass and velocity, as well as solar magnetic field intensity and gradient. 
Instead, we observe a longitudinal deflection component at distances greater than 4\,R\,$_\sun$ only for these two events following the core of the CME structure instead of the leading edge of the CME front, as is usually found in literature. 

The apparent speed on the Metis PoS derived from the slopes in the height-time plots shown in the middle panel of Fig.\,\ref{fig:polarang_position_all} are listed in Tab.\,\ref{table:events_measured_info}.
The estimated radial (de-projected) velocities range from 180 to 440\,km/s. Except for the fastest event, the one on October 28$^\mathrm{th}$-SE, all others have velocities compatible with the slow-speed solar wind range as reported in \cite{Abbo_slow_solar_wind_rev_2016SSRv..201...55A} and \cite{first_light_2021A&A...656A..32R} (see also reviews in \cite{Antonucci_slow_wind_2023PhPl...30b2905A} and \cite{Antonucci_corona_review_2020SSRv..216..117A}.  
The PoS velocities vary between 150 and 375 km/s, and they are comparable for the events of October 25$^\mathrm{th}$ and April 25$^\mathrm{th}$, as well as for October 2$^\mathrm{nd}$ and 28$^\mathrm{th}$-NW. These values are also compatible with the statistics of CME observations performed with SOHO/LASCO (in the catalog from CDAW Data Center by NASA and The Catholic University of America) and reported in \cite{Yashiro_CME_SOHO_catalogue_2004JGRA..109.7105Y} at solar minimum phase.

The bottom panel of Fig.\,\ref{fig:polarang_position_all} shows a significant acceleration for the September 11$^\mathrm{th}$ and both October 28$^\mathrm{th}$ eruptions; the acceleration for the April 25$^\mathrm{th}$ is on the other hand very small (see also measured de-projected values in Tab.\,\ref{table:events_measured_info}). 

We can also compare the measured radial velocity and acceleration profiles with typical values observed during the solar cycle minimum phase as reported by the statistical analysis of 95 impulsive CME events that occurred during the Solar Cycle No. 23/24 observed in STEREO EUVI, COR1 and COR2 in the work of \cite{Bein_acceleration_2011ApJ...738..191B}. Our analyses are in agreement with the general behavior that at distances greater than two solar radii the velocities are nearly constant and that if there are residual accelerations, they occur within a solar radius at the beginning of the eruption \citep[see also similar results from][]{Joshi_CME_acceleration_2011ApJ...739....8J}.

In many theoretical models and low corona observations, CMEs are expected to show rotation on the basis that shear or helical instabilities are involved in the eruption initiation \citep{CME_rotation_2009ApJ...697.1918L}. In white-light coronagraph images, the rotation is usually detected as a deviation from radial expansion, which we find for the September 11$^\mathrm{th}$ and October 28$^\mathrm{th}$ events in the first panel of Fig.\,\ref{fig:polarang_position_all}. But this is not the only element to consider to determine an obvious rotation. The projected angular width is also expected to increase \citep{Vourlidas_rotating_cme_2011ApJ...733L..23V}, as shown in Fig.\,\ref{fig:UV_velocity_track_all} for example for the October 25$^\mathrm{th}$ event. Our evidence of the presence of a rotation of the structures is only qualitative. The rotation parameters could also be determined with the standard analysis procedure for 3D reconstruction of SECCHI data using the Graduated Cylindrical Shell (GCS) model \citep{Thernisien_GCS_2009SoPh..256..111T} on the CME front phase. The result could in principle be validated with three S/C observations, but in our cases, the procedure is not applicable either because the CME front is too faint to be clearly identified, or because there are no simultaneous images taken with other coronagraphs.


\subsection{Intensity radial profiles}\label{subsec:intensity}

Tracing the UV \ion{H}{i}~\Lya\ emission of these structures may provide useful information for diagnosing physical plasma parameters such as density, temperature and outflow velocity relative to the source of the photons at the base of the corona \citep{Withbroe_1982SSRv...33...17W,Noci_1987ApJ...315..706N,Dolei_outflow_2018A&A...612A..84D}.

In the optically thin regime, the observed \ion{H}{i}~\Lya~radiance is simply the integral along the line of sight of the line emissivity, $j(\lambda,\vec{x},\vec{n})$, where $\vec{x}$ is the vector of the coordinate of the emitting plasma and $\vec{n}$ is the direction of emission of the \Lya\ photons.  It is customary to assume that in corona the excited levels are populated only from the ground state; hence, the line emissivity can be shown to be the sum of a component excited by collision with electrons, $j_c$, and a component due to resonant scattering of \Lya\ photons emitted near the solar surface, $j_r$ \citep{Gabriel_1971SoPh...21..392G,Withbroe_1982SSRv...33...17W,Noci_1987ApJ...315..706N}.  Detailed expressions of the \Lya\ emissivity can be found in literature \citep[e.g.][]{Dolei_2015}; here we are mainly concerned with the dependence of the observed radiance on height, and therefore it suffices to highlight the main parameters the line emissivity depends on.  In particular, the collisional component depends quadratically on the local density:
\begin{equation}
  j_c \sim n_\mathrm{H} \, n_\mathrm{e} G(T_\mathrm{e}) \; , \label{eq:Lya_C}
\end{equation}
where $n_\mathrm{H}$ is the neutral hydrogen density, $n_\mathrm{e}$ is the electron density, and $G(T)$ is a function that takes into account the dependence on atomic parameters and the electron temperature, $T_\mathrm{e}$. The radiative component, on the other hand, depends linearly on density but also on the geometry of the scattering process, on temperature and, of course, on the mean radiance of the solar disk,
$\bar{I}_\mathrm{disk}$:
\begin{equation}
  j_r \sim n_\mathrm{H} \, D(v,T_\mathrm{e}) \, W(r) \, \bar{I}_\mathrm{disk,\Lya} \; , \label{eq:Lya_R} 
\end{equation}
For convenience, we have highlighted the so-called Doppler dimming effect through a factor, $D(v,T_\mathrm{e})$, dependent on the radial velocity, $v$, of the plasma and on temperature. The dependence of emissivity on the distance from the disk center, $r$, comes mainly from the angular size of the photoexciting source (the solar disk) as seen from prominence hydrogen atoms, often referred to as the ``dilution factor'', $W(r)$:
\begin{equation}
  W(r) = \frac{1}{2} \; \{ 1 - [1 - (R_\sun/r)^2]^{1/2} \} \label{eq:W}
\end{equation}
\citep[see, e.g.:][chapters 19 and 20]{Hubeny-Mihalas:2015}.  In the above expression, we have assumed that the temperature of hydrogen atoms is isotropic and equal to the electron temperature, but this assumption is often relaxed.  More general expressions for the Doppler dimming factor, which include all relevant details such as limb brightening of the solar disk at the \Lya\ wavelength, the scattering geometry etc., have been evaluated and discussed by various authors \citep[e.g.:][]{Dolei_2015,Dolei_outflow_2018A&A...612A..84D}.  

It is useful to compare the above expressions for the UV \ion{H}{i}~\,\Lya\ emissivity to the corresponding expression for the polarized VL emission, pB, from a static corona as detected simultaneously by Metis. It can be shown that the pB emissivity is proportional to electron density and depends on the geometry of the Thomson scattering process \citep[e.g.][and references therein]{Vourlidas-Howard_2006,Howard-DeForest_2012}; to facilitate a comparison with Eq.~\ref{eq:Lya_R}, we choose here to highlight the dependence on the distance from the photoexciting source (the solar disk) through the dilution factor:
\begin{equation}
  j_\mathrm{pB} \sim n_\mathrm{e} \, W(r) \, \bar{I}_\mathrm{disk,pB} \; , \label{eq:pB}
\end{equation}
noting however that the dependence on $r$ through the dilution factor is a good approximation at distances larger than a few solar radii. 
The above equations show that, for instance, the pB emissivity and the emissivity of the \Lya\ radiative component have a similar dependence on height, but the \Lya\ radiative emissivity depends in addition on a Doppler dimming factor and on the ratio $n_\mathrm{H}/n_\mathrm{e}$, which in turn depends on the ionization degree of the plasma.

\begin{figure}
  \centering
  \includegraphics[clip, trim=1cm 1cm 3cm 1cm,width=9.5cm]{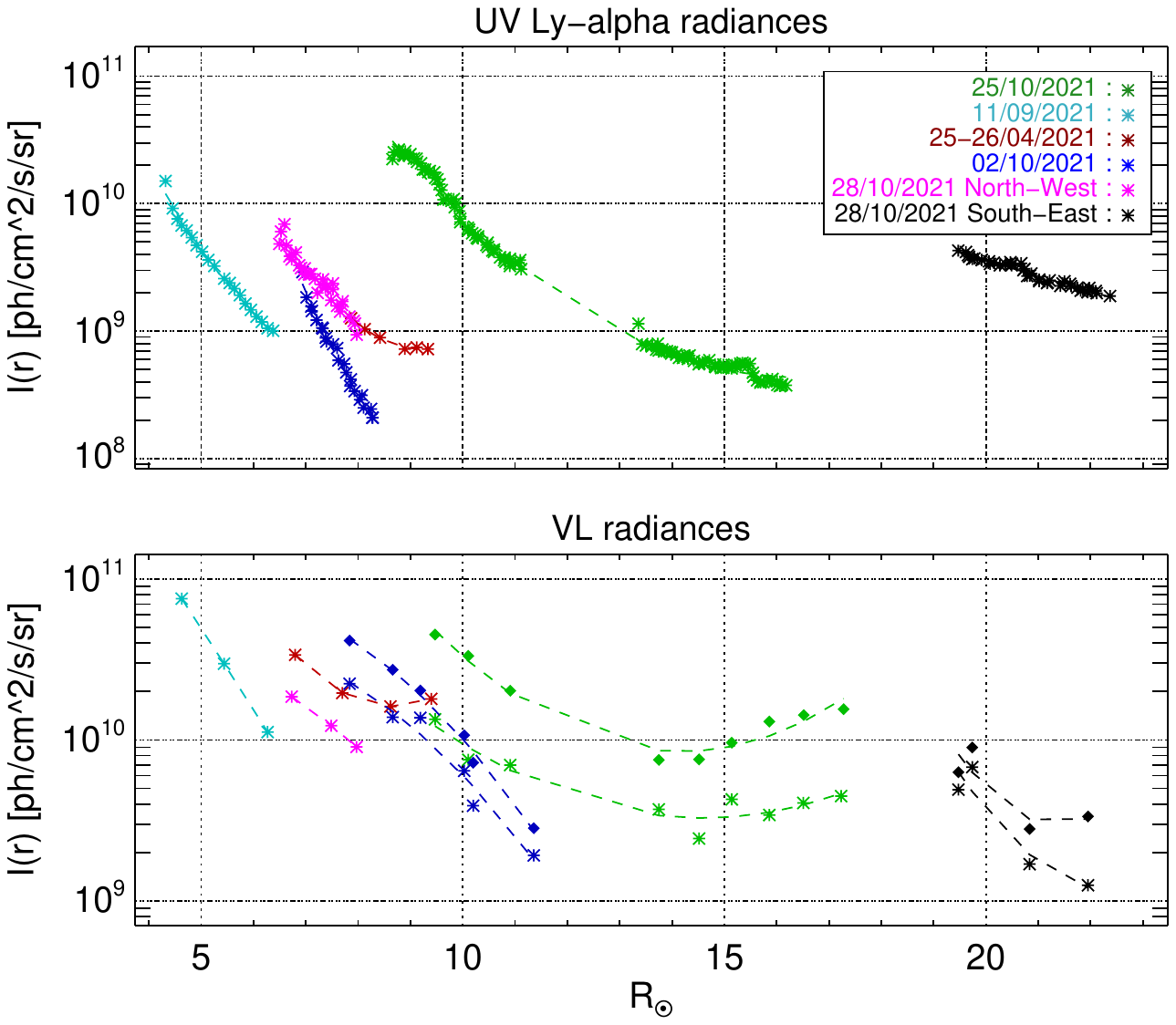}
   \caption{\textit{Top panel}: Mean UV \Lya\ intensity within the area of the observed structures selected for each frame as described in the text (Sec.\,\ref{subsec:intensity}). \textit{Bottom panel}: peak of brightness in the base difference of polarized signal (indicated by stars) and total brightness (indicated by diamonds) at the position corresponding to the highest luminosity peak within the structure area in the respective UV frames (see Sec.\,\ref{subsec:intensity}). \textit{Both panels}: Dashed lines are a quadratic fit to the shown mean radiance as function of de-projected radial distance.}
              \label{fig:area_br_rsun_all}%
   \end{figure}

The top panel of Fig.\,\ref{fig:area_br_rsun_all} shows the mean UV radiance for each observed bright feature as a function of radial distance. To estimate the contribution of the solar foreground and background corona, for each UV frame we consider a box of approximately 70$\times$70 pixels centered around the bright structure and select only the pixels outside the visible feature. We then estimate the background contribution by fitting the radiances of those pixels with a surface parameterized as %
\begin{math} F(x,y) = \sum_{i,j} k_x(j,i) x^i y^j \end{math}. %
The result is finally subtracted from all the pixels in the box, including those at the position of the bright core. After estimating the background, we select the pixels in each frame that exceed the background by at least 10\%, thus defining the area of the UV feature under study.  We show in Fig.\,\ref{fig:area_br_rsun_all} 
the corresponding average UV radiances computed within.

As already discussed in Sec.\,\ref{subsec:11_sep}, a feature smearing effect is usually visible in the VL frames due to the longer exposure time than UV frames. 
For this reason, we chose not to compute average pB values with the procedure described above for the UV channel. In those VL frames that are simultaneous to a UV image, we instead measured the peak pB value within the area of the prominence identified in the corresponding UV frame.  
More specifically, for each pB base difference (base frames are shown in Appendix\,\ref{sec:appendix_source_region}, Fig.\,\ref{fig:pB_base}), we selected the pB profile along the polar angle corresponding to the features identified in the co-temporal UV frames. For each pB profile, we then estimated the background by computing the fifth percentile. 
The peak values above this background are shown in the bottom panel of Fig.\,\ref{fig:area_br_rsun_all}.
For some of the events, as explained at the beginning of Sec.\,\ref{sec:observations}, we chose to analyze total brightness images, and we estimate the peak of \textit{B} value with the same procedure used for the pB; the results are shown in the same figure.

It is worth noting that almost all of the structures under study present a progressive decrease in the UV and pB intensity profiles with the outward expansion of coronal plasma in time as the ejected material travels outwards through the corona, decreasing its emission with the radial distance. Slight deviations or oscillation in some frames of radial profiles are due to background noise in the UV frames, as we will explain later in this section and in Appendix\,\ref{sec:appendix_UVDA_issue}, while cases where pB profiles reach sort of a plateau are discussed on a case-by-case basis.

\begin{figure*}[!ht]
   \centering
   \includegraphics[clip,trim=2cm 1.8cm 2cm 3cm,width=9cm]{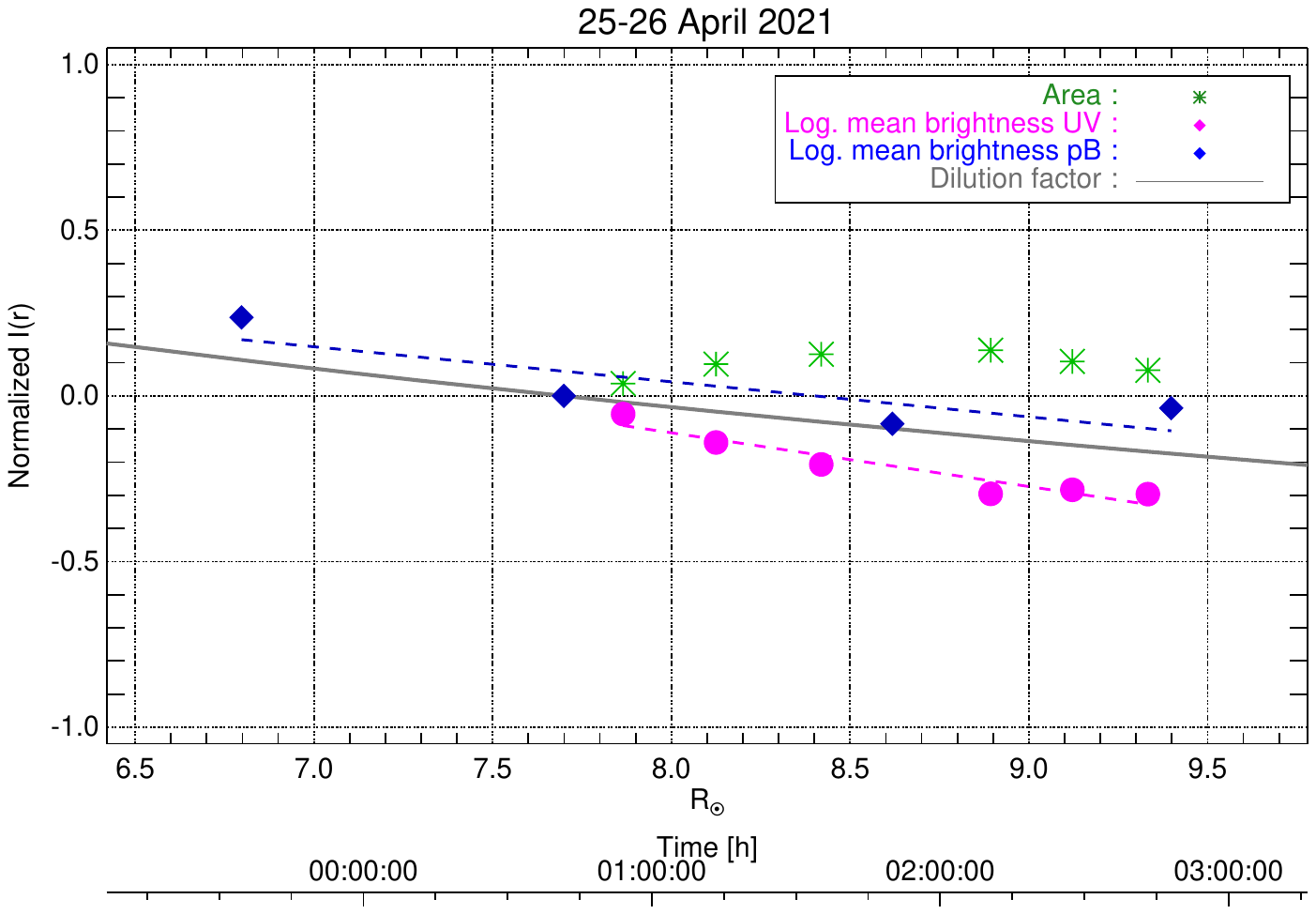}
   \includegraphics[clip,trim=2cm 1.8cm 2cm 3cm,width=9cm]{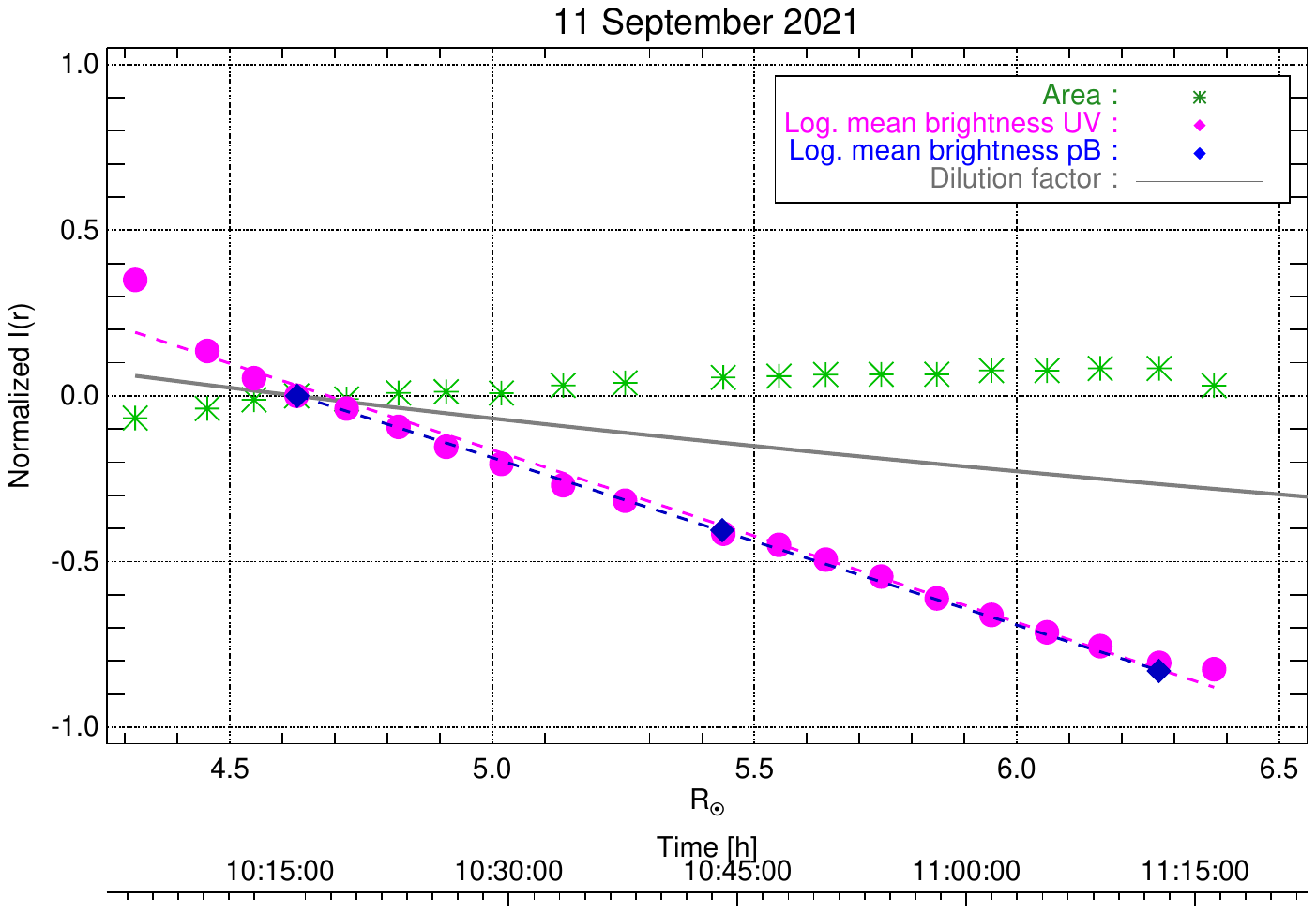}
   \includegraphics[clip,trim=2cm 1.8cm 2cm 3cm,width=9cm]{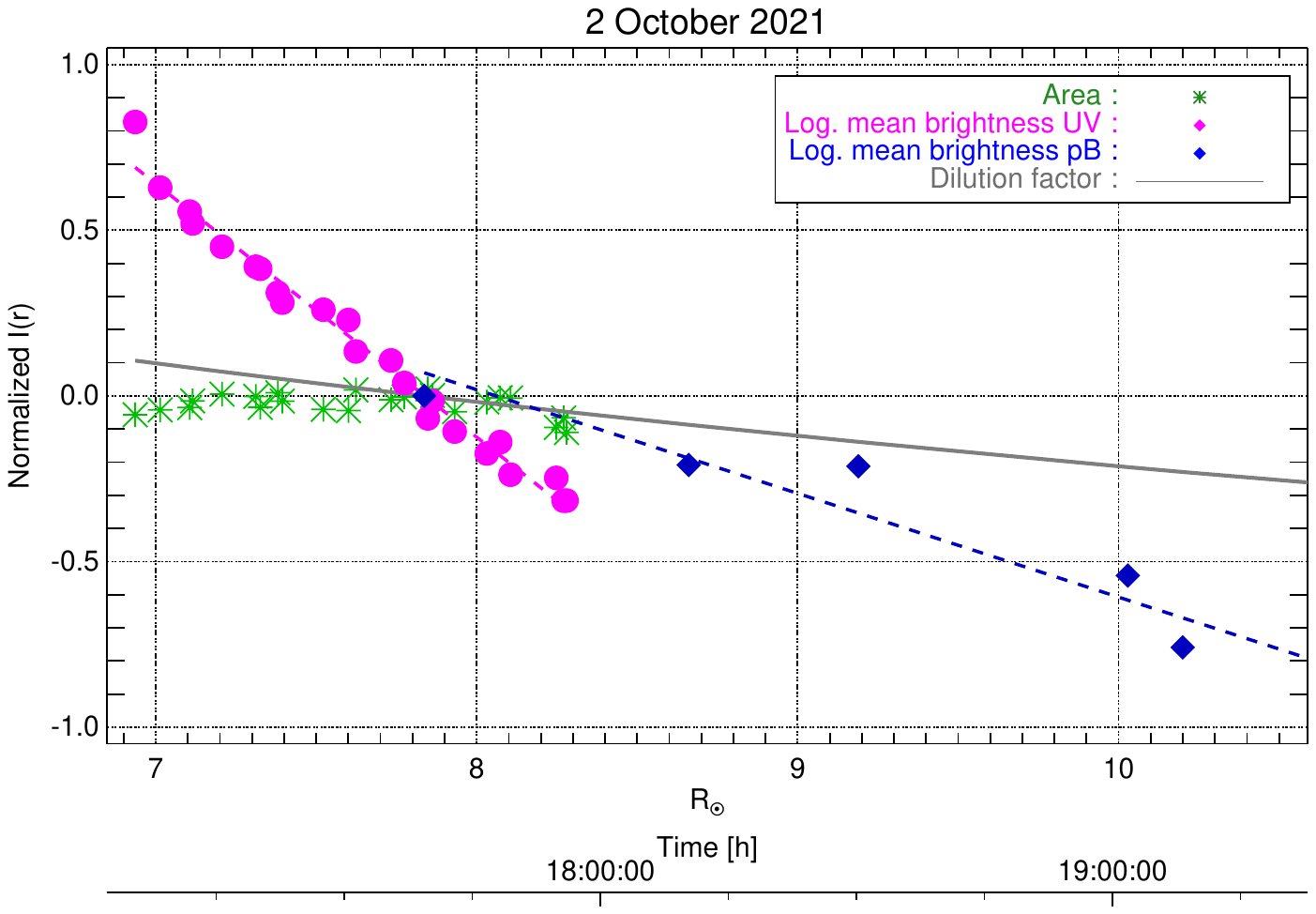}
   \includegraphics[clip,trim=2cm 1.8cm 2cm 3cm,width=9cm]{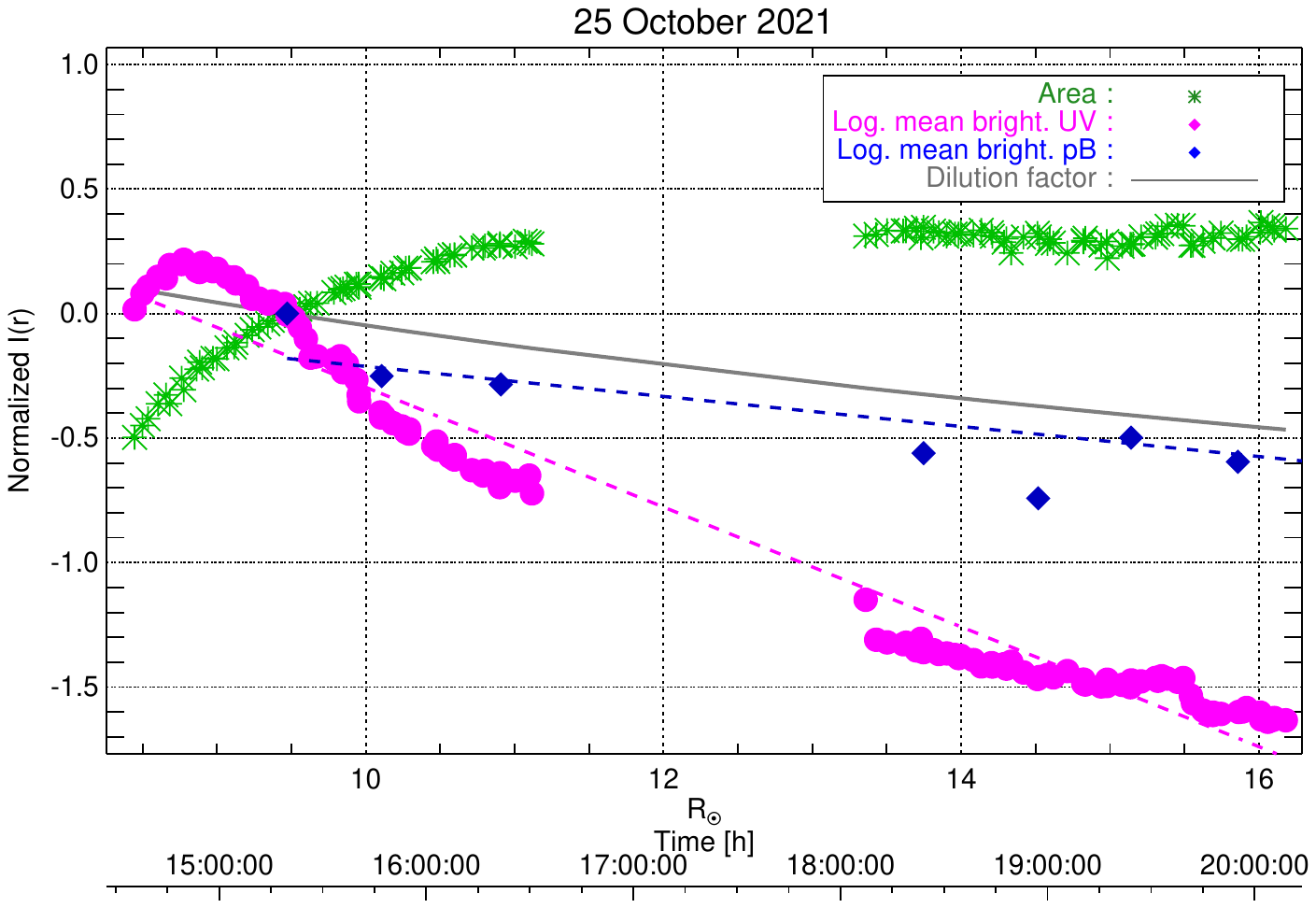}
   \includegraphics[clip,trim=2cm 1.6cm 2cm 3cm,width=9cm]{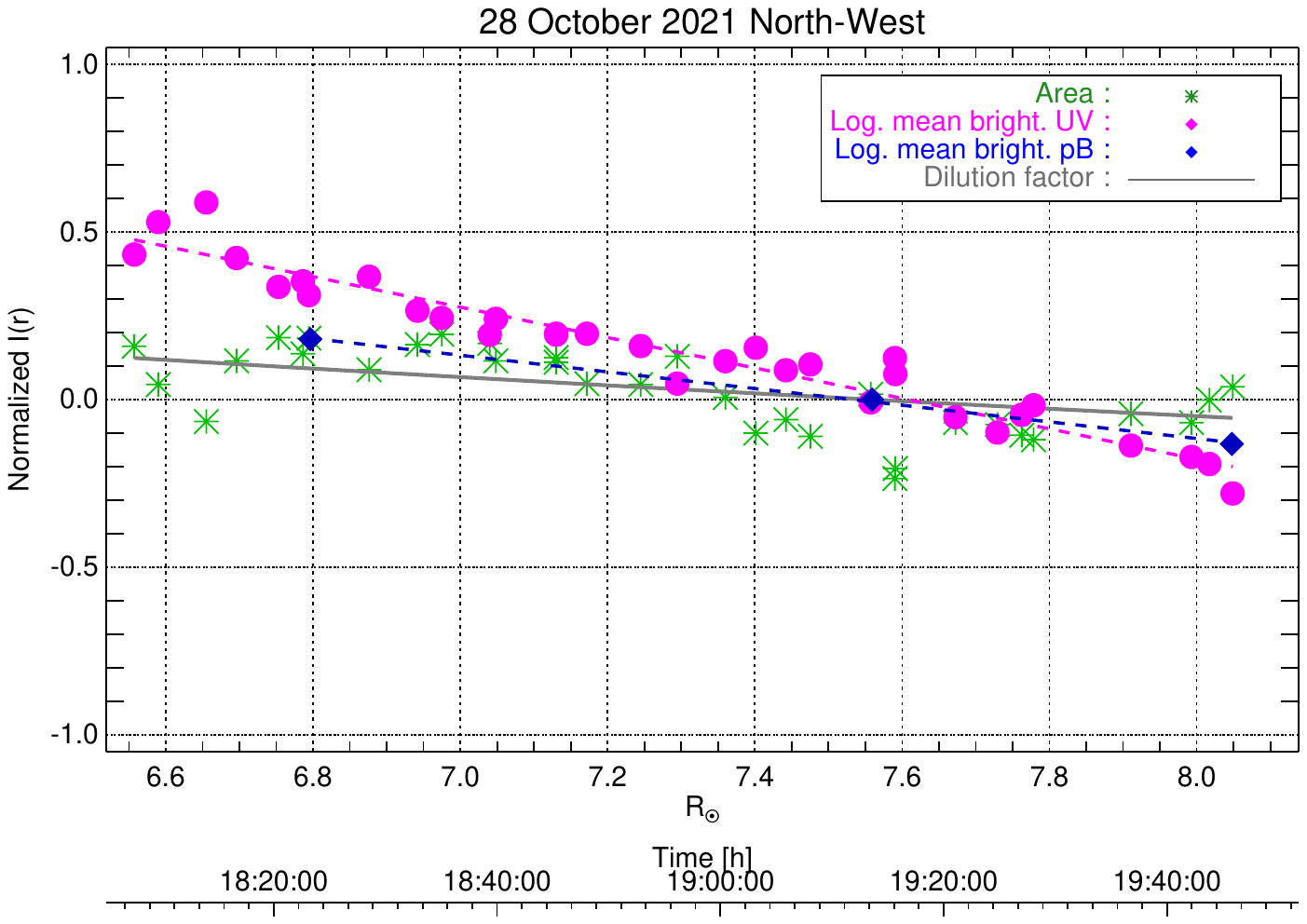}
   \includegraphics[clip,trim=2cm 1.6cm 2cm 3cm,width=9cm]{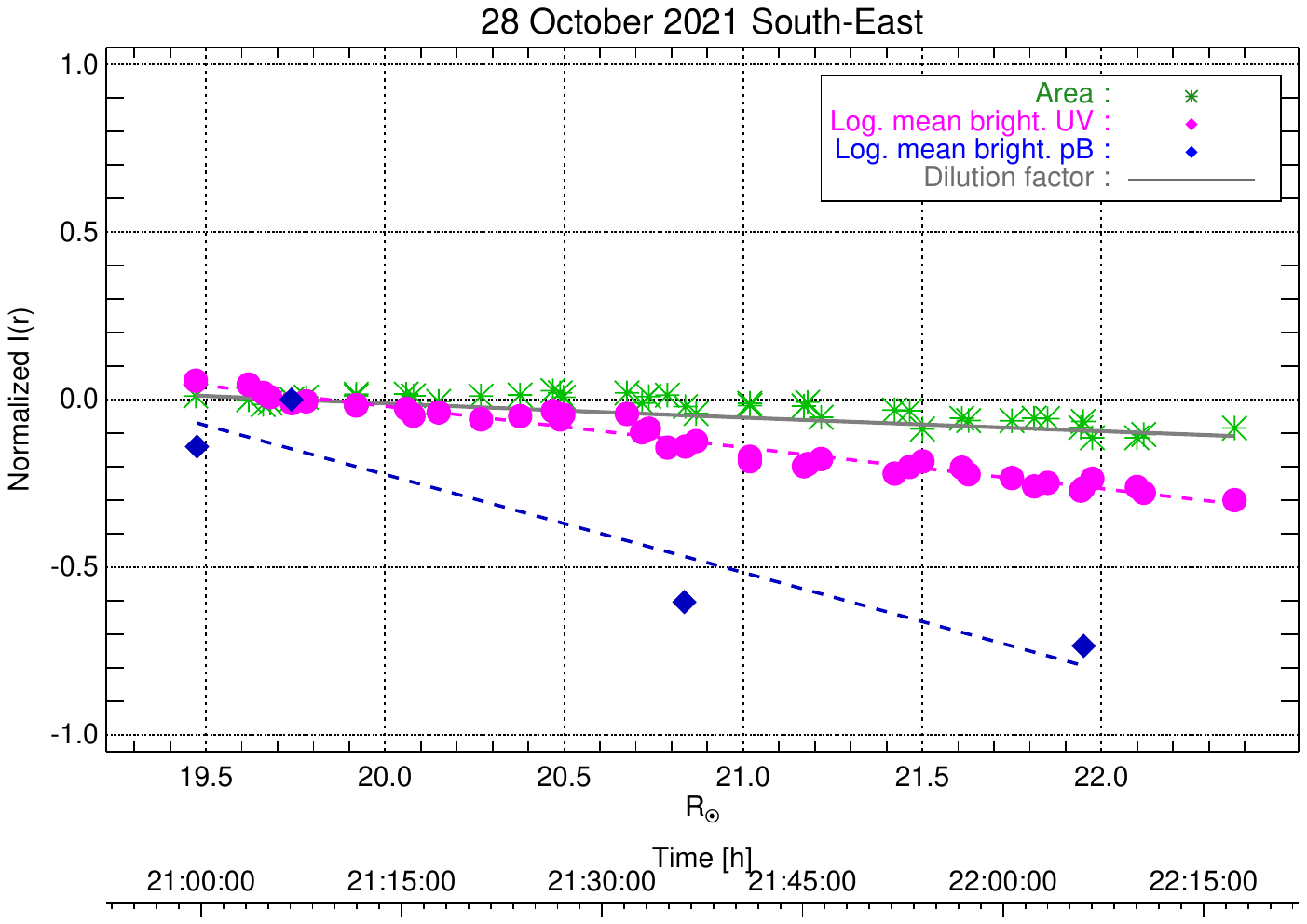}
   \caption{Each panel in temporal order displays: green stars, representing the area of UV structures as a function of radial distance (de-projected) in solar radii and time in hours (UT); magenta points, representing the mean UV \ion{H}{i}~\,\Lya\ intensities of the selected areas for each frame in physical units of ph/cm$^2$/s/sr; blue diamond points, representing the mean brightness of the pB base difference at the pixel position containing the maximum intensity of UV structures for the available frames in physical units of ph/cm$^2$/s/sr; gray solid lines, representing the dilution factor. All quantities are normalized to zero at a reference solar radius, shown in the fourth column of Tab.\,\ref{table:radial_fit_params}. Dashed lines represent a linear fit to UV and pB intensity profiles. Estimated parameters from the fit are reported in Tab.\,\ref{table:radial_fit_params}.}
              \label{fig:I_V_W_R_all}%
    \end{figure*}

In order to highlight the overall behavior of the emission properties of each event, the radial and temporal dependencies of the pB and of UV radiances are displayed in more detail in separate panels of Fig.\,\ref{fig:I_V_W_R_all}.  
In that figure, the observed drop of the radial intensity profiles is more easily compared with the variation with solar distance (de-projected, that is taking into account the distance from the PoS) and time of the dilution factor 
and of the area of the features visible in UV frames. 
To facilitate a comparison with the different radial dependences, the curves in Fig.\,\ref{fig:I_V_W_R_all} are all normalized to a given reference distance from the Sun (i.e. the values in the fourth column of Tab.\,\ref{table:radial_fit_params}), so that only the trend with respect to the dilution factor and then to the radiative excitation term is emphasized. Blue and magenta dashed lines show the linear fit to the UV and pB logarithmic radial profiles as a function of solar distance, $r$: 
\begin{equation}
    \log I (r) = \log I_\circ (r_\circ) - \alpha \times (r - r_\circ)
    \label{eq:radiance_fit}
\end{equation}
where $r_\circ$ is the reference distance from the Sun used for the fit and $I(r)$ can be either the UV radiance or the pB. The parameters $\log I_\circ$, $r_\circ$, and $\alpha$ are listed in columns four to eight in Tab.\,\ref{table:radial_fit_params}.

\begin{table*}
\caption{Eruptions positions and linear fit parameters.}             
\label{table:radial_fit_params}      
\centering          
\begin{tabular}{c | c | c | c | c c | c c }     
\hline\hline       
& & & \multicolumn{5}{c}{\centering Fit parameters}  \\
\cline{4-8}
\multirow{2}{*}{\centering Event date} & Angle from PoS & Polar angle  & \multicolumn{1}{c}{$r_\circ$} & \multicolumn{2}{c}{$\log I_\circ$} & \multicolumn{2}{c}{\centering $\alpha$}  \\ 
 & [deg] & [deg] & \multicolumn{1}{c}{[R$_\sun$]} & \multicolumn{2}{c}{[ph/cm$^2$/s/sr]} & \multicolumn{2}{c}{[R$_\sun^{-1}$]} \\
\cline{4-8}
 & & & \multicolumn{1}{c}{} & pB & \multicolumn{1}{c}{UV} & pB & UV  \\
\hline                    
    25\,-\,26 Apr. 2021 & 19.3 & -89 & 7.7 & 10.3\,$\pm$\,0.8 & 9.1\,$\pm$\,0.3 & 0.1\,$\pm$\,0.1 & 0.16\,$\pm$\,0.03 \\  
   11 Sep. 2021 & 5.4 & 125 & 4.6 & 10.9\,$\pm$\,1.0 & 9.8\,$\pm$\,0.1 & 0.5\,$\pm$\,0.1 & 0.52\,$\pm$\,0.02 \\
   2 Oct. 2021 & 51.5 & -143 & 7.8 & 10.3\,$\pm$\,0.7 & 8.6\,$\pm$\,0.3 & 0.31\,$\pm$\,0.07 & 0.76\,$\pm$\,0.03 \\
   25 Oct. 2021 & 55.2 & 106 & 9.5 & 10.1\,$\pm$\,0.3 & 10.2\,$\pm$\,3.0 & 0.06\,$\pm$\,0.02 & 0.2\,$\pm$\,0.2  \\
   28 Oct. 2021  &\multirow{2}{*}{\centering 19} &\multirow{2}{*}{\centering -67} & \multirow{2}{*}{\centering 7.6} & \multirow{2}{*}{\centering 10.0\,$\pm$\,2.0} & \multirow{2}{*}{\centering 9.3\,$\pm$\,0.1} & \multirow{2}{*}{\centering 0.2\,$\pm$\,0.2} & \multirow{2}{*}{\centering 0.45\,$\pm$\,0.01} \\
   North-west  &  &  &  &   &  &\\
   28 Oct. 2021  & \multirow{2}{*}{\centering 76.7} & \multirow{2}{*}{\centering 142} & \multirow{2}{*}{\centering 19.6} & \multirow{2}{*}{\centering 9.8\,$\pm$\,2.0} & \multirow{2}{*}{\centering 9.57\,$\pm$\,0.06} & \multirow{2}{*}{\centering 0.3\,$\pm$\,0.1} & \multirow{2}{*}{\centering 0.111\,$\pm$\,0.003} \\
   South-east  &  &  &  &   & & \\
\hline                  
\end{tabular}
\tablefoot{For each event listed in the first column, the second column displays the eruption direction angles from the Metis PoS plane on the plane of orbit; the third column shows the polar angles of the trailing features of the structures, with the North of the Sun at 0$^{\circ}$, the east equatorial plane at +90$^{\circ}$, and the west equatorial plane at $-$90$^{\circ}$; the fourth column shows the de-projected reference distance $r_\circ$ of the trailing feature; in columns five to eight, the linear fit parameters to the UV and pB logarithmic radial profiles in Fig.\,\ref{fig:area_br_rsun_all} are displayed as a function of solar radius, calculated from Eq.\,\ref{eq:radiance_fit}, and extrapolated to the PoS distance $r_\circ$ converted in radial distances.}
\end{table*}

Since the structures visible in the UV channel appear compact for the entire duration of the observations, we can assume that the plasma remains confined in a limited volume while the prominence travels through the corona. We therefore assume that the total number of hydrogen atoms and protons is constant and so is the number of electrons --  if there is no change in temperature. Hence, we can suppose that the electron number density is proportional to the inverse of the volume of the prominence emitting in \Lya. From measurements of the apparent area of the structure, $A$, we can obtain a rough estimate of the prominence volume: $V \sim A^{3/2}$, assuming that the aspect ratio on the plane of the sky is the same as along the line of sight. A more quantitative estimate of the electron density can be obtained from the properties of the Thompson scattering process, as described in the following.

Table \ref{tab:electron_densities} displays the electron densities $n_{\mathrm{e}}$ of each feature, calculated from either a pB or \textit{B} reference image taken at the time reported in column two. These electron densities are derived for the same reference frames where the structures have a specific de-projected distance $r_\circ$, as mentioned in the fourth column of Tab.\,\ref{table:radial_fit_params}.
These estimations are derived starting from the formulation of the Thomson scattering process applied to the solar corona as in \cite{Minnaert_1930}, \cite{Billings_1966}  and \cite{Howard-DeForest_2012} 
and following, in particular, section 2.2 in \cite{Howard_Tappin_ICME_theory_2009}. It is possible to formalize the scattered intensities and polarization in a simplified geometry where the scattering process occurs at a position defined by the distance from the Sun center on the PoS (impact distance) and by the angle from the PoS. The equations for the calculation of the expected \textit{B} and pB from solar corona require integration along the LoS. In the case of observed localized features, such as dense prominences, 
after the subtraction of the background coronal contribution, we can factorize the geometrical functions, and compare the expected and observed brightness 
to derive the electron column density, $N_\mathrm{e}$ [cm$^{-2}$]. 
Therefore, by dividing $N_\mathrm{e}$ by the estimated length of the feature we are able to infer the average electron density of the feature itself. This is the common approach adopted in many previous works, e.g. \cite{Colaninno_mass_2009ApJ...698..852C, Susino_CME_params_2016ApJ...830...58S, Bemporad_second_cme_2022A&A...665A...7B}.
We then make the further assumption that the aspect ratio of the feature as observed by Metis is the same on the LoS as on the PoS. Therefore, given the area of the feature measured from the UV frames (see the third column of Tab.\,\ref{tab:electron_densities}), we estimate the length of the feature along the LoS as $A^{1/2}$; hence the average electron density is: 
\begin{equation}
    n_\mathrm{e} = N_\mathrm{e} \; A^{-1/2} \; . \label{eq:density}
\end{equation}

These estimated densities can be considered as lower limits because we assume a unit plasma filling factor: filling factors smaller than unity are reported by e.g. \citep{Susino_filling_fact_2018A&A...617A..21S}. Moreover, for the April 25$^\mathrm{th}$ and the October 2$^\mathrm{nd}$ events in particular, the noticeable elongated shape of the UV features could suggest that a simple $A^{1/2}$ estimate might lead to an overestimate of their LoS length.
Bearing in mind these caveats, the resulting values are listed in the fourth column of Table~\ref{tab:electron_densities}. We note that these values are obtained at large distances from the Sun and are
intermediate between the typical ambient corona densities and prominence densities estimated in CMEs in the lower corona \citep{Heinzel_prominence_2016A&A...589A.128H}. 

 \begin{table}
      \caption[]{Electron densities and areas of the UV structures.}
         \label{tab:electron_densities}
      $ \begin{array}{p{0.25\linewidth}ccl}
            \hline
            \noalign{\smallskip}
            Event date  &  \mathrm{Ref. time / [UT]} & A / {[\mathrm{cm^{2}}]}  &  n_\mathrm{e} / {[\mathrm{cm^{-3}}]}   \\
            \noalign{\smallskip}
            \hline
            \noalign{\smallskip}
            25-26 Apr. 2021 & 00:20 & 1.8 \times 10^{22} & 2.0 \times 10^5    \\
            11 Sep. 2021 & 10:15 & 1.1 \times 10^{22} & 2.8 \times 10^5     \\
            2 Oct. 2021 & 17:33 & 8.9 \times 10^{21} & 2.8 \times 10^5 \, \mathrm{(\textit{B})} \\
            25 Oct. 2021 & 15:19 & 1.3 \times 10^{22} & 3.7 \times 10^5 \, \mathrm{(\textit{B})}        \\
            28 Oct. 2021 NW & 18:49 & 3.9 \times 10^{21} & 1.9 \times 10^5   \\
            28 Oct. 2021 SE & 21:19 & 3.4 \times 10^{22} & 3.9 \times 10^5 \, \mathrm{(\textit{B})}  \\
            \noalign{\smallskip}
            \hline
         \end{array}$
         \tablefoot{Areas $A$ of the selected features obtained from a UV reference frame taken at the reference time in column two, as described for each event in Sec.~\ref{subsec:intensity} and the corresponding electron densities $n_\mathrm{e}$ calculated using Eq.\,\ref{eq:density} from the co-temporal pB or total \textit{B} image.}
   \end{table}

In light of these general findings and considerations, we will now discuss the results pertaining to each event, supported by the plots shown in Fig.\,\ref{fig:I_V_W_R_all}. 

\paragraph{April 25$^\mathrm{th}$ -26$^\mathrm{th}$ event.}
We have chosen six UV frames to calculate the area of the bright core that is entirely within the Metis FoV for this event. The prominence is clearly visible and bright in both Metis channels, making it easy to identify the same features (see Fig.\,\ref{fig:pB_UV_prominence_20210425-26}). In the radial profiles plot it is evident that the trend of both the UV and pB radiances follows the trend of the dilution factor. This suggests a predominantly radiative excitation component of the UV structure \Lya\ emission.

\paragraph{September 11$^\mathrm{th}$ event.}
Both UV and pB radiance profiles for the September 11$^\mathrm{th}$ event deviate from a trend corresponding to a predominantly radiative component, possibly suggesting either a strong collisional component or a significant temperature change. We note that the area of the prominence (represented by green star points) remains nearly constant up to 5.5\,R$_{\sun}$, where a slight deflection is observable, which corresponds to a variation in the UV intensity profile. However, the area determination in this event is affected by UVDA background variation, described in more detail in Appendix\,\ref{sec:appendix_UVDA_issue}. The background surface has a similar deviation for frames after 10:45\,UT, as shown in Fig.\,\ref{fig:UV_box_20210911}. As a result, similar artifacts are also noted in the estimated area trend.

\paragraph{October 2$^\mathrm{nd}$ event.}
On the South edge of the CME observed on October 2$^\mathrm{nd}$, a bright, elongated structure propagating outwards is clearly seen in VL images, but even more so in UV images.
UV images also show that as it propagates this structure appears to stretch longitudinally (see Fig.\,\ref{fig:pB_UV_prominence_20211002}). This suggests that there may be some sort of friction between the flank of the CME and the background corona, thus outlining a potential scenario in which the Kelvin-Helmholtz (KH) instability is at work. The KH instability arises at the interface of two fluids in parallel motion with a velocity shear. This leads the shear sheet to roll up, generating vortex-like structures. A rapid, linear phase in which perturbations grow exponentially with a single characteristic spatial scale is followed by a nonlinear phase that includes a superposition of different modes, i.e., a spectrum (see \cite{Faganello_Califano_2017}, and reference therein). At the interface between the two fluids, turbulence is thus generated, whose dissipation may serve as an additional source of plasma heating \citep{Stawarz_Eriksson_Wilder_2016,Sorriso-Valvo_PRL_2019,Telloni_K-H_2022ApJ...929...98T}. Linear theory \citep{Chandrasekhar_1961hhs..book.....C,Hasegawa_2004Natur.430..755H} predicts that the phase velocity of the vortex structures is half of the CME front speed. 
In Sec.\,\ref{subsec:2_oct}, we reported the front radial velocity of the CME catalog as $V_\mathrm{front}$\,=\,546\,$\pm$\,342\,km/s. This value can be compared with the velocity of the brightest part of the elongating structure obtained through time-distance analysis, which is $V_\mathrm{phase}$\,=\,367\,$\pm$\,22\,km/s. These values are consistent with the expectations from the Kelvin-Helmholtz instability.
The KH instability has already been observed at the flanks of CMEs in the low corona \citep{Foullon_K-H_2011ApJ...729L...8F,Ofman_K-H_2011ApJ...734L..11O} and it is suggested to play a role in the plasma heating in the extended corona \citep{Telloni_K-H_2022ApJ...929...98T}. Further investigation is clearly needed (possibly complemented by a modeling/theory description), but this, however, may be the first evidence of KH instability at the flank of a CME observed in the outer corona.

The KH instability explanation would confirm that the October 2$^\mathrm{nd}$ event is of a different nature than the eruptive prominence interpretation adopted here for the other events, which is also supported by the larger slope of the UV intensity profile shown in the upper panel of Fig.\,\ref{fig:area_br_rsun_all}.
In Fig.\,\ref{fig:I_V_W_R_all} the pB radial profile deviates from a trend common to the photoexcitation term, with a slope that is almost a half of the UV profile (as in Tab.\ref{table:radial_fit_params}). The area represented by the green star points remains almost constant, but there is some variation in the last few points beyond 5\,R$_{\sun}$ because the structure becomes weaker and more difficult to distinguish from the background noise.

\paragraph{October 25$^\mathrm{th}$ event.}
The radial profiles of the eruption prominence on October 25$^\mathrm{th}$ are more complicated because of different factors. 

The area depicted by the green star points and the corresponding magenta UV radial profile present several jumps because the frames, in this case, appear to be affected by the UVDA channel issue as explained in Appendix\,\ref{sec:appendix_UVDA_issue}. 
At approximately 15:40\,UT, 18:20\,UT, and 19:40\,UT, there is a noticeable change in the average intensity of the selected background boxes, as shown in Fig.\,\ref{fig:UV_box_20211025}. This discontinuity coincides with a change in the green and magenta curves in Fig.\,\ref{fig:I_V_W_R_all}.

The area initially increases up to 10\,R$_\sun$ because the structure grows and expands as it moves into the Metis FoV over time.
Between 10\,R$_\sun$ and the start of the data gap at 16:33\,UT, the drift in area and UV intensity could be indicative of a decrease in the plasma temperature, which is condensing over time. Alternatively, it could suggest a strong collisional component. This behavior contrasts with the pB brightness profile of the prominence, which instead follows the dilution factor trend. We will explore this further in the section when discussing the Doppler dimming factor.

After the data gap, the area stabilizes and the structure exhibits the dips already described in Sec.\,\ref{subsec:25_oct}. 

\paragraph{October 28$^\mathrm{th}$ events.}
In October 28$^\mathrm{th}$-NW event, the pB radial profile follows the trend of the dilution factor, delineating a radiative component of the UV emission plus a collisional component due to a significant slope of the UV profile.  
The area is almost constant up to 7.2\,R$_\sun$ (green star points), then the prominence becomes very weak and the background is much noisier due to the proton shower hitting the FoV, as can be noted in Fig.\,\ref{fig:pb_UV_eruption_20211028}. This effect is also responsible for the scatter in green and magenta points.

Similar to the September 11$^\mathrm{th}$ event, the October 28$^\mathrm{th}$-SE pB profile exhibits a deviation from a radiatively excited emission trend. This behavior again suggests the possible presence of a collisional component of the emission \Lya\, or a significant temperature change.
\\

The intensity of the resonant component of the UV \ion{H}{i}~\,\Lya\ emission is also sensitive to the speed of the coronal plasma through the Doppler dimming factor as in Eq.\,\ref{eq:Lya_R}. The Doppler dimming effect causes a lower scattering efficiency which results in a reduction of the intensity of the scattered radiation \citep{Withbroe_1982SSRv...33...17W,Dolei_2015}. We can compute the effective Doppler dimming coefficient $D(\nu,T_{e})$ as an intensity ratio (see Eq.(2) in \cite{Dolei_outflow_2018A&A...612A..84D} and \cite{Heinzel_Rompolt_DD_1987SoPh..110..171H}), considering a uniform chromospheric intensity profile with an analytical shape defined in \cite{Auchere_2005ApJ...622..737A}. The Doppler dimming coefficient depends only on the electron temperature and radial velocity of the eruption. In Tab.\,\ref{table:events_measured_info} we report a pair of Doppler dimming coefficients for each structure, calculated from the electron temperature profiles of the solar corona defined in \cite{Vasquez_2003ApJ...598.1361V} and \cite{Gibson_1999JGR...104.9691G}, respectively, extrapolated to the heliocentric position reported in the fourth column of Tab.\,\ref{table:radial_fit_params}. For all the events, the resonantly scattered component is dimmed by more than 70\,$\%$ due to the high velocity of the structures and the rapid decrease in coronal temperature profiles with height (in the range of 1\,MK at 5\,R\,$_{\sun}$ and 0.2\,MK at 20\,R\,$_{\sun}$). 

These relatively low values of the Doppler dimming factors for the events under study are in contrast with the high values of the observed \Lya\ radiances (about two orders of magnitude higher than the background corona) and could indicate either a significant collisional component, or a higher fractional abundance of neutral hydrogen, or both.  
We postulate that the exceptional brightness of these features could be due to higher densities of neutral hydrogen due to lower temperatures.  From Eq.\,\ref{eq:Lya_R}, the main temperature dependence of the resonant contribution comes from the Doppler dimming factor and the hydrogen ionization fraction.  In order to get the product of the two factors higher than, say, a factor 50 with respect to the surrounding corona, these features should be cooler than about $10^5$~K.  We remark that there are observations \citep{Ding_2017ApJ...842L...7D, kohl_2006, Ciaravella_UVCS_1997ApJ...491L..59C} indicating such low temperatures in ejecta produced in solar eruptions. More accurate estimates will require detailed modeling of these structures, for example with the method based on 1D NLTE modeling of eruptive prominences used in \cite{Heinzel_prominence_2016A&A...589A.128H}. 

\begin{figure}
  \centering
   \includegraphics[clip, trim=1.5cm 1cm 1.6cm 1cm,width=9cm]{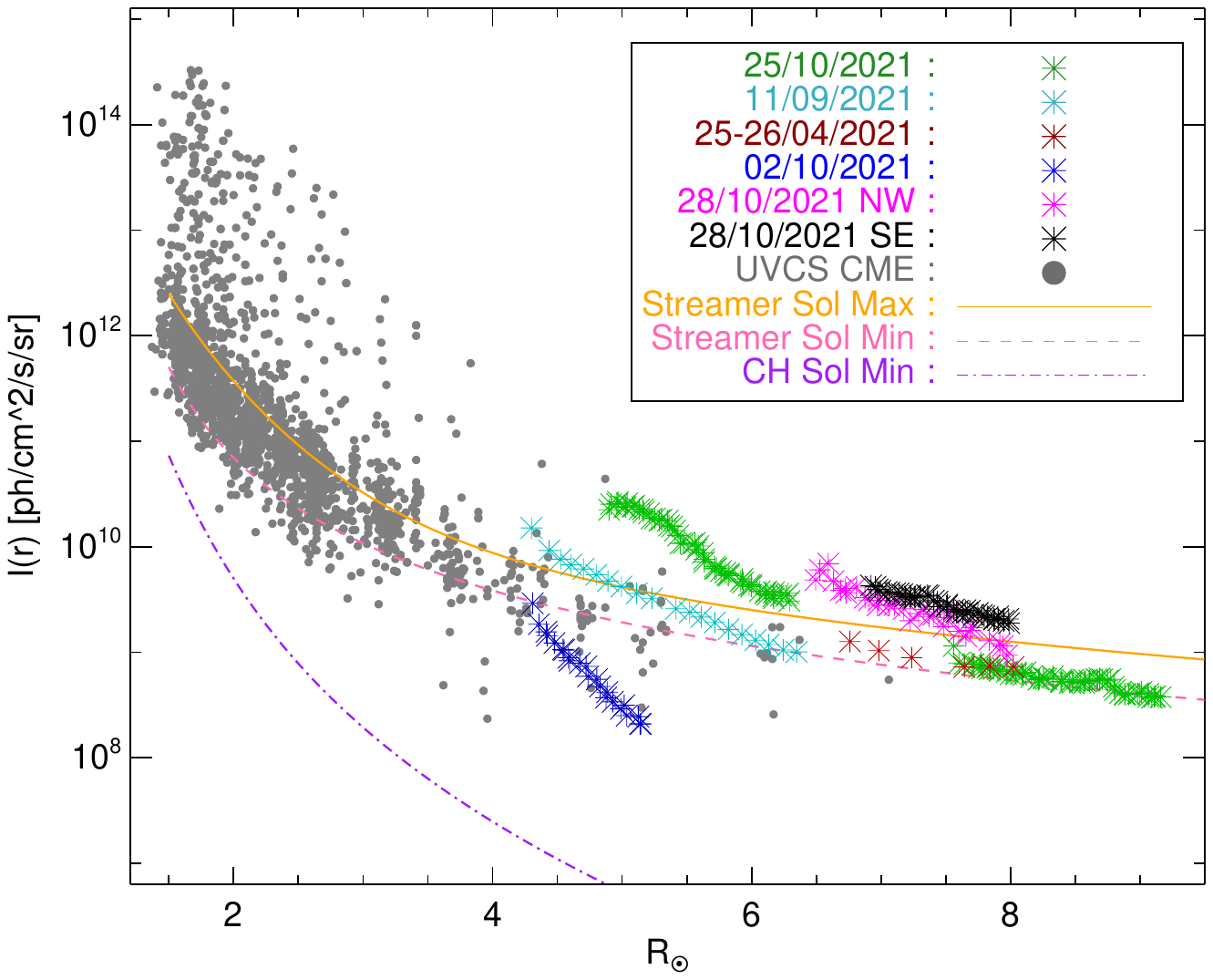}
   \caption{UV \ion{H}{i}~\,\Lya\ mean intensity of the bright cores (colored star points) compared with the SOHO/UVCS CMEs intensity (gray points) taken from Fig.\,15\,(a) of \cite{Giordano_UVCS_2013} as a function of heliocentric heights on the PoS.}
              \label{fig:br_VS_dist_giordano}%
   \end{figure}

Although there are no spectroscopic measurements of the events presented in this work, it is helpful to compare the Metis UV \ion{H}{i}~\,\Lya\ emission profiles with the statistical distribution of CME \Lya\ intensities reported in the SOHO/UVCS catalog as in Fig.\,\ref{fig:br_VS_dist_giordano} \citep[see Fig.\,15\,(a) in][]{Giordano_UVCS_2013}. Here, we use the plane of sky radial distance of the bright structures to compare the data with SOHO/UVCS CMEs directly. For comparison, the typical intensities in streamers at solar minimum and maximum and in a coronal hole are also plotted. The eruptions under study lie in the lower tail of the distribution with a general trend comparable to the streamers in the range of solar distance above 4\,R$_{\sun}$. As already pointed out, the event on October 2$^\mathrm{nd}$ shows a completely different trend (see also Fig.\,\ref{fig:area_br_rsun_all}), whose interpretation is still under investigation. 

Moreover, the contribution of the interplanetary \Lya\ of $\sim$10$^7$\,ph/cm$^2$/s/sr is well below the measured profiles that asymptotically approach that value only at very large distances (beyond 10\,R$_\sun$). On the other hand, the intensities of the October 2$^\mathrm{nd}$ event may have already reached that limit at ~7\,R$_\sun$ on the PoS.

\section{Summary}
\label{sec:summary}

In this work, we present a set of imaging observations of eruptive events obtained with the Metis coronagraph both in a narrow band centered around the \ion{H}{i}~\,\Lya\ 121.6\,nm line and in visible-light polarized brightness. The selected set of observations includes all the events observed by Metis during the Solar Orbiter cruise phase, i.e. until the end of 2021, that are characterized by remarkably bright emission in \Lya.  The observed features remained bright even at heliocentric distances as large as 22 solar radii. 

For all the observed UV-bright structures, we determined the main kinematical parameters, including PoS velocity and propagation direction, as well as estimates of the acceleration and deflection from the radial direction.  In addition, we were able to determine the 3D velocity vector, which allowed us to estimate the Doppler dimming factor, among other parameters.  

To summarize, our findings indicate that only four out of six events (April 25$^\mathrm{th}$, September 11$^\mathrm{th}$, October 25$^\mathrm{th}$ and 28$^\mathrm{th}$-NW) exhibit a clear three-part structure in their associated coronal mass ejections. 
For these occurrences, we can assume that the bright UV feature is the bright core of the expanding CME. The UV structure for the April 25$^\mathrm{th}$ event appears distinct from the part that could be considered the core, as in Fig.\,\ref{fig:pB_UV_prominence_20210425-26}, panels (d) and (n).

The CMEs are difficult to see in Metis pB frames for three of the events (October 2$^\mathrm{nd}$, 25$^\mathrm{th}$, and 28$^\mathrm{th}$-SE) because of the larger angle (above $\sim$ 50\,$^\circ$) between the direction of the CMEs and the Metis PoS. However, they can be clearly seen in Metis total \textit{B} frames and in LASCO-C2 or COR2 frames due to the perspective effect.

Considering the October 2$^\mathrm{nd}$ event, the elongated structure appears to be on the outer edges of the main CME structure (see Fig.\,\ref{fig:pB_UV_prominence_20211002}).

In three cases only (April 25$^\mathrm{th}$, September 11$^\mathrm{th}$ and October 28$^\mathrm{th}$-SE), the orbit configuration of the Solar Orbiter was favorable for the identification of the eruption's source region on the solar disk or at limb, as indicated in Appendix\,\ref{sec:appendix_source_region}.

Three eruptions out of six (September 11$^\mathrm{th}$, October 2$^\mathrm{nd}$ and 28$^\mathrm{th}$-SE) show a nearly constant shape while the April 25$^\mathrm{th}$,  September 11$^\mathrm{th}$ and October 28$^\mathrm{th}$-NW events reveal a deviation from the radial propagation. The evidence of a rotation is in the slight growth of the projected angular width of only two structures (October 25$^\mathrm{th}$ and 28$^\mathrm{th}$-SE).

We furthermore examined the dependence with distance of pB and UV radiances with the aim of obtaining clues on the formation mechanism of the observed \Lya\ emission. With the exception of the April 25$^\mathrm{th}$ and October 28$^\mathrm{th}$-NW events, the radial profiles deviate from the behavior expected from a purely resonant scattering line excitation.  
We also estimated electron densities for these bright features, obtaining values in excess of $10^5$ cm$^{-3}$ for all the events in the brightest UV regions.
Noting that the de-projected, measured velocities imply low values of the Doppler dimming coefficients, these observations and measurements suggest that collisional emission processes should be taken into account in the formation of \Lya\ emission in such bright coronal features, even at large heliocentric distances.  In addition, or as an alternative, the observed \Lya\ emission could be due to large \ion{H}{i} densities in relatively cool plasma, at temperatures of the order of $10^5$~K or less.

To better understand the physical characteristics of the events discussed in this article, suitable models can be developed in future studies. 
For example, several plasma diagnostic techniques can be developed to extrapolate the physical parameters of the plasma embedded in CMEs, as was done with numerical simulations in the work of \cite{Bemporad_Pagano_2018A&A...619A..25B} to estimate the CME plasma electron density and temperature taking into account Doppler dimming effect from coronagraphic UV and VL observations. In this numerical test, it is also evident that the inner part of the CME appears much brighter in the UV \Lya\ emission with respect to the VL, with a relatively faint CME front emission. Other simulation approaches are discussed in \cite{stealth_cme_sim_2021A&A...652A.160Y} and \cite{Pagano_sim_2018JSWSC...8A..26P}. These papers present methods for simulating the coronal magnetic field and plasma parameters of stealth CME eruptions. They also introduce new techniques to provide interplanetary space weather forecasting models with accurate time-dependent boundary conditions of erupting magnetic flux ropes in the upper solar corona.

In the Sec.~\ref{sec:discussion}, we made the assumption that the aspect ratio of UV-bright structures is the same on the PoS and the LoS. However, alternative, model-dependent techniques like the one used in \cite{Susino_filling_fact_2018A&A...617A..21S}, could be employed to estimate the line-of-sight geometrical filling factor and improve the estimation of the dimensions of structures along the LoS. Alternatively, the polarization-ratio technique \citep{polarizatio-ratio_technique_2004Sci...305...66M} can be applied to infer the 3D structure of a CME, as also proposed by \cite{Susino_CME_params_2016ApJ...830...58S} among other techniques used to determine the physical parameters of solar eruptions from a combination of polarized VL and UV observations.

A more detailed analysis of the VL polarized signal is already the subject of a separate analysis in \cite{Heinzel_D3_2023}, which specifically focuses on using the Metis VL signal from the April 25$^\mathrm{th}$ event to identify the D$_3$ line emission and polarization signatures.

\begin{acknowledgements}
      Solar Orbiter is a space mission of international collaboration between ESA and NASA, operated by ESA.  Metis was built and operated with funding from the Italian Space Agency (ASI), under contracts to the National Institute of Astrophysics (INAF) and industrial partners. Metis was built with hardware contributions from Germany (Bundesministerium für Wirtschaft und Energie through DLR), from the Czech Republic (PRODEX) and from ESA.  
      Metis team thanks the former PI, Ester Antonucci, for leading the development of Metis until the final delivery to ESA.
      The work of A.L. was conducted under the NASA contract 80NM0018D0004.
      S.J. acknowledges the support from the Slovenian Research Agency No. P1-0188.
      P.H. and S.J. acknowledge support from the grant 22-34841S of the Czech Science Foundation. P.H. was also supported by the program ’Excellence Initiative - Research University’ for years 2020-2026 at the University of Wroclaw, project no. BPIDUB.4610.96.2021.KG.
      The authors thank Dr. H. Cremades for the fruitful discussion.
\end{acknowledgements}

\bibliographystyle{aa} 
\bibliography{biblio_main} 

\begin{appendix} 

\section{UVDA channel issues}
\label{sec:appendix_UVDA_issue}

As will be discussed in more detail in a dedicated paper on to the UV channel calibration (\cite{De_Leo_2023b_UV}, the UVDA channel shows response instabilities with non-linear temporal and spatial variations that can range from a few percent to even a factor of two. Therefore, it is needed to examine the detector's local behavior within the vicinity of the frames containing the prominent features of the presented events. This investigation aims to ascertain the significance of these effects for the current analysis.

\begin{figure}[!ht]
  \centering
   \includegraphics[clip, trim=2cm 1cm 0.9cm 1cm,width=9cm]{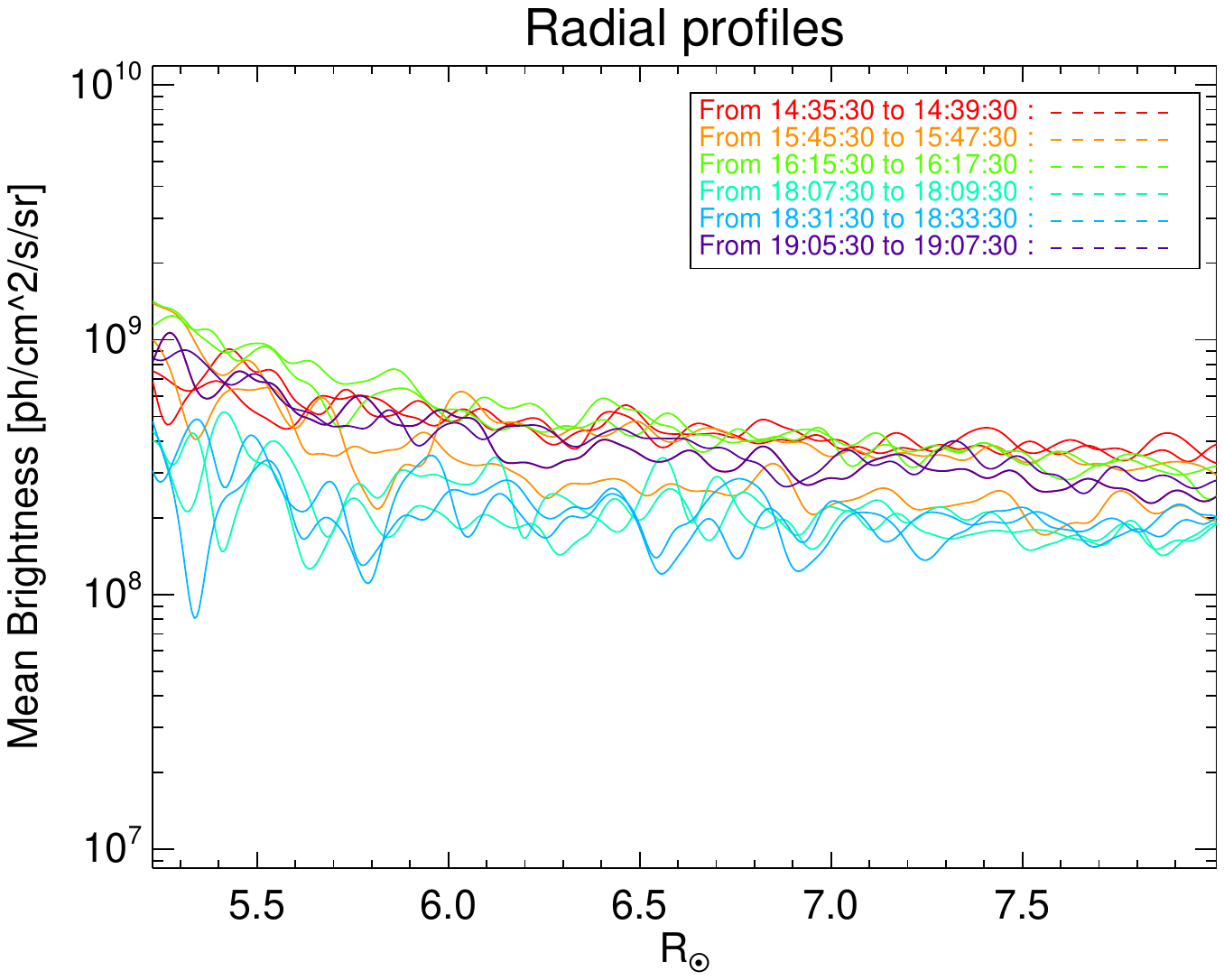}
   \caption{Intensity radial profiles at the north polar angle with respect to the eruption prominence on October 25$^\mathrm{th}$. Different colors correspond to a different time range as indicated in the legend.}
              \label{fig:UV_rad_profiles_20211025}%
\end{figure}

\begin{figure}[!ht]
  \centering
   \includegraphics[clip, trim=3cm 4cm 0.9cm 4cm,width=9cm]{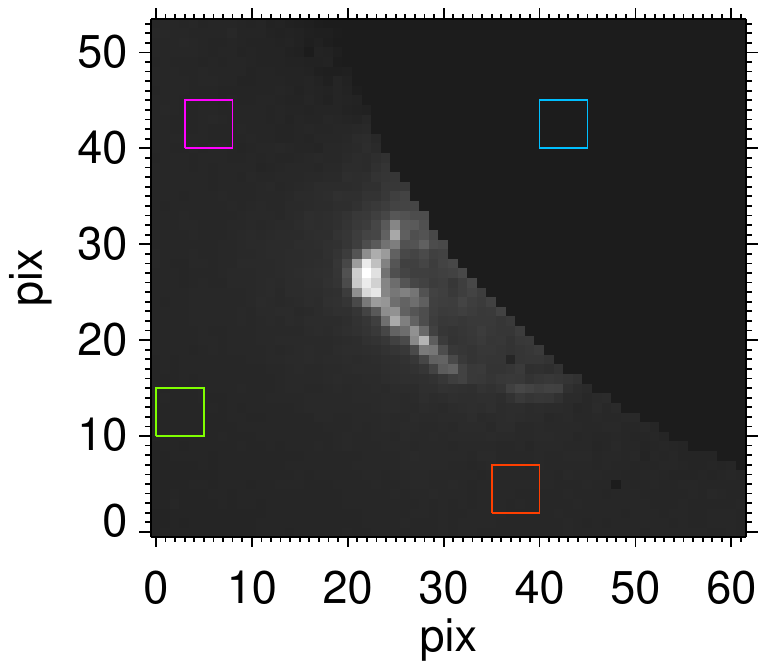}
   \includegraphics[clip, trim=3cm 3cm 0.9cm 3.8cm,width=9cm]{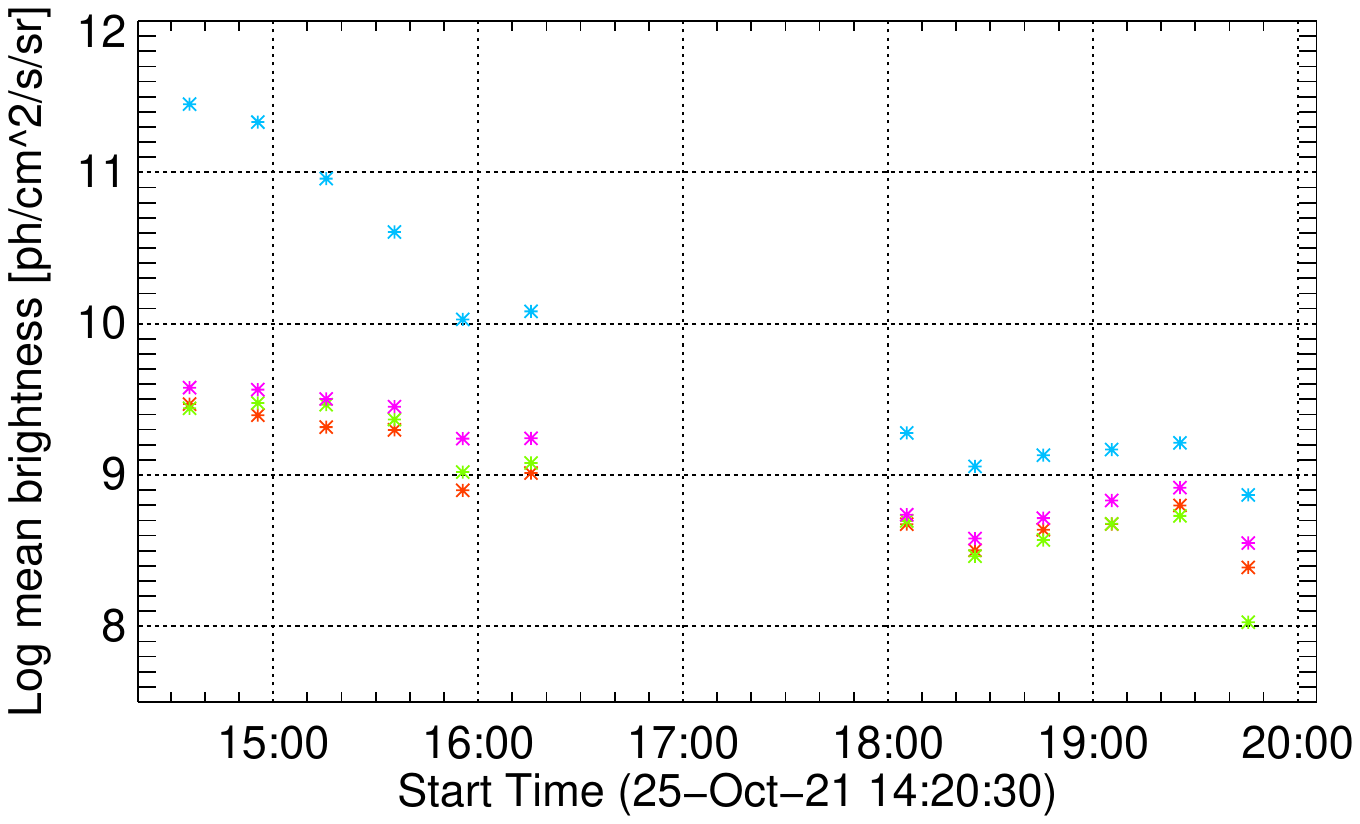}
   \caption{\textit{Top panel}: one of the UV frames of the event on October 25$^\mathrm{th}$, where colored boxes useful for the analysis are highlighted. \textit{Bottom panel}: each colored point represents the logarithm of the mean intensity of the background in the corresponding box visible in the top panel. We chose one frame out of every ten.}
              \label{fig:UV_box_20211025}%
\end{figure}

\begin{figure}[!ht]
  \centering
   \includegraphics[clip, trim=3cm 4cm 0.9cm 4cm,width=9cm]{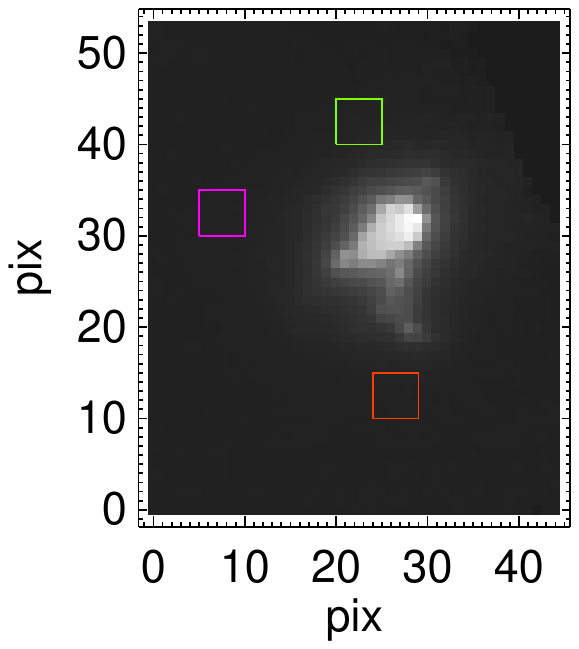}
   \includegraphics[clip, trim=3cm 3cm 0.9cm 3.8cm,width=9cm]{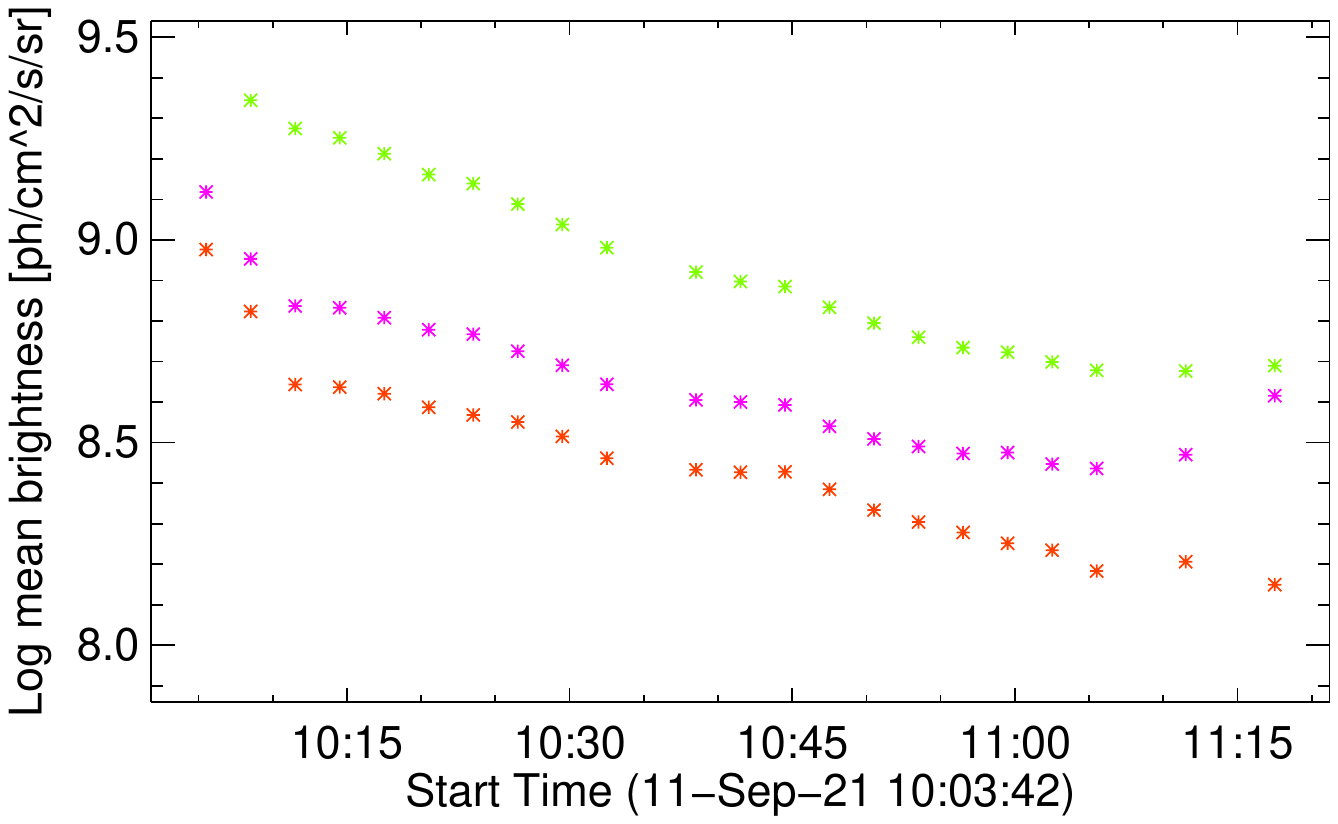}
   \caption{\textit{Top panel}: one of the UV frames of the event on September 11$^\mathrm{th}$, where colored boxes are highlighted. \textit{Bottom panel}: each colored point represents the logarithm of the mean intensity of the background in the corresponding box visible in the top panel for each frame.}
              \label{fig:UV_box_20210911}%
\end{figure}

One of the tests is to trace radial intensity profiles at polar angles north and south with respect to the event structure for each frame in order to observe an intensity variation that is not consistent with the static background corona. For the selected time range of each event under study, the radial profiles are compatible within the signal fluctuations for each frame with the exception of October 25$^\mathrm{th}$ and September 11$^\mathrm{th}$. \\
In the case of October 25$^\mathrm{th}$, as shown in Fig.\,\ref{fig:UV_rad_profiles_20211025}, we notice a variation in the intensity profiles between two levels. The mean threshold goes down around 15:45\,UT for half an hour and after the data gap at 18:00\,UT for almost a couple of hours, to come back to the original value at the end of the selected period.\\
This variation is consistent with the background level behavior in different boxes selected in the FoV adjacent to the prominence position, as it can be noted in Fig.\,\ref{fig:UV_box_20211025}. We notice two changes in the global behavior, one around 15:45\,UT and one at the end of the selected time range around 19:40\,UT. \\
Something similar is also visible for the event of September 11$^\mathrm{th}$ in Fig.\,\ref{fig:UV_box_20210911}, for a few frames before 10:10\,UT and after 11:10\,UT. This behavior reflects in the UV mean intensity trend shown in Fig.\,\ref{fig:I_V_W_R_all}. For this event, no signal threshold variation is evident in the radial profiles.

The origin of the signal threshold variation is still under investigation, but for the events under study, it affects only the event of October 25$^\mathrm{th}$ producing the fluctuation visible in the intensity of the areas in the green curve of Fig.\,\ref{fig:area_br_rsun_all}.  

\FloatBarrier

\section{Possible source regions}
\label{sec:appendix_source_region}

In this Appendix we report on possible source region or low corona counterpart of the eruptions leaving the Sun's surface identified thanks to the estimation of the propagation direction discussed in Sec.\,\ref{sec:observations}. 

Thanks to the estimated Carrington coordinates of the direction of the events we can mark the possible source region in synoptic maps (as magnetograms).
Extrapolating the time of the eruption at the solar disk from fitting the height-time profiles, assuming a constant velocity, we recognize low corona features in frames from disk imagers like SolO/EUI-FSI or STEREO-A/EUVI.

\begin{figure*}[!ht]
  \centering
   \includegraphics[clip, angle=-90, trim=2cm 0.5cm 4cm 3cm,width=9cm]{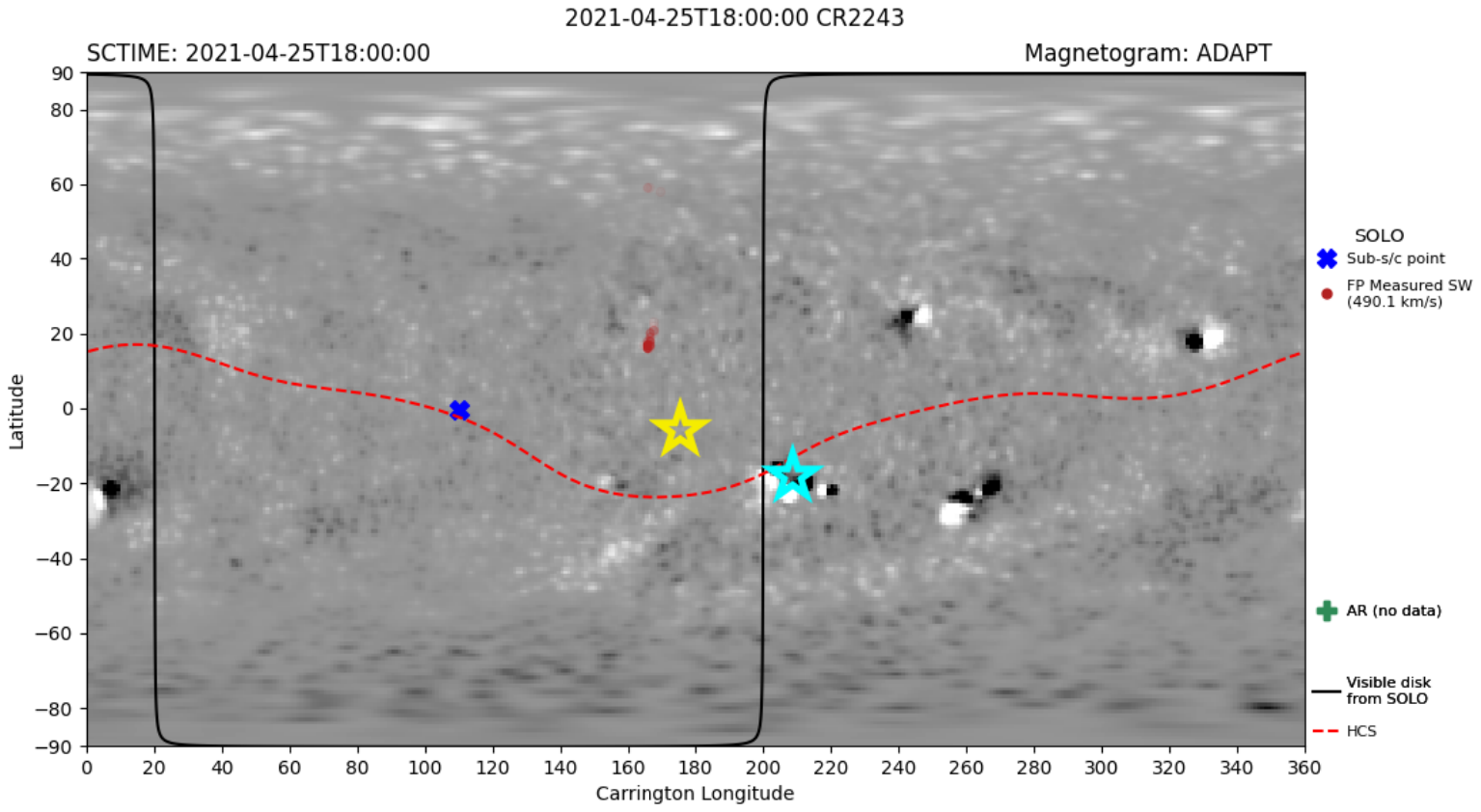}
   \includegraphics[clip, angle=-90, trim=2cm 0.5cm 3cm 3cm,width=9cm]{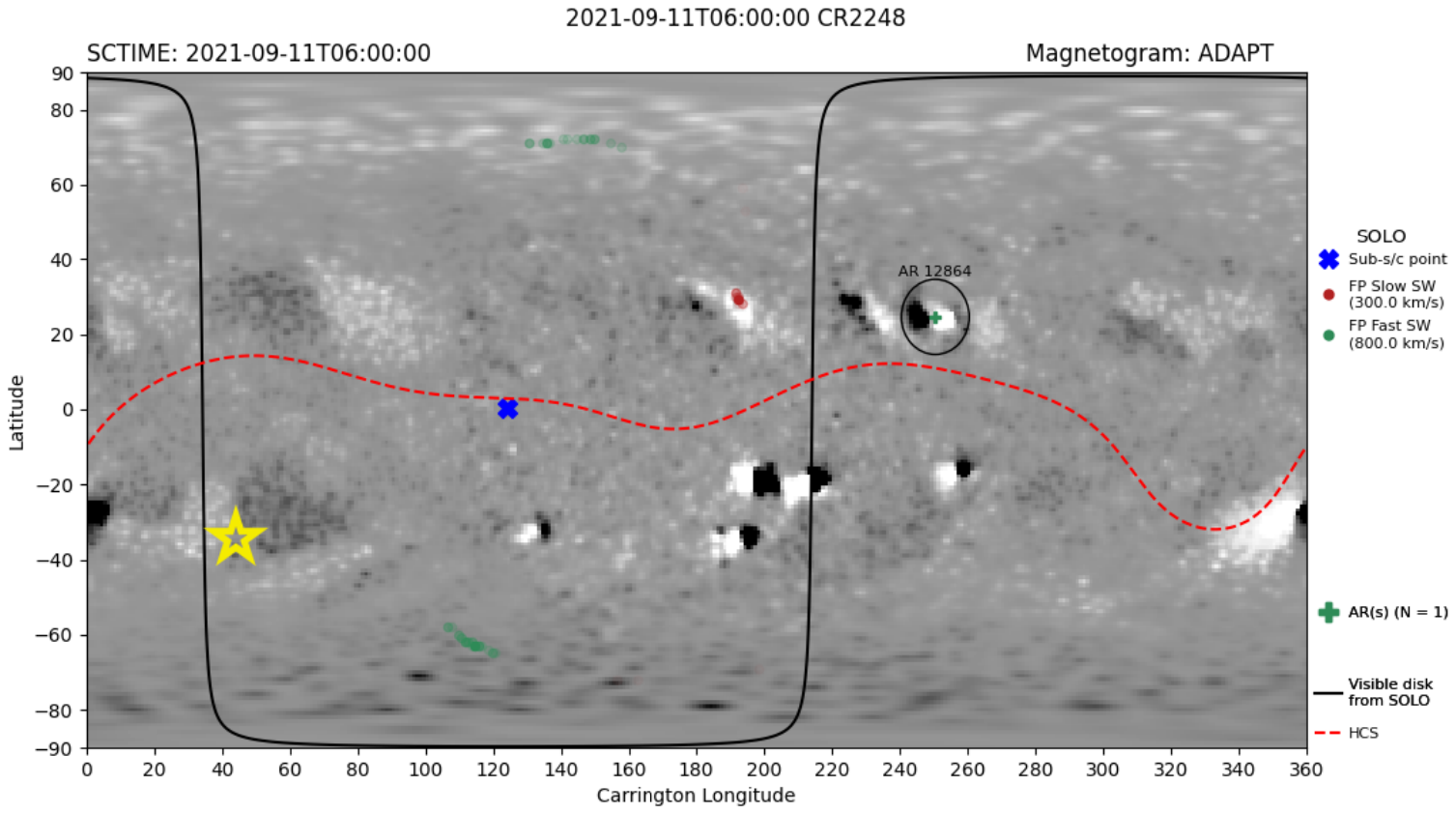}
   \includegraphics[clip, angle=-90, trim=2cm 0.5cm 3cm 3cm,width=9cm]{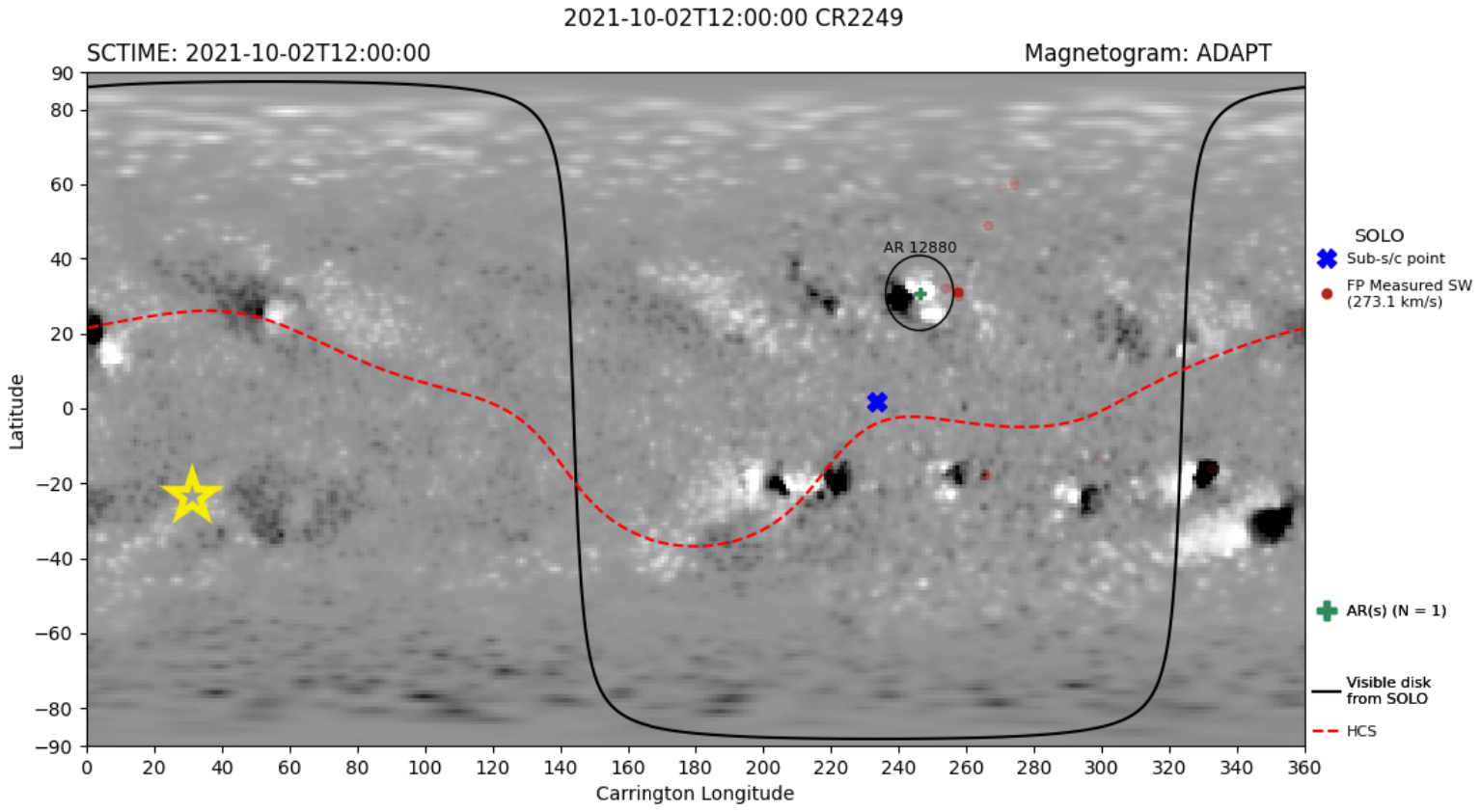}
   \includegraphics[clip, angle=-90, trim=2cm 0.5cm 3cm 3cm,width=9cm]{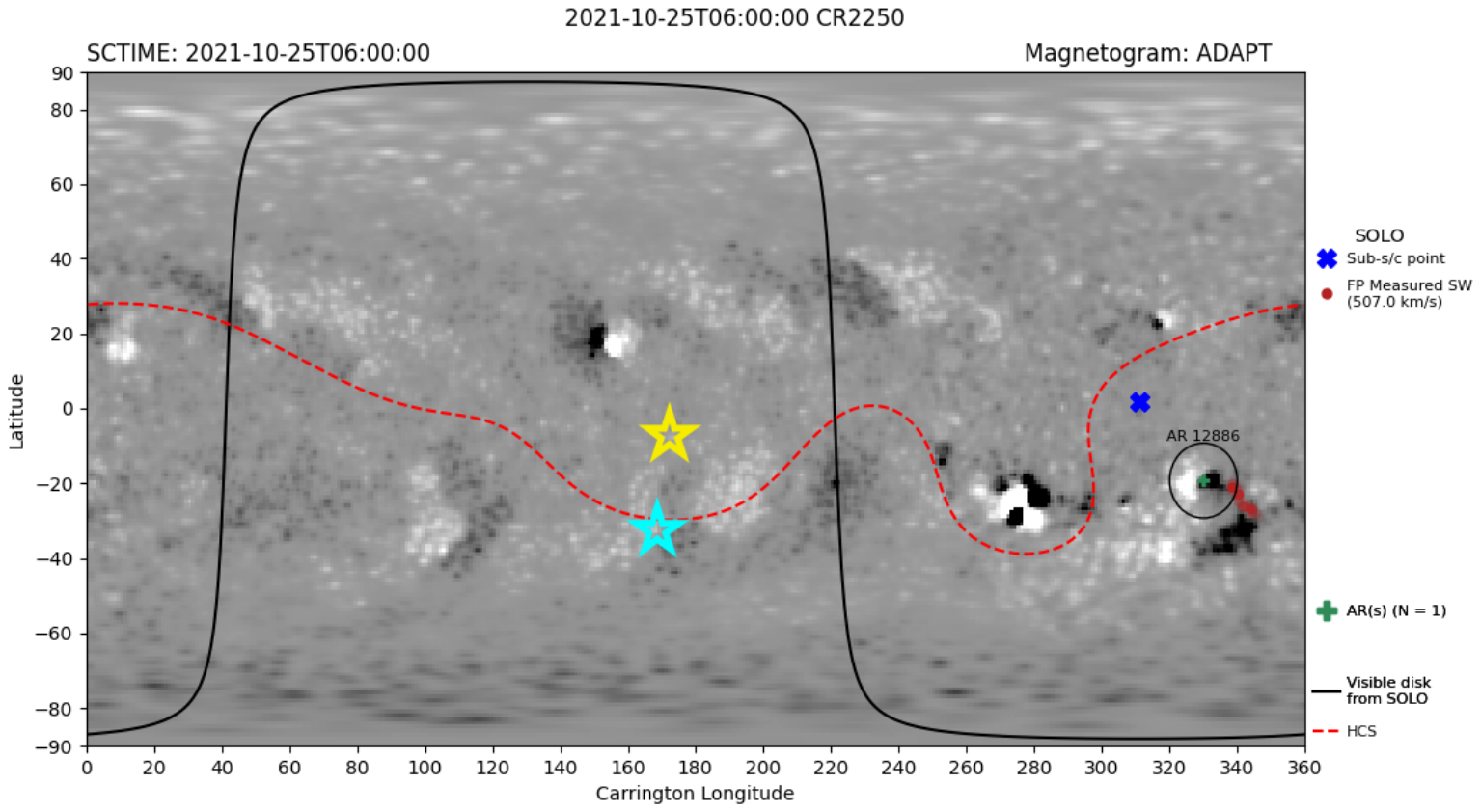}
   \includegraphics[clip, angle=-90, trim=2cm 0.5cm 3cm 3cm,width=9cm]{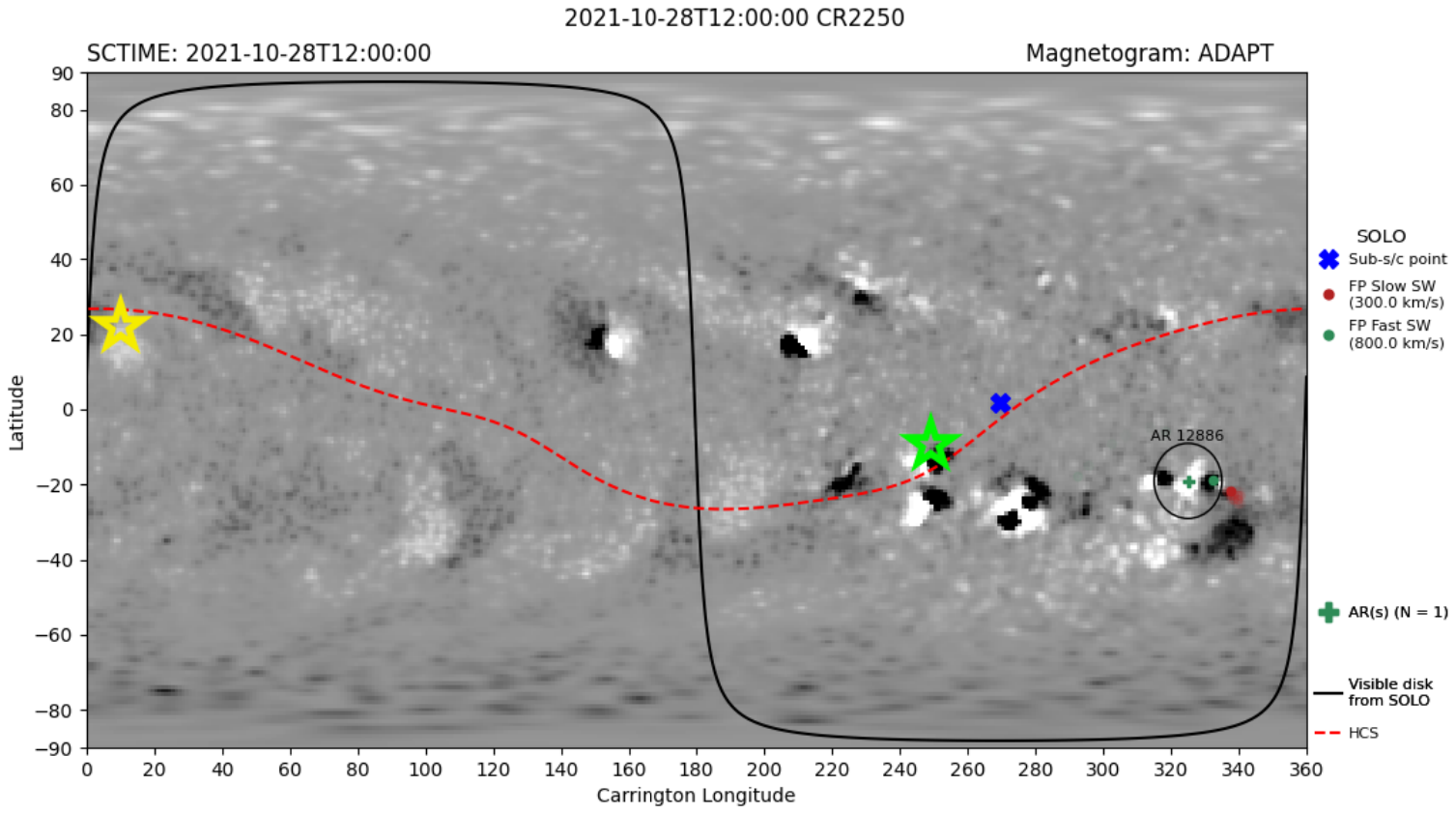}
   \caption{Synoptic maps of all the events for Carrington rotation numbers 2243, 2248, 2249, and 2250 as provided by the \href{http://connect-tool.irap.omp.eu/}{Magnetic Connectivity Tool} \citep{Connectivity_tool_2020A&A...642A...2R} at the start time of the eruption on the solar disk, as extrapolated from the height-time fit. The yellow stars indicate the Carrington coordinates of the eruptions as estimated by triangulation method in the high corona and listed in Tab.\,\ref{table:events_measured_info}. Cyan stars mark the source region of the eruptions on the disk. The green star in the last panel for the October 28$^\mathrm{th}$ magnetogram refers to the SE event and coincides with its source region.} 
              \label{fig:magnetograms}%
\end{figure*}

In Fig.\,\ref{fig:magnetograms}, we report the magnetograms for the Carrington rotations of each event as provided from the \href{http://connect-tool.irap.omp.eu/}{Magnetic Connectivity Tool} \citep{Connectivity_tool_2020A&A...642A...2R} where the corona has been modeled with the PFSS model using ADAPT maps. The black solid lines indicate the solar disk visible from SolO on the day of the eruption. The red dashed lines refer to the heliocentric current sheet (HSC). The yellow stars indicate the estimated Carrington coordinates of the eruptions as in Tab.\,\ref{table:events_measured_info}, i.e. the position of the structure in the high corona as estimated by the triangulation method. In the magnetogram of October 28$^\mathrm{th}$, the yellow star refers to the NW event, while the green star refers to the SE event. The cyan stars indicate the location of the eruption source regions on the Sun's surface, where it was possible to infer their position.

The position of the stars on the magnetogram is only indicative of the actual position from which the eruption starts on the disk, since it is extrapolated by considering that the structure is moving without any deflection or rotation and at a constant speed.

\begin{figure}
  \centering
   \includegraphics[clip, trim=4cm 1cm 4cm 0.5cm,width=7cm]{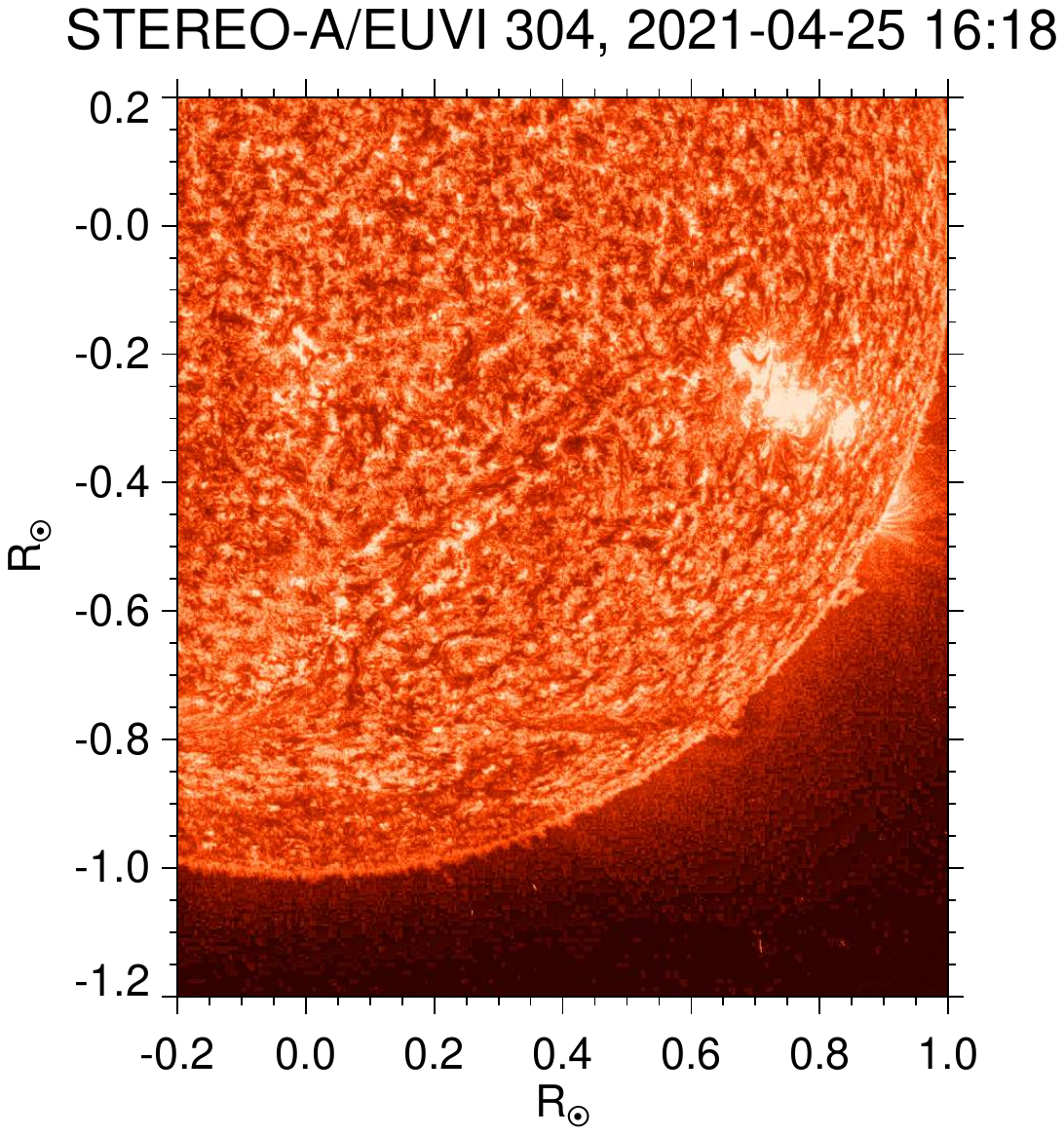}
   \caption{Solar disk at the estimated starting time of the eruption on April 25$^\mathrm{th}$ as seen by EUVI on board STEREO-A in the wavelength 304\,\r{A}. The Active Region 12820 is visible on the right and a long filament stands out darkly on the left.}
              \label{fig:EUVI_20210425-26}%
\end{figure}

\begin{figure}
  \centering
   \includegraphics[clip, trim=4cm 1cm 4cm 0.5cm,width=8cm]{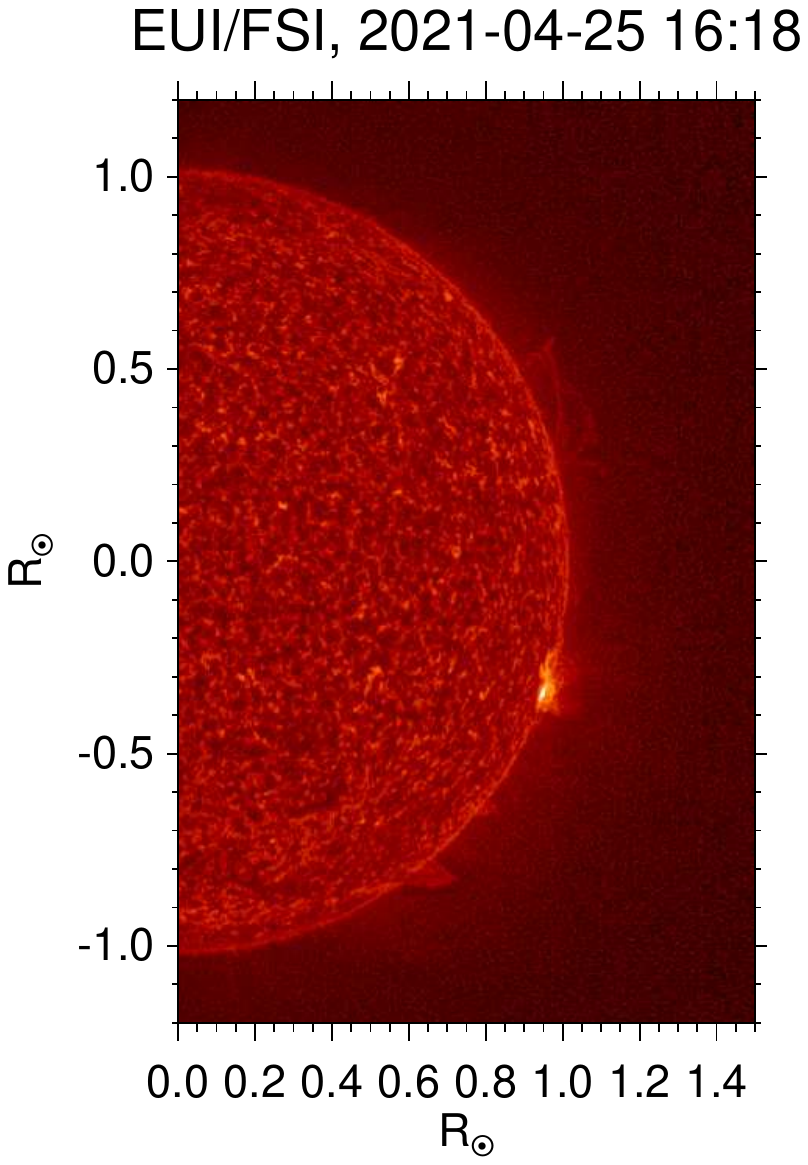}
   \caption{Solar disk at the estimated starting time of the eruption on April 25$^\mathrm{th}$ as seen by EUI-FSI in the wavelength 304\,\r{A}. The Active Region NOAA Ar.12820 is visible on the right at the solar limb activating at 16:55\,UT}
              \label{fig:EUI_FSI_20210425-26}%
\end{figure}

\begin{figure}
  \centering
   \includegraphics[clip, trim=2cm 1cm 2cm 1cm,width=8cm]{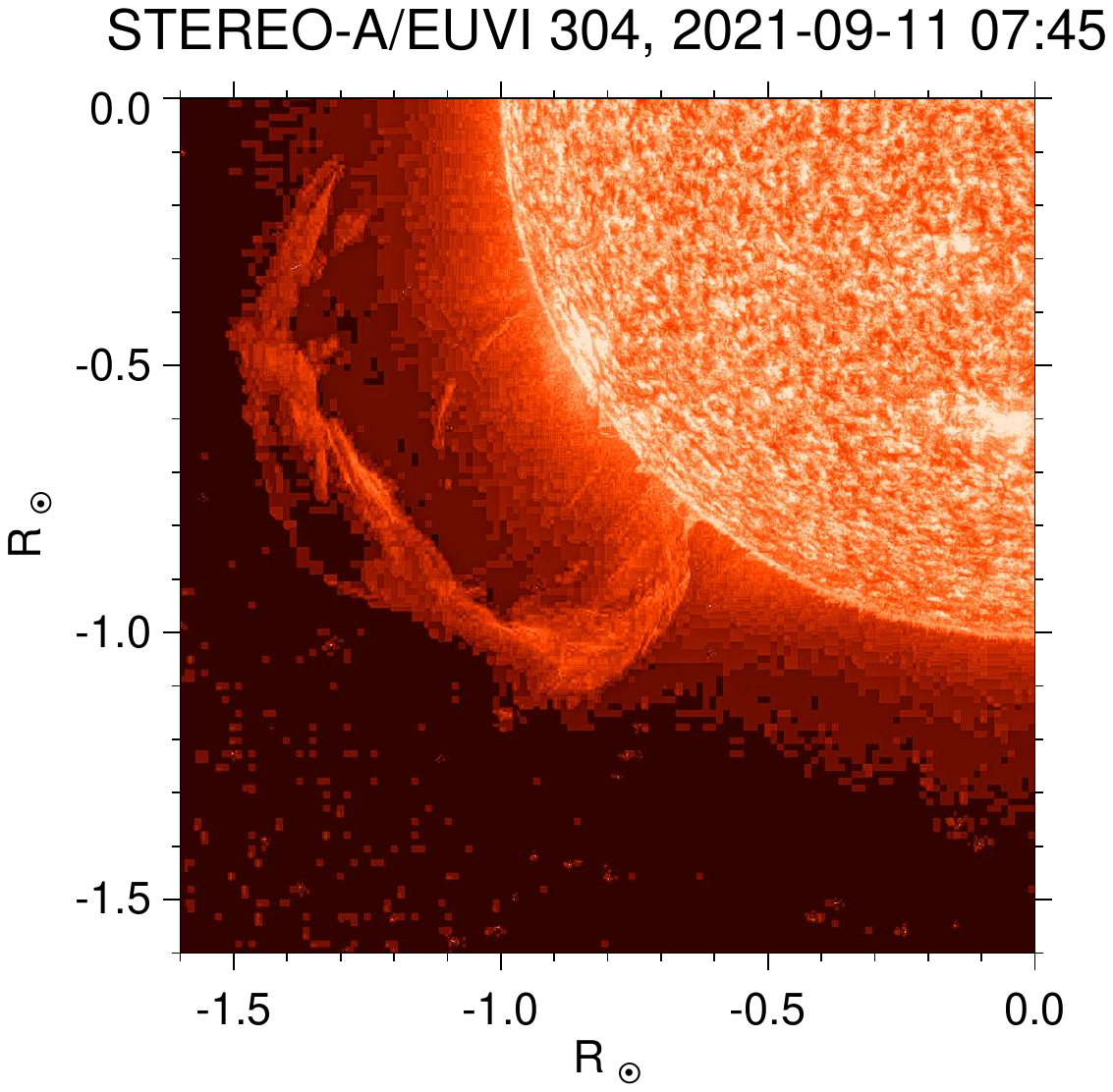}
   \caption{Solar disk at the estimated starting time of the eruption on September 11$^\mathrm{th}$ as seen by STEREO-A/EUVI in the wavelength 304\,\r{A}.}
              \label{fig:EUVI_20210911}%
\end{figure}

\begin{figure}
  \centering
   \includegraphics[clip, trim=4cm 1cm 3cm 1cm,width=7.5cm]{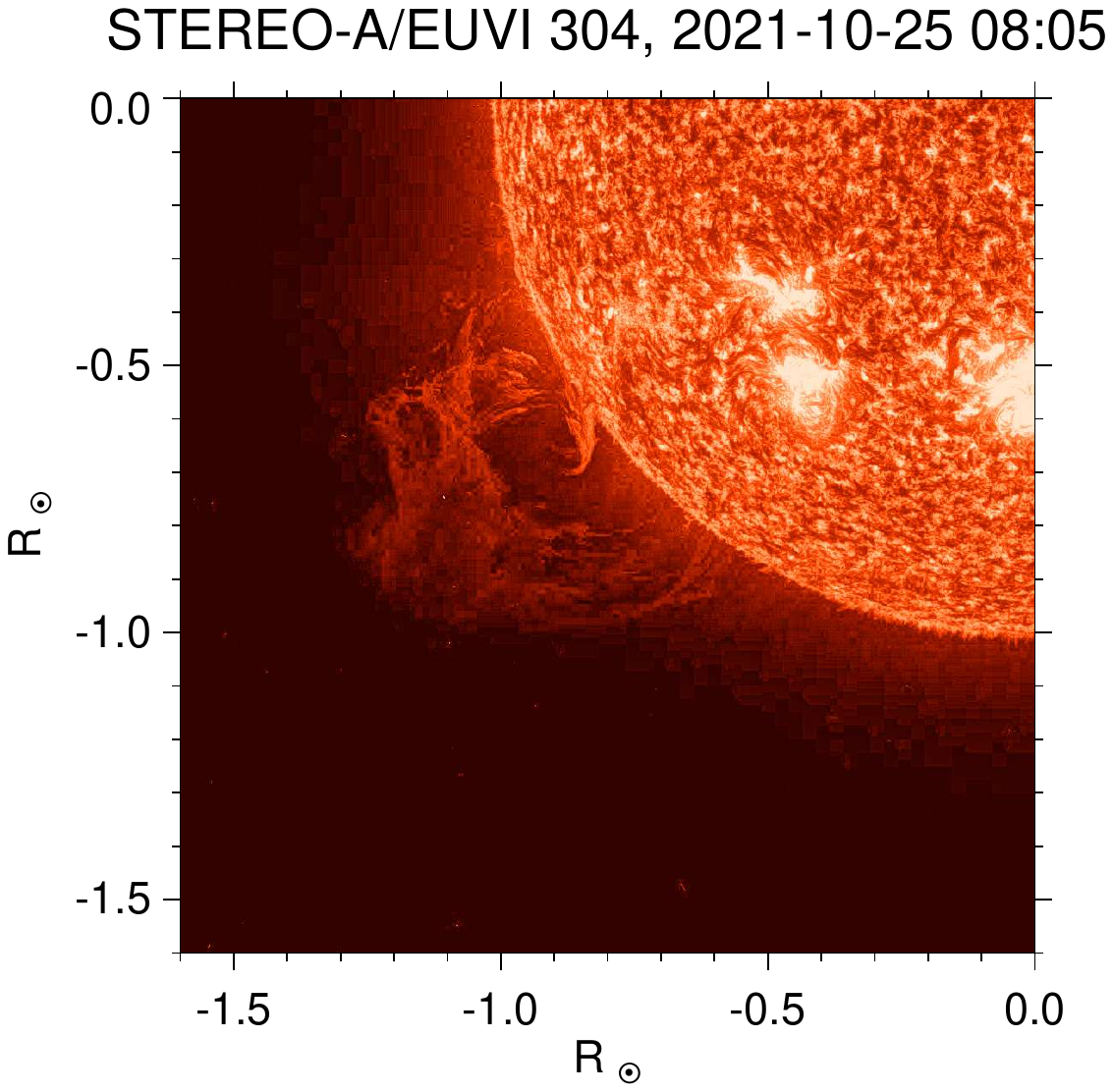}
   \caption{Solar disk at the estimated starting time of the eruption on October 25$^\mathrm{th}$ as seen by STEREO/EUVI in the wavelength 304\,\r{A}.}
              \label{fig:EUVI_20211025}%
\end{figure}

For the events on April 25$^\mathrm{th}$, September 11$^\mathrm{th}$ and October 25$^\mathrm{th}$, we identified different features at the time of the eruption at the solar disk, which could be associated with filament or prominence at the very beginning of the eruptions, as also discussed in Sec.\,\ref{sec:observations}. All the frames in Figg.\,\ref{fig:EUVI_20210425-26}, \ref{fig:EUI_FSI_20210425-26}, \ref{fig:EUVI_20210911} and \ref{fig:EUVI_20211025}, are processed with Multiscale Gaussian Normalization \citep[MGN,][]{MGN} algorithm to highlight the bright features above the solar limb.

\begin{figure*}[!ht]
  \centering
   \includegraphics[clip, trim=2cm 0.5cm 4cm 0.5cm,width=9cm]{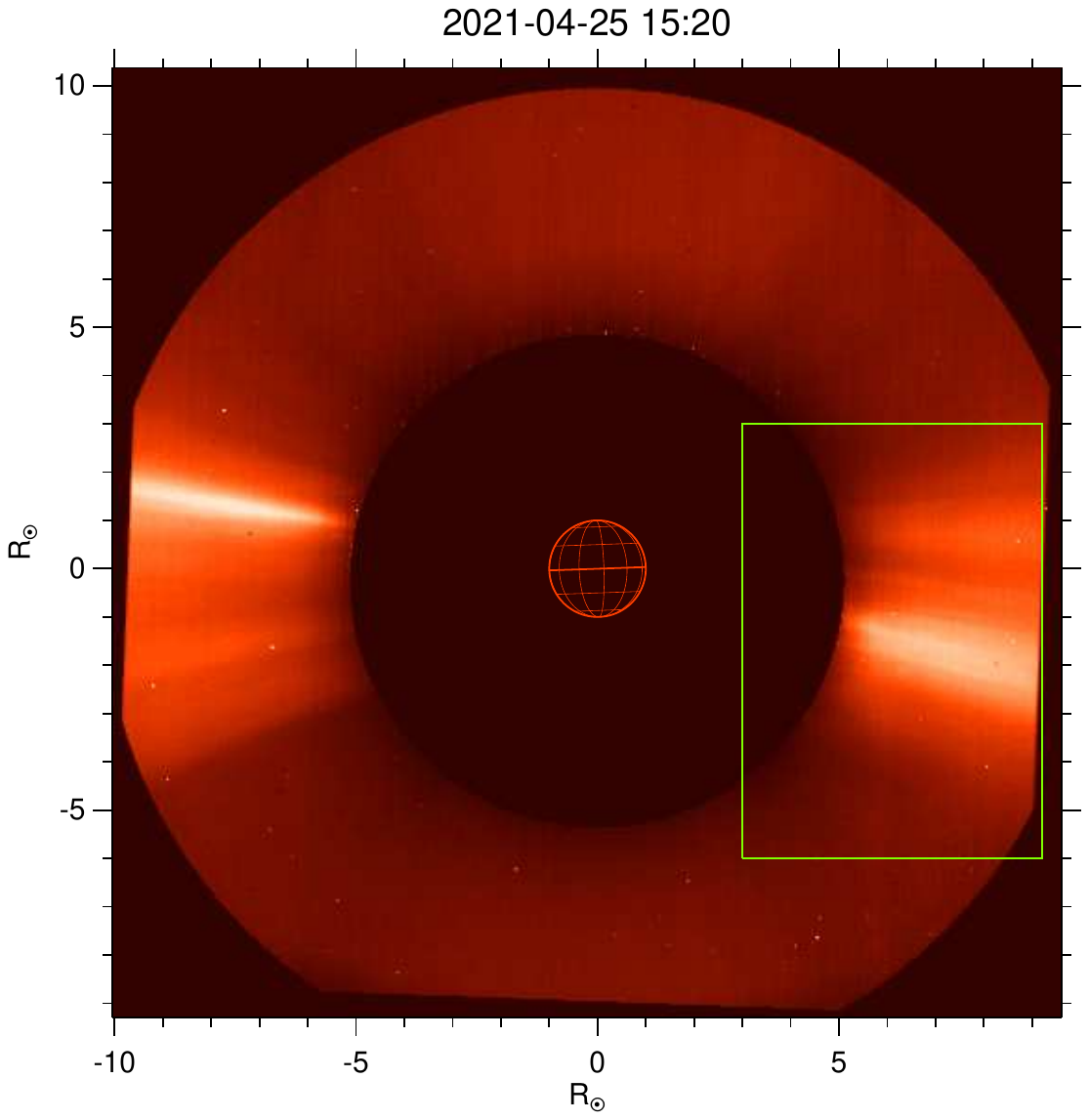} 
   \includegraphics[clip, trim=2cm 0.5cm 4cm 0.5cm,width=9cm]{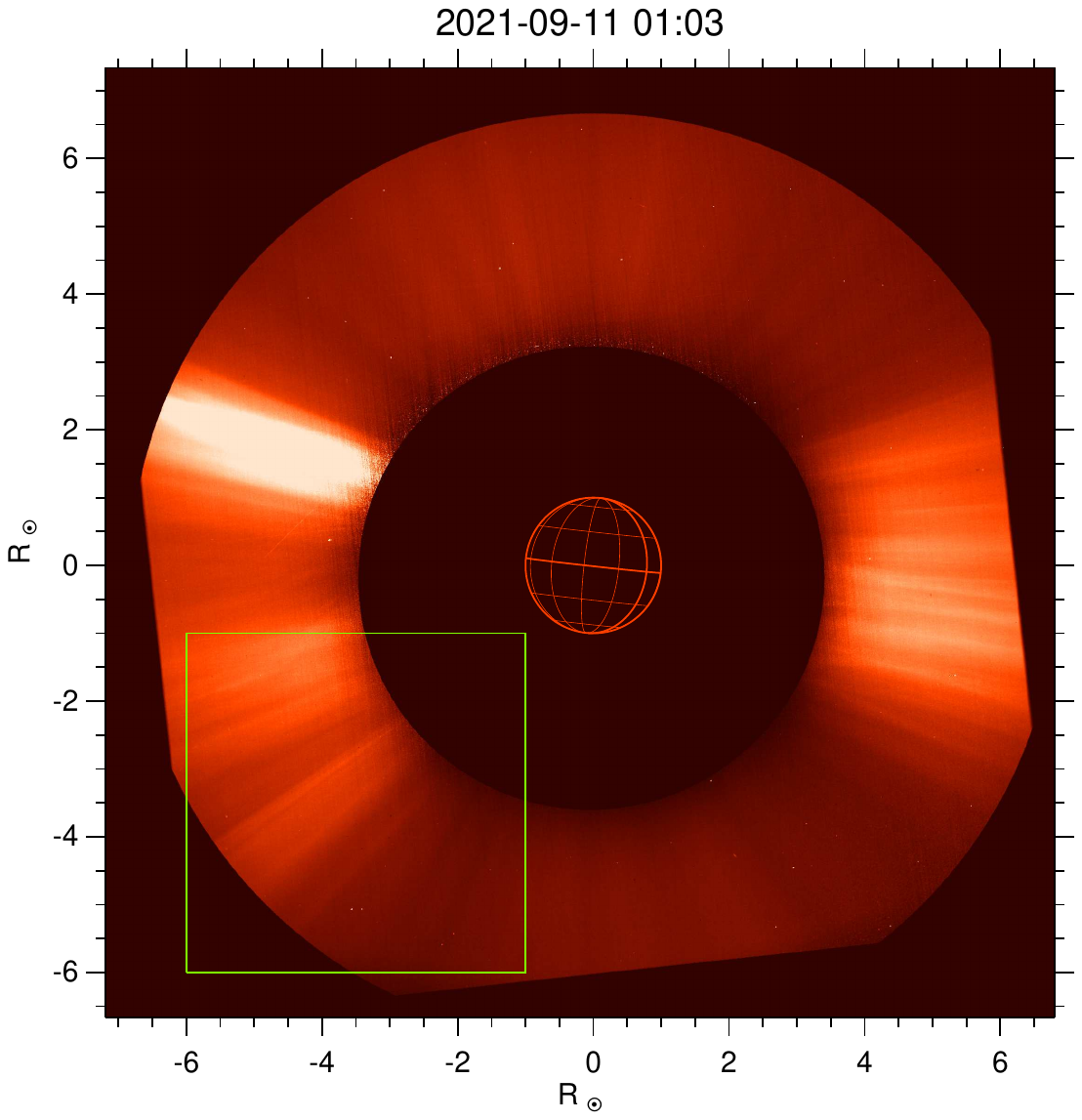} 
   \includegraphics[clip, trim=2cm 0.5cm 4cm 0.5cm,width=9cm]{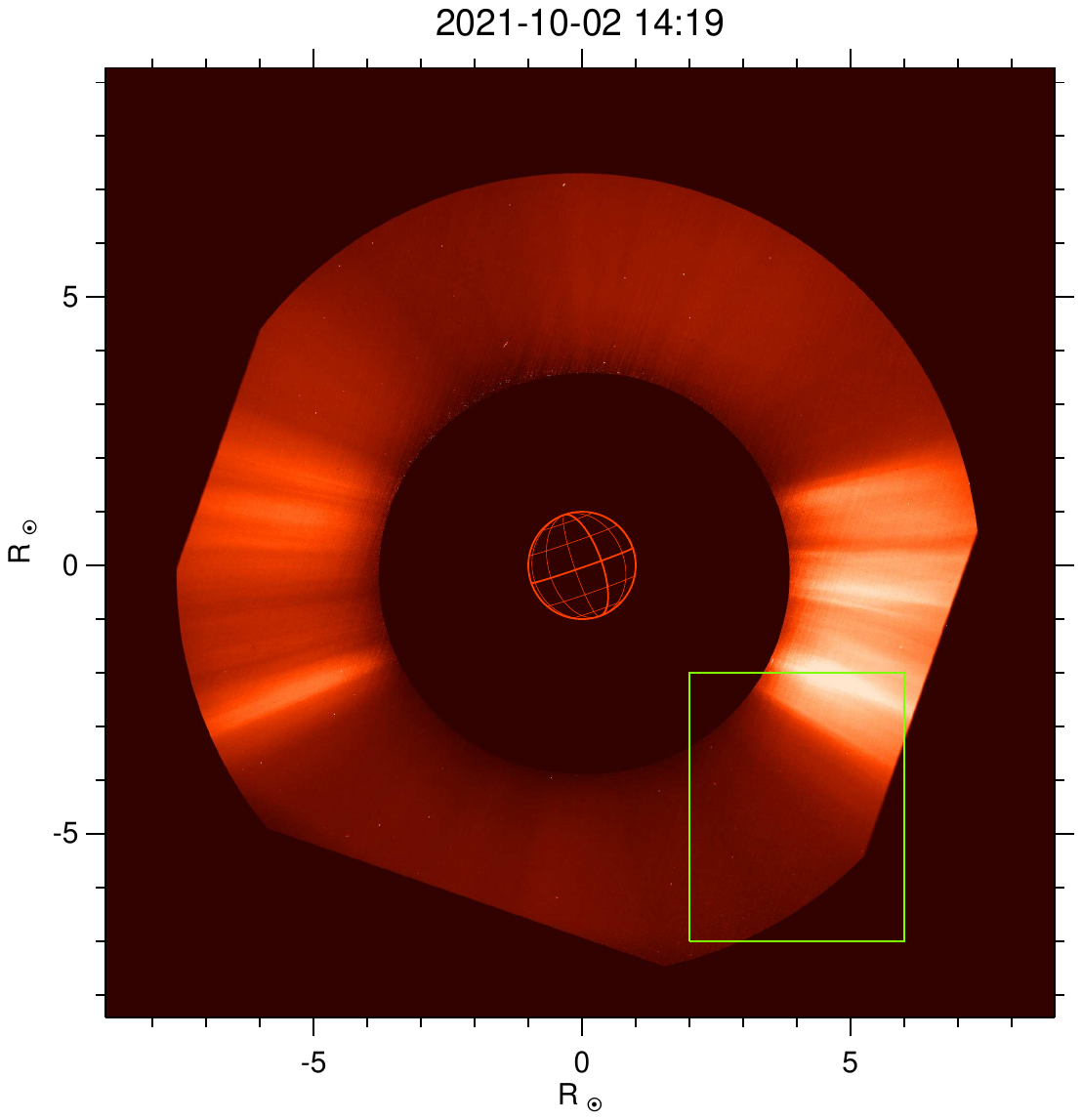}
   \includegraphics[clip, trim=2cm 0.5cm 4cm 0.5cm,width=9cm]{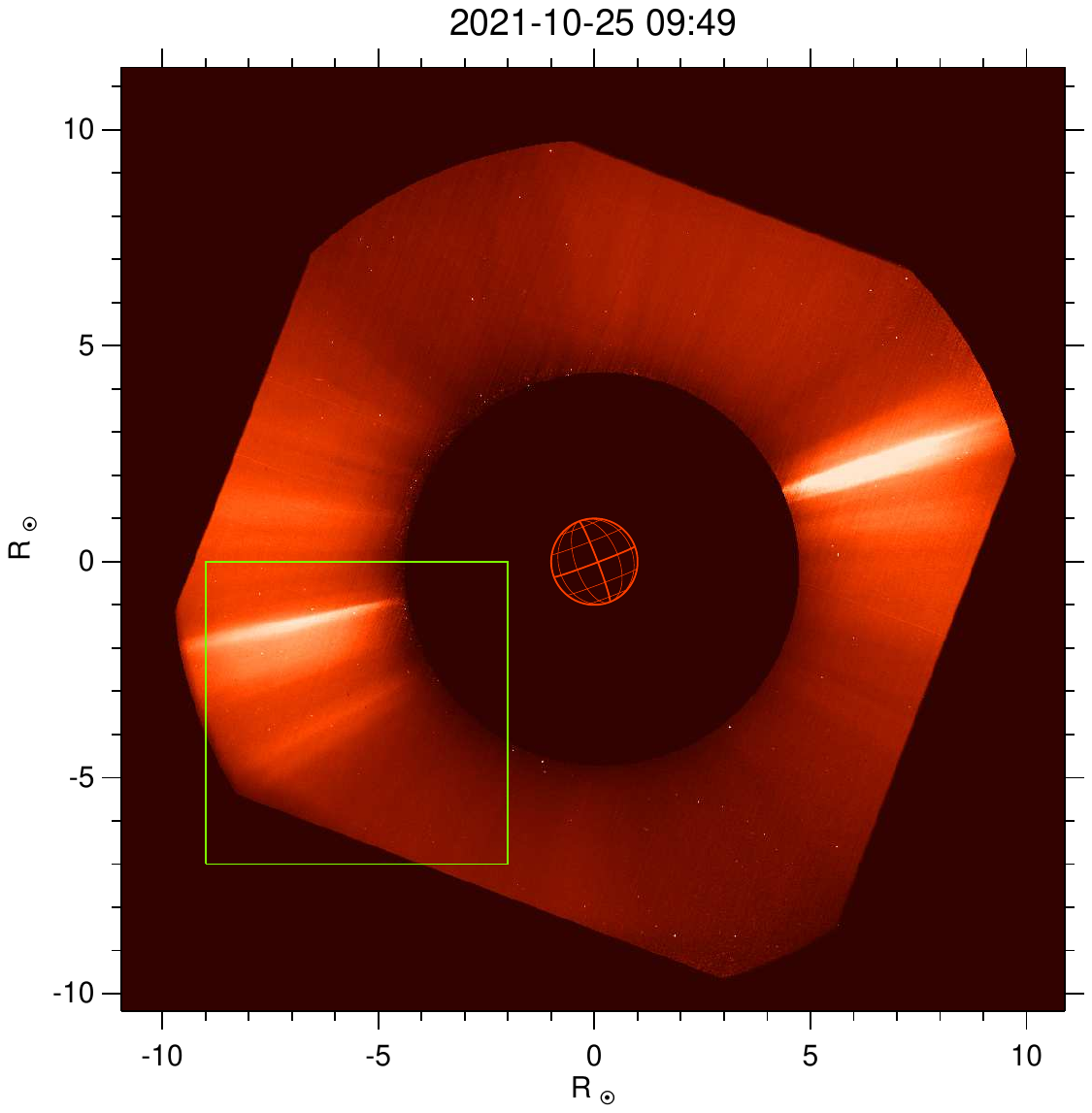}
   \includegraphics[clip, trim=2cm 0.5cm 4cm 0.5cm,width=9cm]{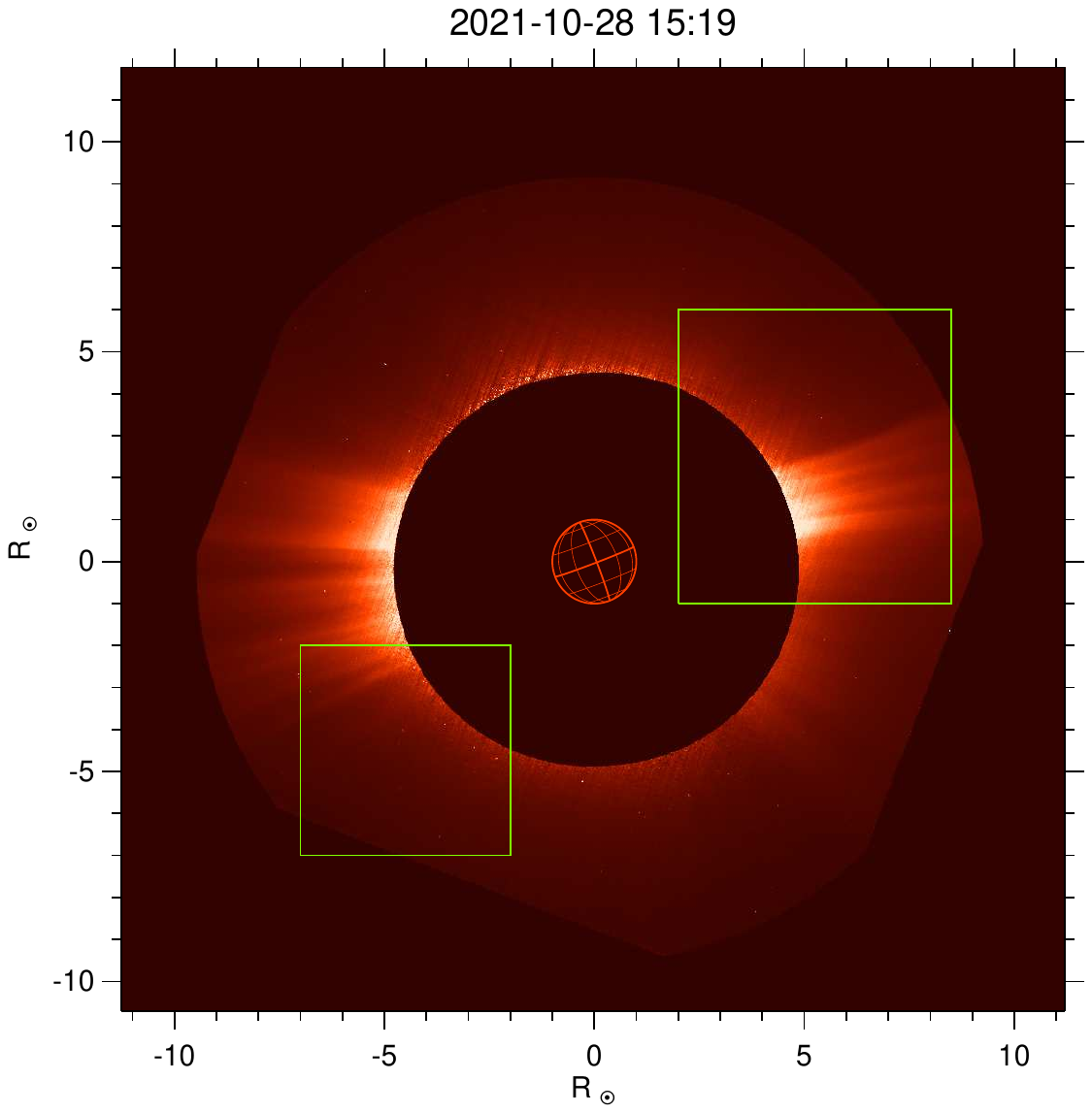}
   \caption{pB base frames for each event. These frames were used to subtract the contribution of the K-corona in the base differences shown in this work. For the October 28$^\mathrm{th}$ events the frame at 15:19\,UT is the same.} 
              \label{fig:pB_base}%
\end{figure*}

Finally, Fig.\,\ref{fig:pB_base} shows the base frame used to produce the base difference images, highlighting the region of interest where the events occurred.

%

\end{appendix}

\end{document}